\documentclass[amssymb,aps,prd,superscriptaddress,floatfix,nofootinbib]{revtex4}
\usepackage[intlimits]{amsmath}
\usepackage{amssymb}
\usepackage{graphicx}
\usepackage{tensor}
\usepackage[x11names]{xcolor}
\usepackage[colorlinks=true,linkcolor=blue,citecolor=Blue4,urlcolor=Blue4]{hyperref}
\usepackage{orcidlink}
\usepackage[caption=false]{subfig}

\usepackage[normalem]{ulem}
\usepackage{amsthm}
\usepackage{physics}
\usepackage{mathtools}

\newcommand{\sgn}[1]{\text{sgn}\left(#1\right)}

\newlength{\threefig}
\newlength{\twofig}
\renewcommand{\d}{\mathrm{d}}

\begin{document}

\title{General-relativistic Vlasov disks in external magnetic fields around Kerr black holes}

\author{Pawe{\l} Doruchowski \protect\orcidlink{0009-0002-0901-8285}}
\email{pawel.doruchowski@doctoral.uj.edu.pl}
\affiliation{Szko{\l}a Doktorska Nauk \'{S}cis{\l}ych i Przyrodniczych, Uniwersytet Jagiello\'{n}ski, \L ojasiewicza 11, 30-348 Krak\'ow, Poland}
\affiliation{Instytut Fizyki Teoretycznej, Uniwersytet Jagiello\'{n}ski, \L ojasiewicza 11, 30-348 Krak\'ow, Poland}
\author{Tomasz Grewenda}
\email{tomasz.grewenda@student.uj.edu.pl}
\affiliation{Instytut Fizyki Teoretycznej, Uniwersytet Jagiello\'{n}ski, \L ojasiewicza 11, 30-348 Krak\'ow, Poland}
\author{David Kofro\v{n} \protect\orcidlink{0000-0002-0278-7009}}
\email{d.kofron@gmail.com}
\affiliation{Institute of Theoretical Physics, Faculty of Mathematics and Physics, Charles University, V Hole\v{s}ovi\v{c}k\'{a}ch 2, 180\,00 Prague 8, Czech Republic}
\affiliation{Astronomical Institute of the Czech Academy of Sciences, Bo\v{c}n\'{\i} II 1401/1a, 141 00 Prague 4, Czech Republic}
\author{Patryk Mach \protect\orcidlink{0000-0003-1493-8668}}
\email{patryk.mach@uj.edu.pl}
\affiliation{Instytut Fizyki Teoretycznej, Uniwersytet Jagiello\'{n}ski, \L ojasiewicza 11, 30-348 Krak\'ow, Poland}

\begin{abstract}
We investigate stationary general-relativistic kinetic models of a collisionless gas around the Kerr black hole immersed in an external magnetic field. The main focus is put on razor-thin disks confined to the equatorial plane, but geometrically thick disks are also briefly mentioned. We consider distribution functions corresponding to constant canonical energy and constant canonical angular momentum of particles within the disk and obtain analytic expressions for the electric current associated with the gas. More general solutions characterized by distribution functions depending on the energy and the angular momentum are obtained assuming the Schwarzschild background metric. In explicit examples, we investigate disks in the Wald magnetosphere, describing an asymptotically uniform magnetic field.
\end{abstract}

\maketitle

\section{Introduction}

The general-relativistic kinetic theory was introduced in the late 1950s and early 1960s in the works by Synge, Tauber, Weinberg, Israel, and Lindquist \cite{Synge1957,Tauber1961,Israel1963,Lindquist1966}. Its geometrical framework was later developed by Ehlers \cite{Ehlers1971a,Ehlers1971b,Ehlers1973}. In the astrophysical context, the general-relativistic kinetic theory can be applied to model dilute gases in strong gravitational fields (plasma, dark matter \cite{Dominguez2017,Mach2021a,Mach2026a}), but also stellar systems and galaxies \cite{Shapiro1993a,Shapiro1993b}. In mathematical General Relativity, the kinetic theory provides a reliable and physically well-motivated matter model \cite{Andreasson2011a}.

The central object of the general-relativistic kinetic theory is the so-called distribution function, defined on the phase space with the aim to model the density of particles at a given location in the spacetime and a given momentum. Depending on whether the collisions between individual particles are taken into account or not, the distribution function satisfies the general-relativistic Boltzmann or Vlasov equations, respectively. Gravitational (Einstein) and electromagnetic (Maxwell) fields can be consistently taken into account \cite{Acuna2022}.

General-relativistic astrophysical plasma can also be modeled by Particle-In-Cell (PIC) simulations without making an explicit reference to the distribution function, although it is, in principle, possible to cast the resulting set of equations into a kinetic form (see, e.g.\ \cite{Touati2022} for a discussion in the non-relativistic context). Sample results of modern, general-relativistic PIC simulations of collisionless plasma can be found in \cite{Parfrey2019,Hirotani2021,Bransgrove2021,Crinquand2022,Galishnikova2023,Yuan2025,Meringolo2025,Galishnikova2026,Mehlhaff2026,Kolos2026}. Astrophysical models of collisionless plasma are particularly relevant for the description of matter around supermassive black holes, in which case the mean free path for electrons colliding with ions can be much larger than the black hole gravitational radius (see, e.g., \cite{Luepker2025} for rough estimates for M87$^\ast$).

In this work, we deal with a collisionless (Vlasov) gas in a fixed spacetime and a fixed electromagnetic field. In particular, we construct stationary kinetic models of disks around Kerr black holes embedded in a given magnetic field. In specific examples, this electromagnetic field is taken as the Wald magnetosphere \cite{Wald1974}, which, depending on the chosen form of the distribution function, leads to analytically tractable systems of equations. We consider either infinitely thin disks confined to the equatorial plane, or geometrically thick configurations. Since the former case admits a sufficient number of constants of the electrogeodesic motion, it can also be analyzed in terms of suitably defined action-angle variables.

For simplicity, in all our examples we only consider the so-called simple gas of identical particles of the same mass and charge. On the other hand, astrophysical plasma is composed of different types of particles (electrons and ions). Since we neglect the self-gravity of the gas and assume a fixed Kerr background metric, solutions describing a plasma composed of electrons and ions can, in principle, be obtained by superposing simple-gas solutions analyzed in this paper. Because a considerable effort is required to obtain quasi-neutral configurations, we postpone the discussion of this construction to future work.

Many elements of our analysis are directly related to the properties of electrogeodesics in Kerr spacetime. In particular, the chaotic motion of charged particles in the Kerr spacetime immersed in the Wald magnetosphere was analyzed in \cite{esteban1990,takahashi2009, kopacek2010,stuchlik2016}. Most of the techniques used in this work were previously developed for stationary, non-magnetized models of Vlasov gases in curved spacetimes, including models of disks around black holes. A useful overview of the tangent-bundle formulation of general-relativistic kinetic theory can be found in \cite{Sarbach2014}. Stationary kinetic disks around Schwarzschild black holes were recently constructed in \cite{Gabarrete2022,Gabarrete2023,Gabarrete2026}. Thin Vlasov disks in the equatorial plane of the Kerr black hole were studied in \cite{Rioseco2018,Cieslik2022,Mach2023,Khan2025}. An example of a thick constant angular momentum disk in the Kerr spacetime was given in \cite{Luepker2025}. Aspects of the kinetic theory in the Kerr spacetime, relevant to this work, were also covered in \cite{RS2024,Mach2026b}.

Some elements of the construction of stationary and axially symmetric Vlasov--Maxwell models can be found in the work of Thaller \cite{Thaller2020}, who proved the existence of solutions to the Einstein--Vlasov--Maxwell system (i.e., with the additional coupling to Einstein equations).

The order of this paper is as follows. In Section \ref{sec:MaxwellVlasov} we introduce elements of the description of the Vlasov--Maxwell systems in curved spacetimes. Wald's magnetosphere in the Kerr spacetime is introduced in Sec.\ \ref{sec:WaldMagnetosphere}. In Section \ref{sec:thickdisks} we construct models of thick disks characterized by either a constant energy or a constant angular momentum of the particles for arbitrary stationary axially and reflectively symmetric electromagnetic test fields on Kerr background. Specific examples are then obtained for Wald's magnetosphere. In Section \ref{sec:Thindisks} we develop models of thin disks confined to the equatorial plane of the Kerr spacetime. Since the electrogeodesic motion in the equatorial plane is integrable, this case allows for an introduction of suitably defined action-angle variables and a more precise control of the symmetries of the distribution function. Similarly to thick disks investigated in Sec.\ \ref{sec:thickdisks}, we construct models of thin disks of constant energy or constant angular momentum of the particles. Explicit examples of thin disks in the Wald magnetosphere are given in Sec.\ \ref{sec:thinWald}. In Section \ref{sec:SchwarzschildLimit} we discuss thin disks in the Schwarzschild limit. In this case, the effective radial potential associated with the motion of charged particles in Wald's electromagnetic field is sufficiently simple to allow for a qualitative analysis, which in turn allows for a construction of disks composed of particles with varying energies and angular momenta. Section \ref{sec:Conclusions} contains concluding remarks.

\section{Maxwell--Vlasov systems}
\label{sec:MaxwellVlasov}

\subsection{Electrogeodesic motion}

The equations of motion of a charged spinless test particle moving in a given spacetime $(\mathcal M, g)$ and the electromagnetic field (the so-called electrogeodesic equations) can be written in the Hamiltonian form
\begin{equation}
\label{hamilton}
\frac{\d x^\mu}{\d \tau} = \frac{\partial H}{\partial \Pi_\mu}, \qquad \frac{\d \Pi_\nu}{\d \tau} = - \frac{\partial H}{\partial x^\nu},
\end{equation}
where the Hamiltonian is defined as
\begin{equation}
\label{hamiltonian}
H \coloneqq \frac{1}{2} g^{\mu \nu}(x) [\Pi_\mu - q A_\mu(x)] [\Pi_\nu - q A_\nu(x)]. \end{equation}
Here, $q$, $A_\mu$, and $\Pi_\mu$ denote the particle charge, the components of the electromagnetic vector potential, and the canonical momenta, respectively. The kinetic momentum components $p_\mu$ are defined as $p_\mu \coloneqq \Pi_\mu - q A_\mu(x)$, such that
\begin{equation} H = \frac{1}{2}g^{\mu \nu}(x) p_\mu p_\nu. \end{equation}
In Equations (\ref{hamilton}) we assume that  electrogeodesics are parametrized as $\tau \mapsto x(\tau) \in \mathcal M$, and the affine parameter $\tau$ is related to the proper time $s$ by $\tau = s/m$, where $m$ denotes the rest mass of the particle. Thus, $H = - m^2/2$, and $p^\mu = m \d x^\mu/\d s$.

In explicit terms, Eqs.\ (\ref{hamilton}) read
\begin{eqnarray}
\frac{\d x^\mu}{\d\tau} & = & g^{\mu \alpha} p_\alpha, \\
\frac{\d\Pi_\nu}{\d \tau} & = & - \frac{1}{2} \frac{\partial g^{\alpha \beta}}{\partial x^\nu} p_\alpha p_\beta + q g^{\alpha \beta} \frac{\partial A_\alpha}{\partial x^\nu} p_\beta.
\end{eqnarray}
A straightforward calculation shows that they imply the following second-order equation:
\begin{equation}
\label{eqsmotion}
\frac{\d^2 x^\mu}{\d\tau^2} + \Gamma\indices{^\mu_{\alpha \beta}}\frac{\d x^\alpha}{\d\tau} \frac{\d x^\beta}{\d \tau} = q F\indices{^\mu_\alpha} \frac{\d x^\alpha}{\d \tau},
\end{equation}
where
\begin{equation}
F_{\mu \nu} \coloneqq \frac{\partial A_\nu}{\partial x^\mu} - \frac{\partial A_\mu}{\partial x^\nu}
\end{equation}
is the electromagnetic tensor.

\subsection{One-particle distribution function}

In the kinetic framework, the state of the gas is described by the distribution function, defined on the one-particle phase space, which, depending on whether momenta are represented as vectors or covectors, can be modeled by the tangent or cotangent bundle on $(\mathcal M, g)$. In this work, we prefer to use the cotangent-bundle formulation, as it is more suited to the Hamiltonian framework introduced above.

The cotangent bundle on $(\mathcal{M}, g)$ is defined as
\begin{equation}
T^\ast \mathcal M \coloneqq \{ (x, p) \colon x \in \mathcal{M}, \,  p \in T_x^\ast \mathcal M \},  \end{equation}
where $T_x^\ast \mathcal M$ denotes the cotangent space to $\mathcal{M}$ at the point $x$. The one-particle distribution function can be defined as $\mathcal F \colon U \to [0,+\infty)$, where $U \subseteq T^\ast \mathcal M$. It has the following statistical interpretation. Let $S$ be a three-dimensional spacelike surface in $\mathcal M$. Define $\mathcal N[S]$ as an averaged number of particle trajectories in $U$, whose projections on $\mathcal M$ intersect the surface $S$. One can show that (cf.\ \cite{Acuna2022}, Eqs.\ (94))
\begin{equation}
\mathcal N[S] = - \int_S \left[ \int_{P_x^+} \mathcal F(x, p) p_\mu s^\mu \mathrm{dvol}_x( p) \right] \eta_S,
\end{equation}
where $ s$ is a unit future-directed vector normal to $S$, $\eta_S$ denotes the three-dimensional volume element induced on $S$,
\begin{equation}
P_x^+ \coloneqq \{ p \in T_x^\ast \mathcal M \colon g^{\mu \nu} p_\mu p_\nu < 0, \,  p \text{ is future-directed} \},
\end{equation}
and $\mathrm{dvol}_x( p)$ denote the volume element in $P_x^+$, given by
\begin{equation}\label{dvolp}
\mathrm{dvol}_x( p) \coloneqq \sqrt{- \mathrm{det} g^{\mu \nu}(x)} \d p_0 \d p_1 \d p_2 \d p_3 \equiv \sqrt{- \mathrm{det} g^{\mu \nu}(x)} \d^4 p.
\end{equation}
Defining the so-called particle current density
\begin{equation}
\label{Jmu}
\mathcal J_\mu(x) \coloneqq \int_{P_x^+} \mathcal F(x, p) p_\mu \mathrm{dvol}_x( p),
\end{equation}
one can write the expression for $\mathcal N[S]$ as
\begin{equation}
\label{eq:ParticleNrCrossing}
\mathcal N[S] = - \int_S \mathcal J_\mu s^\mu \eta_S,
\end{equation}
i.e., the flux of the particle current density through $S$. The energy-momentum tensor of the Vlasov gas is defined as
\begin{equation}
\label{vlasovemtensor}
\mathcal T_{\mu \nu} (x) \coloneqq \int_{P_x^+} \mathcal F(x, p) p_\mu p_\nu \mathrm{dvol}_x( p). 
\end{equation}

In what follows, we will restrict ourselves to the so-called simple gas composed of same-mass particles satisfying the mass shell condition. In this case, it is sometimes convenient to introduce the volume element induced on the future mass hyperboloid 
\begin{equation}
P_x^+(m) \coloneqq \{ p \in T_x^\ast \mathcal M \colon g^{\mu \nu} p_\mu p_\nu = -m^2, \,  p \text{ is future-directed} \}.
\end{equation}
The restriction of momenta to $P_x^+(m)$ can be done by assuming
\begin{equation}
\mathcal F(x;p_0,p_1,p_2,p_3) = \delta \left( \sqrt{-p_\mu p^\mu} - m \right) \mathcal F_0(x;p_1,p_2,p_3), \end{equation}
for some function $\mathcal F_0(x;p_1,p_2,p_3)$. In this case, integrals of the form
\begin{equation}
I = \int_{P_x^+} \Phi(x;p_0,p_1,p_2,p_3) \mathcal F (x;p_0,p_1,p_2,p_3) \mathrm{dvol}_x( p),
\end{equation}
where $\Phi(x, p)$ denotes a test function, 
can be expressed as
\begin{eqnarray}
I & = & \int_{P_x^+} \Phi(x;p_0,p_1,p_2,p_3) \mathcal F_0 (x;p_1,p_2,p_3) \frac{m \delta(p_0 - \bar p_0)}{|\bar p^0|} \mathrm{dvol}_x( p) \nonumber \\
& = & \int_{P_x^+(m)} \Phi(x;\bar p_0,p_1,p_2,p_3) \mathcal F_0 (x;p_1,p_2,p_3) m \sqrt{- \mathrm{det} \, g^{\mu \nu}} \frac{\d p_1 \d p_2 \d p_3}{|\bar p^0|}.
\end{eqnarray}
Here $\bar p_0$ denotes the time component of the future-directed momentum four-vector, satisfying
\begin{equation}
g^{00}(\bar p_0)^2 + 2 g^{0i}\bar p_0 p_i + g^{ij}p_ip_j = -m^2.
\end{equation}
For simplicity, $\bar p_0$ will be further denoted simply as $p_0$. Thus, the volume element on $P_x^+(m)$ can be written as
\begin{equation}\label{eq:volElementMomenta}
\mathrm{dvol}^{(m)}_x( p) =  m \sqrt{- \mathrm{det} \, g^{\mu \nu}} \frac{\d p_1 \d p_2 \d p_3}{|p^0|} = m \sqrt{- \mathrm{\det} \, g_{\mu \nu}} \frac{\d p^1 \d p^2 \d p^3}{|p_0|}.
\end{equation}
This result can also be obtained in an elegant geometrical fashion. The derivation can be found in \cite{Acuna2022}. For completeness, we give it also in Appendix \ref{appvolume}.

\subsection{Vlasov equation}

If collisions between individual particles of the gas can be neglected, the distribution function satisfies the following (Vlasov) equation
\begin{equation}
\label{vlasovgeneral}
\frac{\d \mathcal F(x(\tau),\Pi(\tau))}{\d \tau} = \frac{\partial \mathcal F}{\partial x^\mu} \frac{\d x^\mu}{\d \tau} + \frac{\partial \mathcal F}{\partial \Pi_\nu} \frac{\d \Pi_\nu}{\d \tau} = \frac{\partial \mathcal F}{\partial x^\mu} \frac{\partial H}{\partial \Pi_\mu} -\frac{\partial \mathcal F}{\partial \Pi_\nu} \frac{\partial H}{\partial x^\nu} = 0,
\end{equation}
expressing the requirement that $\mathcal F$ should remain constant along electrogeodesics (i.e., characteristic curves of Eq.\ (\ref{vlasovgeneral}) coincide with solutions of Eq.\ (\ref{hamilton})). Note that both $(x^\mu,p_\nu)$ and $(x^\mu,\Pi_\nu)$ can be regarded as local coordinates on $T^\ast \mathcal M$. In the coordinates $(x^\mu,p_\nu)$, one obtains
\begin{equation}
\label{vlasovem}
\frac{\d \mathcal F(x(\tau),p(\tau))}{\d \tau} = \frac{\partial \mathcal F}{\partial x^\mu} \frac{\d x^\mu}{\d \tau} + \frac{\partial \mathcal F}{\partial p_\nu} \frac{\d p_\nu}{\d \tau} = g^{\mu \nu} p_\nu \frac{\partial \mathcal F}{\partial x^\mu} - \frac{1}{2}  \frac{\partial g^{\alpha \beta}}{\partial x^\mu} p_\alpha p_\beta \frac{\partial \mathcal F}{\partial p_\mu} + q \frac{\partial \mathcal F}{\partial p_\mu} g^{\alpha \beta} p_\beta F_{\mu \alpha} = 0.
\end{equation}
We note that any distribution function $\mathcal F$ that depends on the coordinates $(x^\mu,\Pi_\nu)$ via constants of the electrogeodesic motion $I_1(x,\Pi), \dots, I_s(x,\Pi)$, i.e., $\mathcal F = \mathcal F_0(I_1(x,\Pi),\dots,I_s(x,\Pi))$ for some function $\mathcal F_0$, satisfies the Vlasov equation (\ref{vlasovgeneral}).

It follows from Eq.\ (\ref{vlasovgeneral}) that the particle current density defined in Eq.\ (\ref{Jmu}) is conserved, i.e., that $\nabla_\mu \mathcal J^\mu = 0$ (even in the presence of the electromagnetic field). The sketch of the proof can be found in Appendix \ref{appconservation}. The conservation of the particle current density allows us to define the electric current vector as
\begin{equation}
\label{elcurrent}
j_\mu(x) \coloneqq q \mathcal J_\mu(x),
\end{equation}
so that the electromagnetic field tensor $F_{\mu\nu}$ satisfies the Maxwell equations\footnote{The Hodge star of a  2-form $\omega$ is defined as
\[ (\ast \omega)_{\alpha \beta} \coloneqq \frac{1}{2} \omega_{\mu \nu} \epsilon\indices{^\mu^\nu_\alpha_\beta}, \] where $\epsilon_{\mu\nu\alpha\beta} \coloneqq \sqrt{-\mathrm{det} \, g_{\rho \sigma}}[\mu,\nu,\alpha,\beta]$ denote the components of the Levi-Civita alternating tensor in the right-handed basis, and $[\mu,\nu,\alpha,\beta]$ is equal to 1 (respectively, $-1$) if $(\mu,\nu,\alpha,\beta)$ is an even (respectively, odd) permutation of $(0,1,2,3)$, and 0 otherwise.}
\begin{equation}
\nabla_\nu \, {\ast F}^{\mu \nu} = 0, \qquad \nabla_\nu F^{\mu \nu} = 4 \pi j^\mu.
\end{equation}

Taking into account the self-gravity of the system (coupling with the Einstein equations) requires the energy-momentum tensor of the Vlasov--Maxwell system, satisfying $\nabla_\nu T^{\mu \nu} = 0$. It can be shown (see again Appendix \ref{appconservation}) that
\begin{equation}
\nabla_\nu \mathcal T^{\mu\nu} = F^{\mu\nu} j_\nu.
\end{equation}
This fact allows for a consistent coupling with the Einstein equations. For the energy-momentum tensor of the electromagnetic field,
\begin{equation}
T^\mathrm{(EM)}_{\mu \nu} \coloneqq \frac{1}{4 \pi} \left(F\indices{_\mu^\alpha} F_{\nu \alpha} - \frac{1}{4} g_{\mu \nu} F^{\alpha \beta} F_{\alpha \beta} \right),
\end{equation}
one has
\begin{equation}
\nabla_\nu T_\mathrm{(EM)}^{\mu \nu} = j_\nu F\indices{^\nu^\mu}.
\end{equation}
Thus, defining the total energy momentum tensor as
\begin{equation}
T^\mathrm{(tot)}_{\mu\nu} \coloneqq \mathcal T_{\mu\nu} + T^\mathrm{(EM)}_{\mu\nu},
\end{equation}
we obtain $\nabla_\nu T_\mathrm{(tot)}^{\mu \nu} = 0$.

\section{Wald's magnetosphere}
\label{sec:WaldMagnetosphere}

\subsection{Kerr spacetime}

We will work in the Boyer--Lindquist coordinates $(t,r,\vartheta,\varphi)$, in which the Kerr metric can be written in the form
\begin{equation}
\label{kerr4d}
    g = g_{\mu \nu} \d x^\mu \d x^\nu = g_{t t} \d t^2 + 2 g_{t \varphi} \d t \d\varphi + g_{r r} \d r^2 + g_{\vartheta \vartheta} \d\vartheta^2 + g_{\varphi \varphi} \d\varphi^2,
\end{equation}
with
\begin{subequations}
\begin{eqnarray}
g_{t t} & = & -1 + \frac{2Mr}{\rho^2}, \\
g_{t \varphi} & = & - \frac{2 M a r \sin^2 \vartheta}{\rho^2}, \\
g_{r r} & = & \frac{\rho^2}{\Delta}, \\
g_{\vartheta \vartheta} & = & \rho^2, \\
g_{\varphi \varphi} & = & \left( r^2 + a^2 + \frac{2 M a^2 r \sin^2 \vartheta}{\rho^2} \right) \sin^2 \vartheta,
\end{eqnarray}
\end{subequations}
where
\begin{subequations}
\begin{eqnarray}
\Delta & \coloneqq & r^2 - 2 M r + a^2, \\
\rho^2 & \coloneqq & r^2 + a^2 \cos^2 \vartheta,
\end{eqnarray}
\end{subequations}
and where $M$ and $a$ denote the black hole mass and the spin parameter, respectively. The black hole angular momentum reads $J = a M$. The contravariant metric components have the form
\begin{eqnarray}
    g^{tt} & = & - \frac{1}{\Delta} \left( r^2 + a^2 + \frac{2 M r a^2 \sin^2 \vartheta}{\rho^2} \right), \\
    g^{t \varphi} & = & - \frac{2 M a r}{\Delta \rho^2}, \\
    g^{rr} & = & \frac{\Delta}{\rho^2}, \\
    g^{\vartheta \vartheta} & = & \frac{1}{\rho^2}, \\
    g^{\varphi \varphi} & = & \frac{1}{\Delta \rho^2} \left( -a^2 + \frac{\Delta}{\sin^2 \vartheta} \right).
\end{eqnarray}
We assume $M > 0$ and $-M < a < M$. Zeros of $\Delta$, corresponding to the locations of the inner and outer Kerr horizons, will be denoted by $r_\pm \coloneqq M \pm \sqrt{M^2 - a^2}$.

A vector
\begin{equation}
\label{vecN}
    N \coloneqq -g^{t\mu}\partial_\mu = \frac{1}{\Delta} \left( r^2 + a^2 + \frac{2 M r a^2 \sin^2 \vartheta}{\rho^2} \right) \partial_t + \frac{2 M r a}{\Delta \rho^2} \partial_\varphi
\end{equation}
is normal to hypersurfaces of constant time $t = \mathrm{const}$. The corresponding covector has the components $N_\mu = (-1,0,0,0)$, so that
\begin{equation}
    N_\mu N^\mu = g^{tt} = - \frac{1}{\Delta} \left( r^2 + a^2 + \frac{2 M r a^2 \sin^2 \vartheta}{\rho^2} \right).
\end{equation}
Thus, $N$ is timelike for $r > r_+$, and it can be used to define the time orientation outside the black hole horizon. We assume that $N$ is directed to the future for $r > r_+$.

The Kerr metric admits two Killing vectors, expressed in the Boyer--Lindquist coordinates $(t,r,\vartheta,\varphi)$ as $k \coloneqq \partial_t$ and $\eta \coloneqq \partial_\varphi$, or equivalently,
\begin{equation}
\label{kerrkilling}
    k^\mu = (1,0,0,0), \qquad \eta^\mu = (0,0,0,1).
\end{equation}

The Komar charges associated with the vectors $k$ and $\eta$ can be identified with the mass $M$ and the angular momentum $J = a M$. More precisely \cite{WaldBook},
\begin{equation}
\label{komar}
    M = - \frac{1}{8 \pi} \int_S \ast \d k, \qquad J = \frac{1}{16 \pi} \int_S \ast \d \eta,
\end{equation}
where $S$ denotes a closed 2-surface embedded in a hypersurface of constant time $t$ and enclosing the black hole.

In the following, it will be convenient to work in an orthonormal frame, adapted to Boyer--Lindquist coordinates. We consider, in particular, the following tetrad, introduced by Bardeen \cite{Bardeen1972}:
\begin{equation}
\theta_{(t)} \coloneqq \frac{1}{\rho}\sqrt{\frac{A}{\Delta}} \partial_t + \frac{2 M a r}{\rho \sqrt{A \Delta}} \partial_\varphi, \qquad \theta_{(r)} \coloneqq \frac{\sqrt{\Delta}}{\rho} \partial_r, \qquad \theta_{(\vartheta)} \coloneqq \frac{1}{\rho} \partial_\vartheta, \qquad \theta_{(\varphi)} \coloneqq \frac{1}{\sin \vartheta} \frac{\rho}{\sqrt{A}} \partial_\varphi,
\label{eq:frame}
\end{equation}
where $A \coloneqq (r^2 + a^2)^2 - a^2 \Delta \sin^2 \vartheta$. Note that the vector $\theta_{(t)} = N/\sqrt{-N_\mu N^\mu}$ is future-directed. Contravariant momentum components with respect to this frame read:
\begin{equation}
p^{(t)} = \rho \sqrt{\frac{\Delta}{A}} p^t, \quad p^{(r)} = \frac{\rho}{\sqrt{\Delta}} p^r, \quad p^{(\vartheta)} = \rho p^\vartheta, \quad p^{(\varphi)} = - \frac{2 a M r \sin \vartheta}{\rho \sqrt{A}} p^t + \frac{\sin \vartheta \sqrt{A}}{\rho} p^\varphi = \frac{\sin \vartheta}{\rho \sqrt{A}} \left( - 2 a M r p^t + A p^\varphi \right).
\end{equation}

\subsection{Wald's magnetosphere in the Kerr spacetime}

The Wald magnetosphere belongs to a family of solutions to the Maxwell equations in vacuum spacetimes that can be constructed out of spacetime symmetries---Killing vectors or Killing--Yano tensors (see, e.g., \cite{Achour2026} for a recent discussion). The Wald magnetosphere is constructed by setting \cite{Wald1974}
\begin{equation}
    A_\mu = \frac{1}{2}B_0(\eta_\mu + \kappa   k_\mu),
\end{equation}
where $B_0$ and $\kappa$ are constant. In vacuum spacetimes, it corresponds to a vanishing electric current $j^\mu \equiv 0$. Indeed, one can show that for $F_{\mu \nu} = \nabla_\mu A_\nu - \nabla_\nu A_\mu$,
\begin{equation}
    4 \pi j^\mu  = \nabla_\nu F^{\mu \nu} = 2 R\indices{^\mu_\nu} A^\nu = 0,
\end{equation}
where $R\indices{^\mu_\nu}$ denotes the Ricci tensor.

The electric charge $Q$ associated with the potential $A_\mu$ can be computed as
\begin{equation}
    4 \pi Q = \int_S \ast \d A = \frac{1}{2} B_0 \int_S \ast \d \eta + \frac{\kappa}{2} B_0 \int_S \ast \d k  = 4 \pi B_0 (2J - \kappa M),
\end{equation}
where in the last step, we have used Eqs.\ (\ref{komar}). Thus,
\begin{equation}
\label{kappacharge}
    \kappa = \frac{2 J}{M} - \frac{Q}{M B_0} = 2a - \frac{Q}{MB_0}.
\end{equation}
Asymptotically, the field approaches a uniform magnetic field of strength $B_0$.

Since for the Kerr metric in Boyer--Lindquist coordinates,
\begin{equation}
    k_\mu = (g_{tt},0,0,g_{t\varphi}), \qquad \eta_\mu = (g_{t\varphi},0,0,g_{\varphi \varphi}),
\end{equation}
we have
\begin{equation}
    A_\mu = \frac{1}{2} B_0 (g_{t\varphi} + \kappa g_{tt},0,0,g_{\varphi \varphi} + \kappa g_{t \varphi})
\end{equation}
or
\begin{equation}
    A_t = \frac{1}{2} B_0 \left( g_{t \varphi} + 2 a g_{tt} \right) - \frac{Q}{2 M} g_{tt}, \qquad A_\varphi = \frac{1}{2} B_0 \left( g_{\varphi \varphi} + 2 a g_{t \varphi} \right) - \frac{Q}{2M} g_{t \varphi}.
\end{equation}
The explicit expressions for $A_t$ and $A_\varphi$ in the case of the Kerr metric read
\begin{subequations}
\label{WaldKerr}
\begin{eqnarray}
    A_t & = & \frac{a B_0}{\rho^2} \left( -M r \sin^2 \vartheta - \rho^2 + 2 M r \right) - \frac{Q}{2 M} \left( -1 + \frac{2 M r}{\rho^2} \right), \\
    A_\varphi & = & \frac{B_0 \sin^2 \vartheta}{2 \rho^2} \left[ (r^2 + a^2) \rho^2 + 2 M a^2 r \sin^2 \vartheta - 4 a^2 M r \right] + \frac{ Q a r \sin^2 \vartheta}{\rho^2}.
\end{eqnarray}
\end{subequations}

\section{Geometrically thick disks}
\label{sec:thickdisks}

We now turn to a general description of stationary and axially symmetric Vlasov--Maxwell disks in the Kerr spacetime. As immediate examples, we will consider sample thick-disk configurations. Thin disks confined to the equatorial plane will be discussed in subsequent sections.

Consider the vector potential with possibly non-vanishing components $A_t = A_t(r,\vartheta)$ and $A_\varphi = A_\varphi(r,\vartheta)$, and with $A_r = A_\vartheta = 0$. The Hamiltonian \eqref{hamiltonian} has the form
\begin{eqnarray}
    H & = & \frac{1}{2}\left( g^{tt} p_t^2 + 2 g^{t \varphi} p_t p_\varphi + g^{rr}p_r^2 + g^{\vartheta \vartheta} p_\vartheta^2 + g^{\varphi \varphi} p_\varphi^2  \right) \nonumber \\
    & = & \frac{1}{2} \left[ g^{tt} (\Pi_t - q A_t)^2 + 2 g^{t \varphi} (\Pi_t - q A_t) (\Pi_\varphi - q A_\varphi) + g^{rr} \Pi_r^2 + g^{\vartheta \vartheta} \Pi_\vartheta^2 + g^{\varphi \varphi}(\Pi_\varphi - q A_\varphi)^2 \right].
\end{eqnarray}
By standard reasoning, we conclude that there exist 3 constants of the electrogeodesic motion: the energy $E \coloneqq - \Pi_t = -p_t - q A_t$, the angular momentum $L \coloneqq \Pi_\varphi = p_\varphi + q A_\varphi$, and the Hamiltonian function itself, $H = -m^2/2$.

The following construction has been loosely adapted from an analogous one used in \cite{Andreasson2011b,Andreasson2014} for Einstein--Vlasov systems. In terms of the momentum components $p^{(r)}$, $p^{(\vartheta)}$, and $p^{(\varphi)}$ with respect to the Bardeen tetrad (\ref{eq:frame}), the volume element on $P_x^+(m)$ can be written as
\begin{equation}
\mathrm{dvol}_x^{(m)}(p) = m \frac{\d p^{(r)} \d p^{(\vartheta)} \d p^{(\varphi)}}{\sqrt{m^2 + (p^{(r)})^2 + (p^{(\vartheta)})^2 + (p^{(\varphi)})^2}}.
\end{equation}
It is convenient to introduce the following cylindrical
coordinates $\left(\zeta,\chi,p^{(\varphi)}\right)$ in the space $p^{(r)}$, $p^{(\vartheta)}$, $p^{(\varphi)}$:
\begin{equation}
\label{defchi}
p^{(r)} = \zeta \cos \chi, \qquad p^{(\vartheta)} = \zeta \sin \chi, \qquad p^{(\varphi)} =p^{(\varphi)},
\end{equation}
so that $\d p^{(r)} \d p^{(\vartheta)} \d p^{(\varphi)} = \zeta \d \zeta \d \chi \d p^{(\varphi)}$. This gives
\begin{equation}\label{eq:intMeasureThick1}
\mathrm{dvol}_x^{(m)} (p) = m \frac{\zeta \d \zeta \d\chi \d p^{(\varphi)}}{\sqrt{m^2 + \zeta^2 + (p^{(\varphi)})^2}}.
\end{equation}
The two constants of motion $E$ and $L$ are now expressed as
\begin{eqnarray}
L & = & p_\varphi + q A_\varphi = \frac{\sin^2 \vartheta}{\rho^2} \left( - 2 a M r p^t + A p^\varphi  \right) + q A_\varphi = \frac{\sin \vartheta \sqrt{A}}{\rho} p^{(\varphi)} + q A_\varphi, \label{Ltetrad} \\
E & = & -p_t - q A_t =  \left( 1 - \frac{2 M r}{\rho^2} \right) p^t + \frac{2 M a r \sin^2 \vartheta}{\rho^2} p^\varphi - q A_t \nonumber \\
& = & \rho \sqrt{\frac{\Delta}{A}} p^{(t)} + \frac{2 M a r \sin \vartheta}{\sqrt{A} \rho} p^{(\varphi)} - q A_t = \rho \sqrt{\frac{\Delta}{A}} \sqrt{m^2 + \zeta^2 + (p^{(\varphi)})^2} + \frac{2 M a r \sin \vartheta}{\sqrt{A} \rho} p^{(\varphi)} - q A_t.
\end{eqnarray}
Conversely, the covariant momentum components can be written, in terms of $E$ and $L$, as
\begin{equation}
p_t = - E - q A_t, \qquad p_r = \frac{\rho}{\sqrt{\Delta}} \zeta \cos \chi, \qquad p_\vartheta = \rho \zeta \sin \chi, \qquad p_\varphi = L - q A_\varphi,
\end{equation}
where
\begin{equation}\label{eqzeta}
\begin{aligned}
\zeta^2 &= g^{rr}(p_r)^2 + g^{\vartheta \vartheta} (p_\vartheta)^2\\
&= - m^2 -g^{tt}(E + q A_t)^2 + 2 g^{t \varphi} (E + q A_t)(L - q A_\varphi) - g^{\varphi \varphi}(L - q A_\varphi)^2.
\end{aligned}
\end{equation}
We also note that for the motion confined to the equatorial plane, the coordinate $\zeta^2$ coincides with $\mathcal G$, defined below (Eq.\ \eqref{eq:Gdef}).

We will now introduce yet another change of the momentum coordinates $(\zeta,p^{(\varphi)},\chi) \mapsto (E,L,\chi)$. Since
\begin{equation}
\frac{\partial(E,L,\chi)}{\partial(\zeta,p^{(\varphi)},\chi)}  = \frac{\sqrt{\Delta} \sin\vartheta \zeta}{\sqrt{m^2 + \zeta^2 + (p^{(\varphi)})^2}},
\end{equation}
we obtain the following expression for the volume element
\begin{equation}
\label{Pvolume}
\mathrm{dvol}_x^{(m)} (p) = m \frac{\zeta \d \zeta \d\chi \d p^{(\varphi)}}{\sqrt{m^2 + \zeta^2 + (p^{(\varphi)})^2}} = \frac{m \d E \d L \d \chi}{\sqrt{\Delta} \sin \vartheta}.
\end{equation}
This form turns out to be especially convenient for distribution functions depending only on $E$ and $L$, i.e., of the form
\begin{equation}
\mathcal F(x,p) = \mathcal F_0(E(x,p),L(x,p)).
\end{equation}
In this case, one obtains
\begin{equation}
\label{Jthick}
    (\mathcal J_t, \mathcal J_r,\mathcal J_\vartheta,\mathcal J_\varphi) = \frac{m}{\sqrt{\Delta} \sin \vartheta} \int_{\mathcal P(m)} \mathcal F_0(E,L) \left( -E - q A_t, \frac{\rho \zeta}{\sqrt{\Delta}} \cos \chi, \rho \zeta \sin \chi,L - q A_\varphi \right) \d E \d L \d \chi,
\end{equation}
where $\zeta$ is given by Eq.\ (\ref{eqzeta}) and $\mathcal P(m) \subseteq P_x^+(m)$.\footnote{Note that one can have $P_x^+(m) \setminus \mathcal P(m) \neq \emptyset$. In the literature, it is also customary to retain the integration over the entire $P_x^+(m)$ and assume (sometimes tacitly) that $\mathcal F$ vanishes outside $\mathcal P(m)$.} The components $\mathcal J_r$ and $\mathcal J_\vartheta$ vanish, provided that the integration is taken over a region $\mathcal P(m)$ sufficiently symmetric with respect to $\chi$. This happens, in particular, when the integration with respect to $\chi$ can be performed over the entire range $\chi \in [0,2\pi)$ and the limits of the integrals with respect to $E$ and $L$ do not depend on $\chi$. We emphasize that Eqs.\ (\ref{defchi}) mean that the integration with respect to $\chi$ corresponds to the averaging of the momenta in the meridional plane $\left(p^{(r)},p^{(\vartheta)}\right)$. The magnitude of the momentum is bounded by admissible values of $\zeta$.

For a vanishing electromagnetic field, one can relate the coordinates $(E,L,\chi)$ to the constants of the geodesic motion in the Kerr spacetime. We discuss this relation in Appendix \ref{app:kerr}.

\subsection{Constant angular momentum disks}

In the following, we consider a sample model of a stationary disk characterized by the distribution function $\mathcal F_0 (L) = C \delta(L - L_0)$, where $C$ and $L_0$ are constants. We assume that the energies are bounded by the condition $\zeta^2 \ge 0$. This leads to $\mathcal J_r = \mathcal J_\vartheta = 0$ and
\begin{equation}
\label{Jthick2}
    \left( \begin{array}{c} \mathcal J_t \\ \mathcal J_\varphi \end{array}\right) = \frac{2 \pi m C}{\sqrt{\Delta} \sin \vartheta} \int_{\mathcal P(m)} \left( \begin{array}{c} - E - q A_t \\ L_0 - q A_\varphi  \end{array} \right) \d E =  \frac{2 \pi m^3 C}{M \sqrt{\bar \Delta} \sin \vartheta} \int_{\mathcal P(m)} \left( \begin{array}{c} - \varepsilon - \frac{q}{m} A_t \\ M \lambda_0 - \frac{q}{m} A_\varphi \end{array} \right) \d \varepsilon,
\end{equation}
where we have introduced dimensionless quantities
\begin{equation}
\label{dimensionless1}
    \varepsilon \coloneqq E/m, \qquad \lambda_0 \coloneqq L_0/(m M), \qquad \bar \Delta \coloneqq \Delta/M^2.
\end{equation}
In the following, we will also use
\begin{equation}
\label{dimensionless2}
    \xi \coloneqq r/M, \qquad \alpha \coloneqq a/M,
\end{equation}
so that $\bar \Delta = \xi^2 - 2 \xi + \alpha^2$.

The expression \eqref{eqzeta} for $\zeta^2$ can be factorized as
\begin{equation}
    \zeta^2 = -g^{tt}\left[E - \mathcal W_+(L,r,\vartheta)\right]\left[E - \mathcal W_-(L,r,\vartheta)\right],
\end{equation}
where
\begin{equation}
    \mathcal W_\pm(L,r,\vartheta) \coloneqq -\frac{1}{ g^{tt}} \left[ g^{tt} q A_t -  g^{t \varphi} (L -q A_\varphi) \pm \sqrt{\mathcal W_0} \right], \qquad \mathcal W_0 \coloneqq (L-qA_\varphi)^2\left[(g^{t\varphi})^2-g^{tt}g^{\varphi\varphi}\right]-m^2g^{tt}.
\end{equation}
Notice that the term $(g^{t\varphi})^2-g^{tt}g^{\varphi\varphi}$ satisfies
\begin{align}
\label{geominv}
    \frac{1}{(g^{t\varphi})^2-g^{tt}g^{\varphi\varphi}} = -(k_\mu k^\mu)(\eta_\mu\eta^\mu) + (k_\mu\eta^\mu)^2,
\end{align}
where $k$ and $\eta$ denote the Killing vectors (\ref{kerrkilling}),
and it is a geometrical invariant. In Boyer--Lindquist coordinates,
\begin{equation}
    (g^{t\varphi})^2-g^{tt}g^{\varphi\varphi} = \frac{1}{\Delta \sin^2 \vartheta}.
\end{equation}

Since $g^{tt} < 0$ for $r > r_+$, the positivity of $\zeta^2$ implies that either $E > \mathcal W_+(L,r,\vartheta)$ or $E < \mathcal W_-(L,r,\vartheta)$. We now show that the condition $E < \mathcal W_-(L,r,\vartheta)$ is not compatible with future-directed trajectories. To test the time orientation of four-momenta $p_\mu$ with a fixed value of $L$, we use the vector $N$, introduced in Eq.\ (\ref{vecN}).\footnote{Note that for $r > r_+$, $N^\mu = \sqrt{-g^{tt}} \, u^\mu$, where $u^\mu$ denotes the four-velocity of the Zero Angular Momentum Observer (ZAMO).} A direct calculation gives
\begin{equation}
    - N^\mu p_\mu = - g^{tt} \left[ E - \frac{1}{2} \left( \mathcal W_- + \mathcal W_+ \right) \right] = -g^{tt}(E - \mathcal W_+) + \sqrt{\mathcal W_0} = - g^{tt}(E - \mathcal W_-) - \sqrt{\mathcal W_0}.
\end{equation}
A future-directed momentum vector satisfies $-N^\mu p_\mu > 0$. This yields
\begin{subequations}
\begin{align}
    -N^\mu p_\mu = -\left(E-\mathcal{W}_{-}\right)g^{tt}-\sqrt{\mathcal{W}_0}  &>0 \,,\\
    -\left(E-\mathcal{W}_{-}\right)g^{tt}&>\sqrt{\mathcal{W}_0} > 0.
\end{align}
\end{subequations}
Since for $r > r_+$, we have $-g^{tt}>0$, a future-directed momentum four-vector satisfies $E > \mathcal W_-$. This excludes the case with $E < \mathcal W_-$, as claimed. On the other hand, for $r > r_+$, one has $\mathcal W_+ > \mathcal W_-$ (actually $\mathcal{W}_+ -\mathcal{W}_-=-2\sqrt{\mathcal{W}_0}/g^{tt}>0$). Hence, four-momenta with $E > \mathcal W_+$ are both timelike and future-directed.

As an example, we consider the set $\mathcal P$ characterized by energies $\varepsilon$ bounded from below by $w_+(\lambda_0,\xi,\vartheta)$ and from above by a (suitably chosen) constant value $\varepsilon_\mathrm{max}$. Here, the dimensionless effective potentials $w_\pm(\lambda,\xi,\varepsilon)$ are defined as $w_\pm(\lambda,\xi,\vartheta) \coloneqq \mathcal W_\pm(L,r,\vartheta)/m$. In addition, we exclude a region close to the black hole, in which $w_+(\lambda_0,\xi,\vartheta) < \varepsilon_\mathrm{max}$, and which would allow for particle trajectories plunging into the black hole. In other words, we assume the value of $\lambda_0$ to be sufficiently large, so that $w_+(\lambda_0,\xi,\vartheta)$ exhibits a local maximum, corresponding to the centrifugal barrier, and the value of $\varepsilon_\mathrm{max}$ to be sufficiently small, so that the radial range available for the motion of individual particles is bounded both from below and from above. Strictly speaking, the existence of the centrifugal barrier is an assumption on the properties of the effective potential, which in turn depends on $A_t$ and $A_\varphi$. The components of the particle current density $\mathcal J_\mu$ can now be written as
\begin{subequations}
\label{Jthick3}
\begin{eqnarray}
    \mathcal J_t & = &  - \frac{2 \pi m^3 C}{M \sqrt{\bar \Delta} \sin \vartheta} \Theta(\varepsilon_\mathrm{max} - w_+(\lambda_0,\xi,\vartheta))\int_{w_+(\lambda_0,\xi,\vartheta)}^{\varepsilon_\mathrm{max}} \left(\varepsilon + \frac{q}{m} A_t \right) \d \varepsilon \nonumber \\
    & = & - \frac{2 \pi m^3 C}{M \sqrt{\bar \Delta} \sin \vartheta} \Theta(\varepsilon_\mathrm{max} - w_+(\lambda_0,\xi,\vartheta)) \left\{ \frac{1}{2}\left[ \varepsilon_\mathrm{max}^2 - w_+(\lambda_0,\xi,\vartheta)^2 \right] + \frac{q}{m} A_t \left[ \varepsilon_\mathrm{max} - w_+(\lambda_0,\xi,\vartheta) \right] \right\}, \\
    \mathcal J_\varphi & = & \frac{2 \pi m^3 C}{\sqrt{\bar \Delta} \sin \vartheta} \Theta(\varepsilon_\mathrm{max} - w_+(\lambda_0,\xi,\vartheta)) \int_{w_+(\lambda_0,\xi,\vartheta)}^{\varepsilon_\mathrm{max}} \left(\lambda_0 - \frac{q}{m M} A_\varphi \right) \d \varepsilon \nonumber \\
    & = & \frac{2 \pi m^3 C}{\sqrt{\bar \Delta} \sin \vartheta} \Theta(\varepsilon_\mathrm{max} - w_+(\lambda_0,\xi,\vartheta)) \left(\lambda_0 - \frac{q}{m M} A_\varphi \right) \left[ \varepsilon_\mathrm{max} - w_+(\lambda_0,\xi,\vartheta) \right],
\end{eqnarray}
\end{subequations}
where $\Theta$ denotes the Heaviside step function. We emphasize that the lower integration limit $w_+(\lambda_0,\xi,\vartheta)$ explicitly depends on $\xi$ and $\vartheta$. For clarity, the exclusion of the region close to the black hole has not been explicitly written. We should also stress a formal similarity of Eqs.\ (\ref{Jthick3}) with Eq.\ (\ref{CarterIntegration}) derived in Appendix \ref{app:kerr} for the case with a vanishing electromagnetic field, in which the integration in the momentum space was performed with respect to the Carter constant.

\begin{figure}[t]
    \centering
    \includegraphics[width=0.9\twofig]{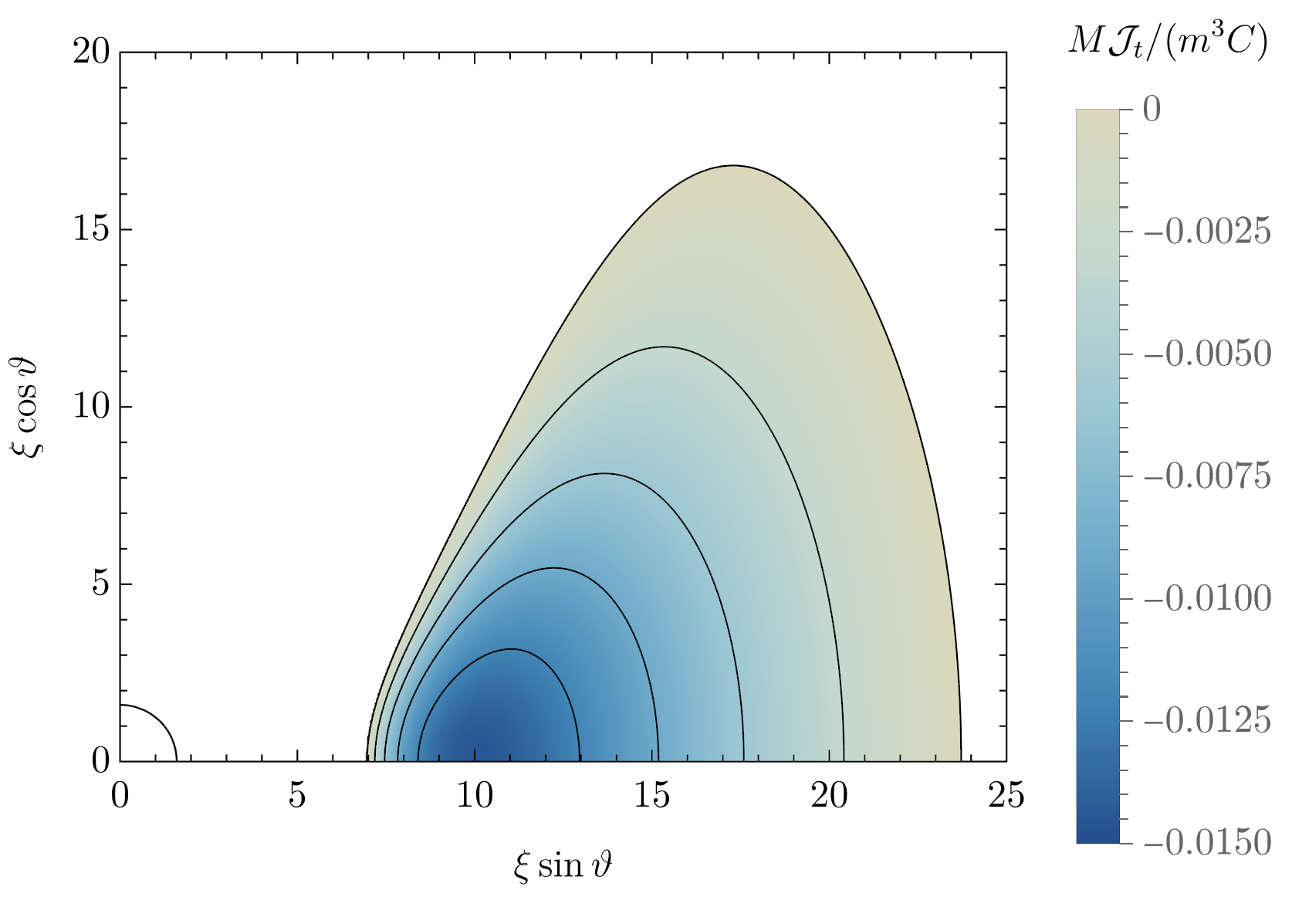}
    \hspace{1em}
    \includegraphics[width=0.9\twofig]{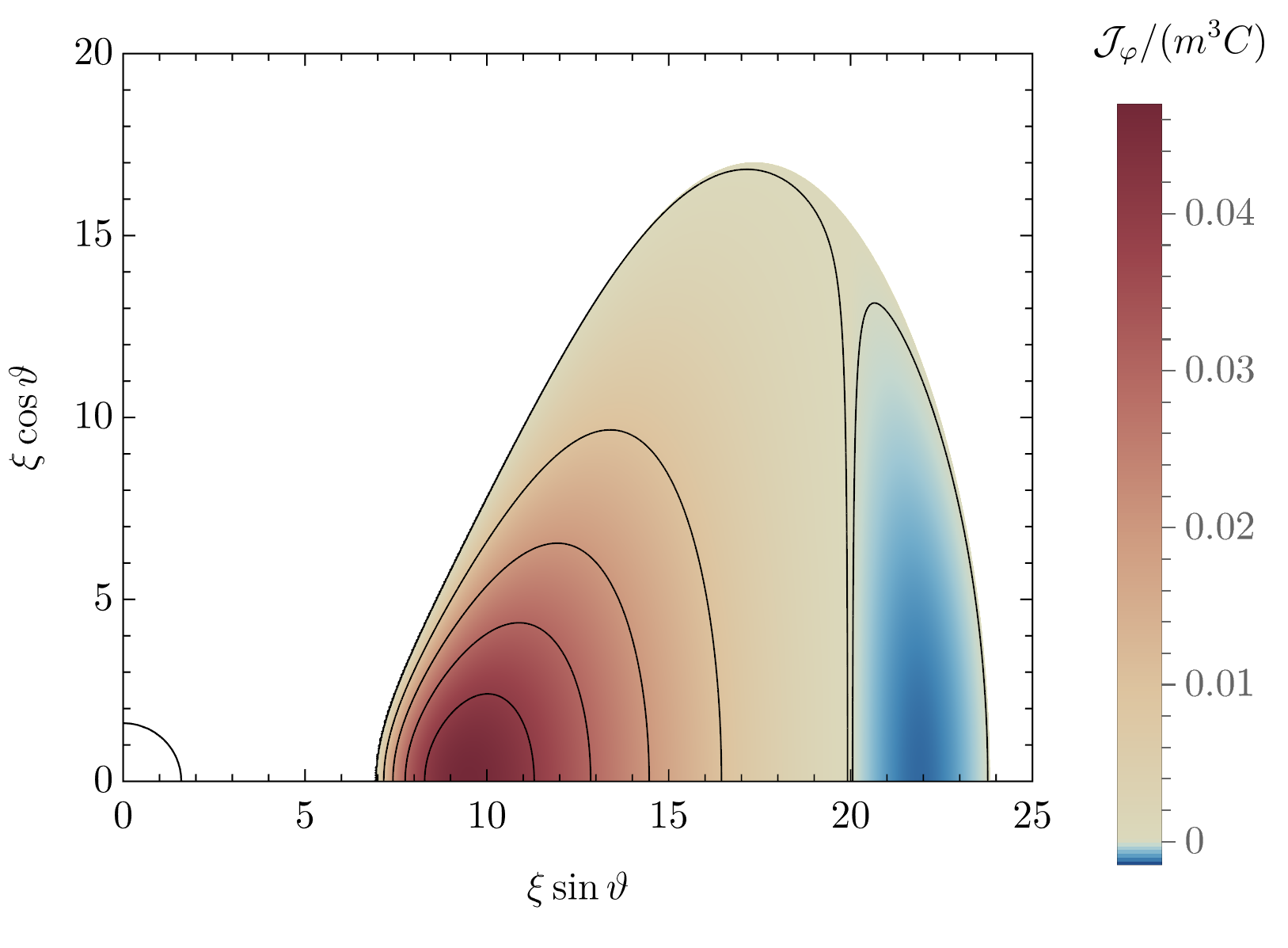}
    
    \caption{Sample thick disk configuration in the Wald magnetosphere with $Q = 0$ for a constant angular momentum (see Eqs.~ \eqref{Jthick3}). We assume $\alpha = 0.8$, $\beta = 0.01$, $\lambda_0 = 4$, and $\varepsilon_\mathrm{max} = \sqrt{0.95}$. Left panel: $M \mathcal J_t/(m^3 C)$; right panel: $\mathcal J_\varphi/(m^3 C)$. The disk turns out to be divided into two parts rotating in opposite directions. The small quarter circle represents the outer Kerr horizon.}    \label{fig:ThickTorus1}
\end{figure}

\begin{figure}[t]
    \centering
    \includegraphics[width=0.9\twofig]{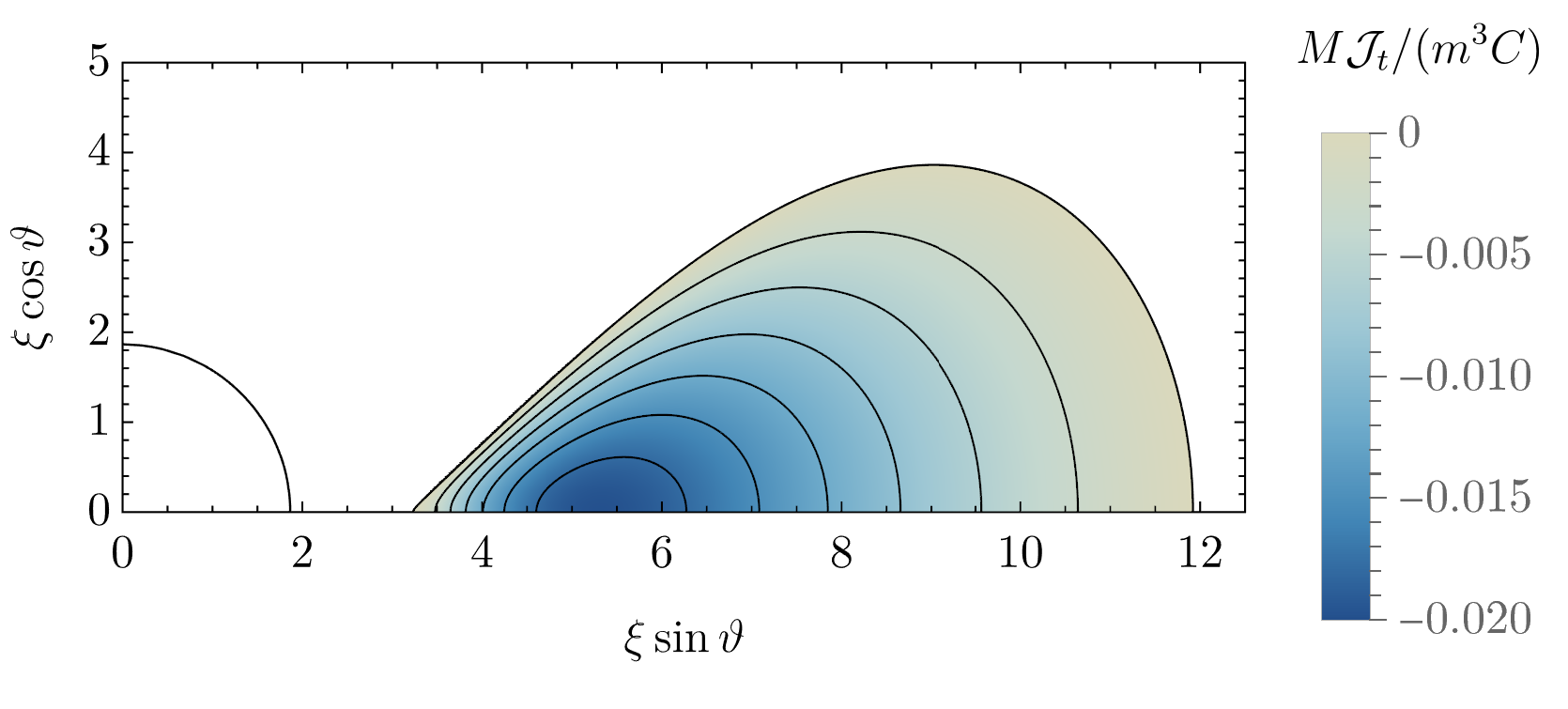}
    \hspace{1em}
    \includegraphics[width=0.9\twofig]{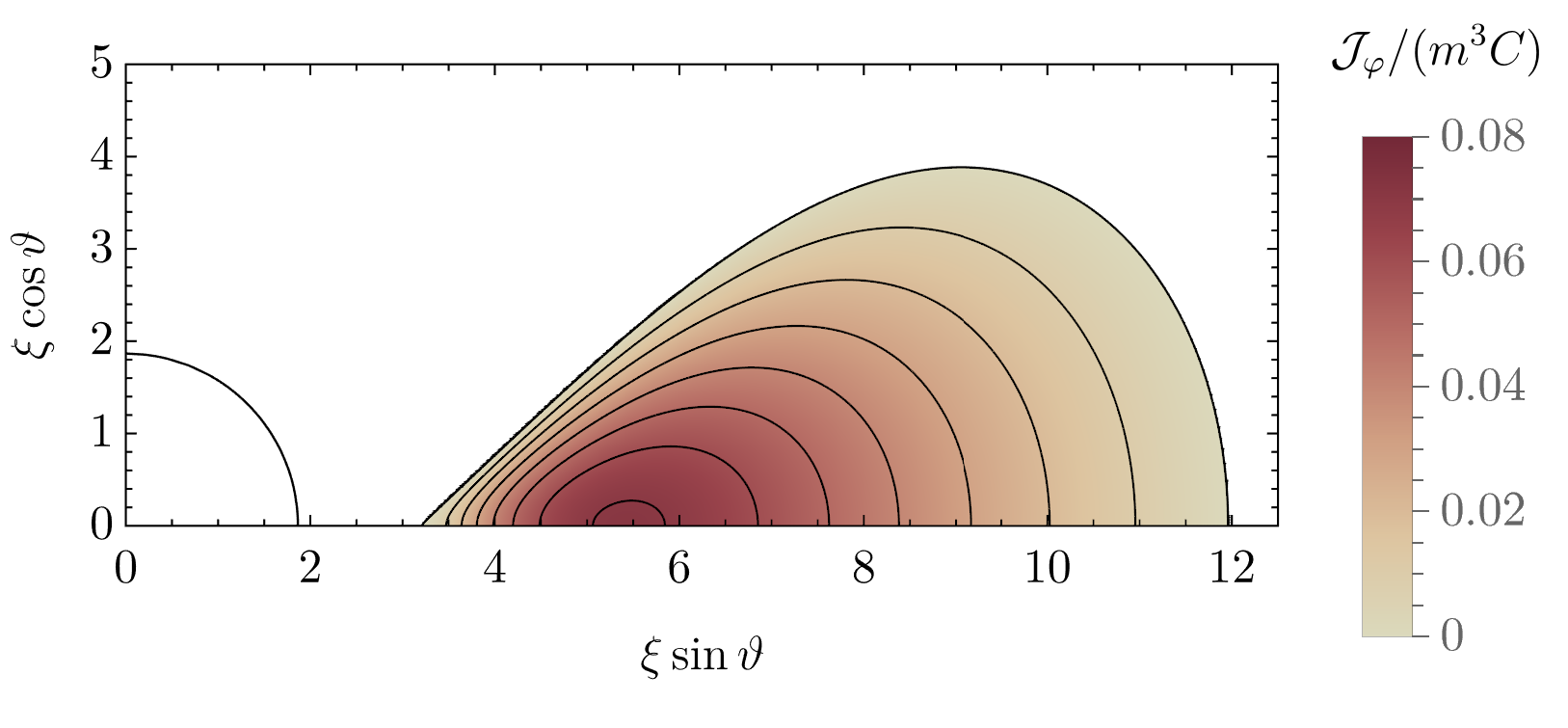}
    \caption{Same as in Fig.\ \ref{fig:ThickTorus1}, but with $\alpha = 0.5$, $\beta = -0.01$, $\lambda_0 = 3.2$, and $\varepsilon_\mathrm{max} = 0.972097$ (chosen as the maximal value of $w_+$ in the equatorial plane).}
    \label{fig:ThickTorus2}
\end{figure}

\begin{figure}[t]
    \centering
    \includegraphics[width=0.9\twofig]{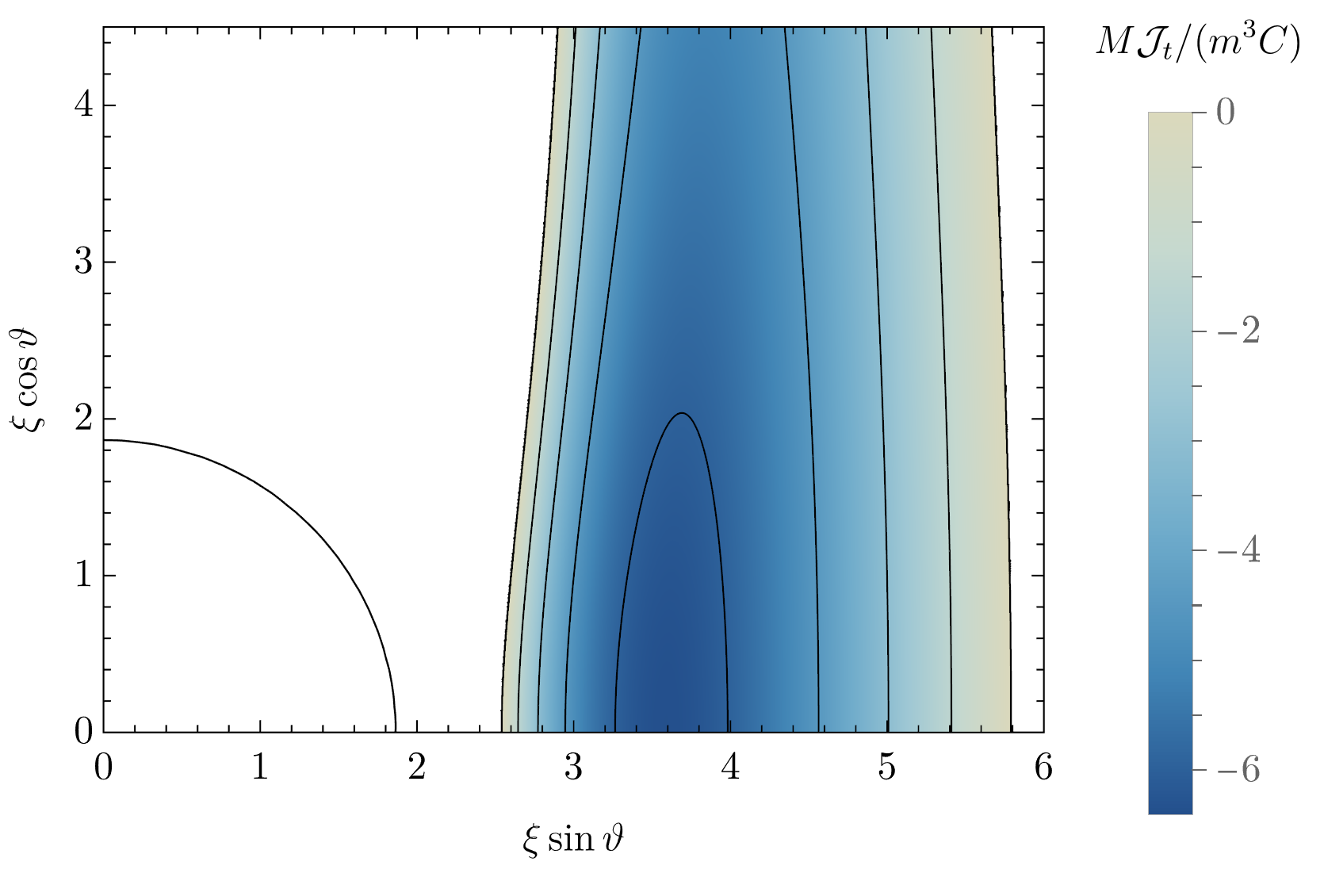}
    \hspace{1em}
    \includegraphics[width=0.9\twofig]{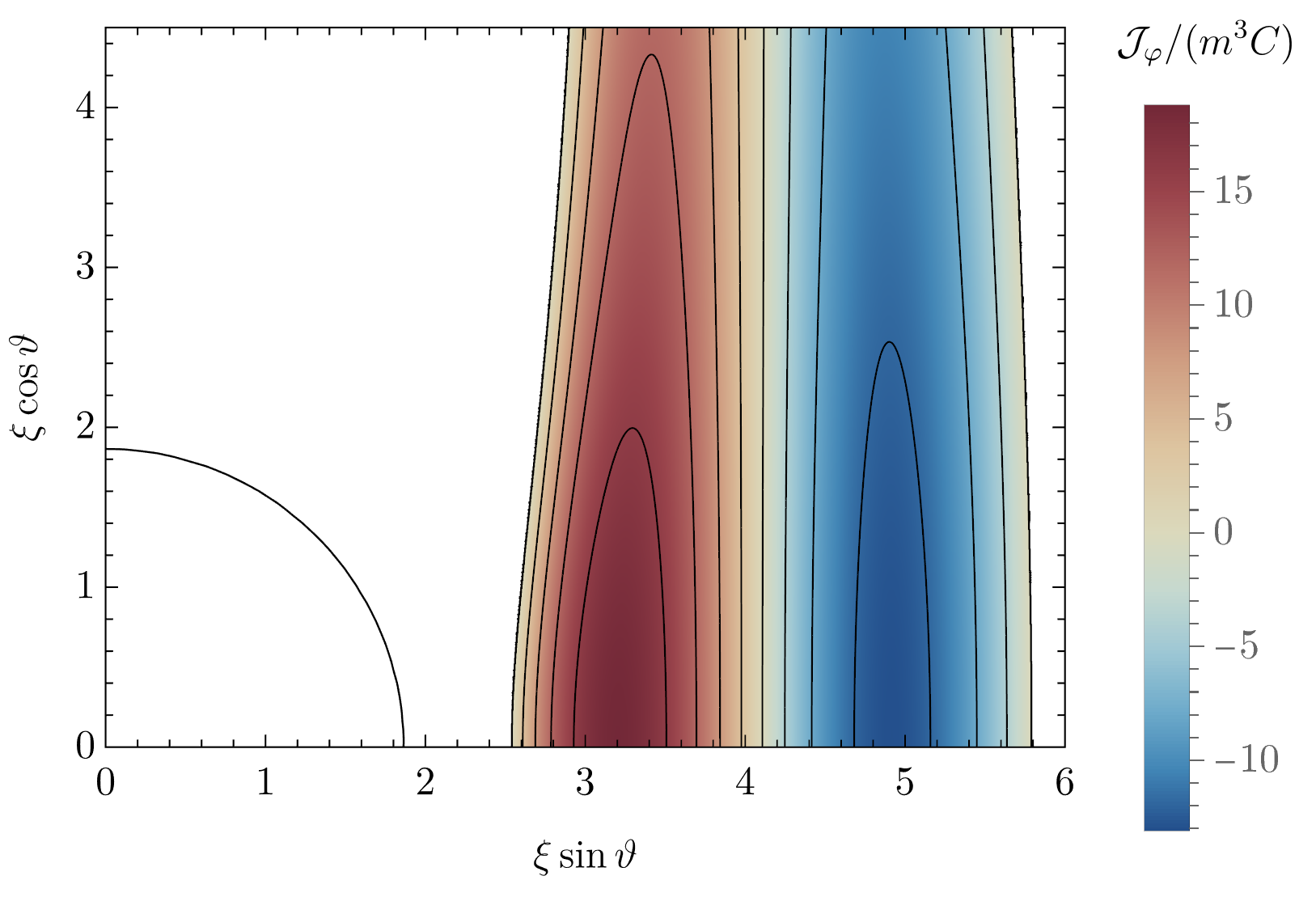}
    \caption{Same as in Figs.\ \ref{fig:ThickTorus1} and \ref{fig:ThickTorus2}, but for $\alpha = 0.5$, $\beta = 1$, $\lambda_0 = 17$, and $\varepsilon_\mathrm{max} = 3.2$. Here the disk extends to the axial infinity, as $\varepsilon_\mathrm{escape} = 2\alpha\beta + 1 = 2 < \varepsilon_\mathrm{max}$ (see Eqs.\ \eqref{w_infty} and \eqref{eps_escape}).}
    \label{fig:ThickTorus3}
\end{figure}

Sample graphs of $\mathcal J_t$ and $\mathcal J_\varphi$ computed according to Eqs.\ (\ref{Jthick3}) are shown in Figs.\ \ref{fig:ThickTorus1}--\ref{fig:ThickTorus3}, assuming the electromagnetic field of the Wald solution with $Q = 0$, i.e., $A_t$ and $A_\varphi$ given by Eqs.\ \eqref{WaldKerr}. In specifying the parameters of these solutions, we use $\beta \coloneqq q B_0 M/(2m)$. We show the covariant components of the particle current density $\mathcal J_t$ and $\mathcal J_\varphi$, as they can be expressed covariantly as $\mathcal J_t = \mathcal J_\mu k^\mu$ and $\mathcal J_\varphi = \mathcal J_\mu \eta^\mu$, where $k$ and $\eta$ denote the Killing vectors \eqref{kerrkilling}. 

A special example is shown in Fig.\ \ref{fig:ThickTorus3}. In this case, the disk becomes cylindrical and extends to infinity in the direction normal to the equatorial plane. This is allowed by the properties of the effective potential $w_+(\lambda_0,\xi,\vartheta)$, associated with the Wald magnetosphere (cf.\ \cite{stuchlik2016}), and can be seen as follows. Consider the expression for $w_+$ with the electromagnetic potential given by Eqs.\ (\ref{WaldKerr}) with $Q = 0$ and substitute standard Cartesian relations $x \coloneqq \xi \sin \vartheta$ and $z \coloneqq \xi \cos \vartheta$. It can be shown that
\begin{equation}\label{w_infty}
    w_\infty(x) \coloneqq \lim_{z \to +\infty} w_+(x,z) = 2 \alpha \beta + \sqrt{1 + x^2 \beta^2 - 2 \beta \lambda_0 + \frac{\lambda_0^2}{x^2}}. 
\end{equation}
As long as this limit remains real, it has a minimum for $x_0 = \sqrt{|\lambda_0/\beta|}$. Moreover,
\begin{equation}\label{eps_escape}
    \varepsilon_\mathrm{escape} \coloneqq w_\infty(x_0) =
    \begin{cases}
        2\alpha\beta + 1, & \beta\lambda_0 > 0,\\
        2\alpha\beta + \sqrt{1 - 4\beta\lambda_0}, & \beta\lambda_0 < 0.
    \end{cases}
\end{equation}
This value can be smaller than $\varepsilon_\mathrm{max}$ allowed in the construction of the disk. In particular, it can be smaller than the value of $w_+$ at its local maximum attained in the equatorial plane.

\subsection{Constant energy disks}

It is also possible to consider disks characterized by a constant energy $E = E_0$, in which case $\mathcal F_0(E) = C \delta(E - E_0)$, for constant $C$ and $E_0$. This yields, in analogy to Eq.\ (\ref{Jthick2}), 
\begin{equation}
\label{Jthick7}
    \left( \begin{array}{c} \mathcal J_t \\ \mathcal J_\varphi \end{array}\right) = \frac{2 \pi m C}{\sqrt{\Delta} \sin \vartheta} \int_{\mathcal P(m)} \left( \begin{array}{c} - E_0 - q A_t \\ L - q A_\varphi  \end{array} \right) \d L =  \frac{2 \pi m^3 C}{\sqrt{\bar \Delta} \sin \vartheta} \int_{\mathcal P(m)} \left( \begin{array}{c} - \varepsilon_0 - \frac{q}{m} A_t \\ M \lambda - \frac{q}{m} A_\varphi \end{array} \right) \d \lambda.
\end{equation}

The quantity $\zeta^2$ is also quadratic in $L$, and hence
\begin{equation}
    \zeta^2 = -g^{\varphi \varphi} \left[ L - \Lambda_+ (E,r,\vartheta) \right] \left[ L - \Lambda_- (E,r,\vartheta) \right],
\end{equation}
where
\begin{equation}
\Lambda_\pm (E,r,\vartheta) \coloneqq \frac{1}{g^{\varphi \varphi}} \left[ (E + q A_t) g^{t \varphi} + q A_\varphi g^{\varphi 
    \varphi} \pm \sqrt{\Lambda_0} \right], \qquad \Lambda_0 \coloneqq  (E + q A_t)^2 \left[ (g^{t \varphi})^2 - g^{tt} g^{\varphi \varphi} \right] - g^{\varphi \varphi} m^2.
\end{equation}
The region $\Lambda_-(E,r,\vartheta) \le L \le \Lambda_+(E,r,\vartheta)$ corresponds to $\zeta^2 \ge 0$. In particular examples, we consider the range $L_\mathrm{min} \le L \le \Lambda_+(E,r,\vartheta)$ with $L_\mathrm{min}$ constant, or equivalently $\lambda_\mathrm{min} \le \lambda \le \lambda_+(\varepsilon,\xi,\vartheta)$, where $\lambda_\mathrm{min} \coloneqq L_\mathrm{min}/(mM)$ and $\lambda_\pm(\varepsilon,\xi,\vartheta) \coloneqq \Lambda_{\pm}(E,r,\vartheta)/(m M)$. This yields
\begin{subequations}
\label{Jthick8}
\begin{eqnarray}
    \mathcal J_t & = &  - \frac{2 \pi m^3 C}{\sqrt{\bar \Delta} \sin \vartheta} \Theta(\lambda_+(\varepsilon_0,\xi,\vartheta) - \lambda_\mathrm{min})\int_{\lambda_\mathrm{min}}^{\lambda_+(\varepsilon_0,\xi,\vartheta)} \left(\varepsilon_0 + \frac{q}{m} A_t \right) \d \lambda \nonumber \\
    & = & - \frac{2 \pi m^3 C}{\sqrt{\bar \Delta} \sin \vartheta} \Theta(\lambda_+(\varepsilon_0,\xi,\vartheta) - \lambda_\mathrm{min})  \left(\varepsilon_0 + \frac{q}{m} A_t \right) \left[  \lambda_+(\varepsilon_0,\xi,\vartheta) - \lambda_\mathrm{min}\right] , \\
    \mathcal J_\varphi & = & \frac{2 \pi m^3 M C}{\sqrt{\bar \Delta} \sin \vartheta} \Theta(\lambda_+(\varepsilon_0,\xi,\vartheta) - \lambda_\mathrm{min}) \int_{\lambda_\mathrm{min}}^{\lambda_+(\varepsilon_0,\xi,\vartheta)} \left(\lambda - \frac{q}{m M} A_\varphi \right) \d \lambda \nonumber \\
    & = & \frac{2 \pi m^3 M C}{\sqrt{\bar \Delta} \sin \vartheta} \Theta(\lambda_+(\varepsilon_0,\xi,\vartheta) - \lambda_\mathrm{min}) \left\{ \frac{1}{2}\left[ \lambda_+(\varepsilon_0,\xi,\vartheta)^2 - \lambda_\mathrm{min}^2 \right] - \frac{q}{m M} A_\varphi \left[ \lambda_+(\varepsilon_0,\xi,\vartheta) - \lambda_\mathrm{min} \right] \right\}.
\end{eqnarray}
\end{subequations}

\begin{figure}
\centering
\includegraphics[width=\twofig]{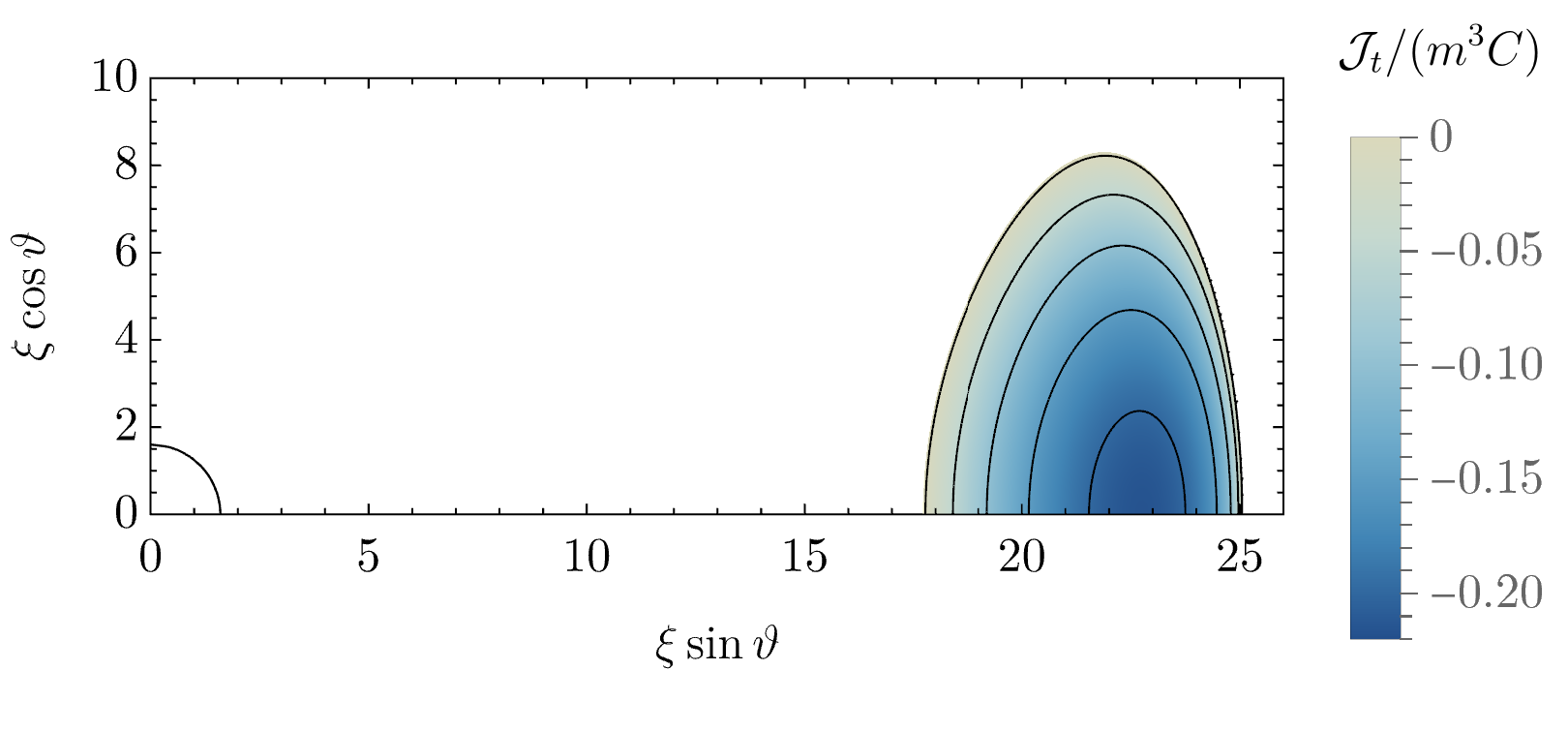} \includegraphics[width=\twofig]{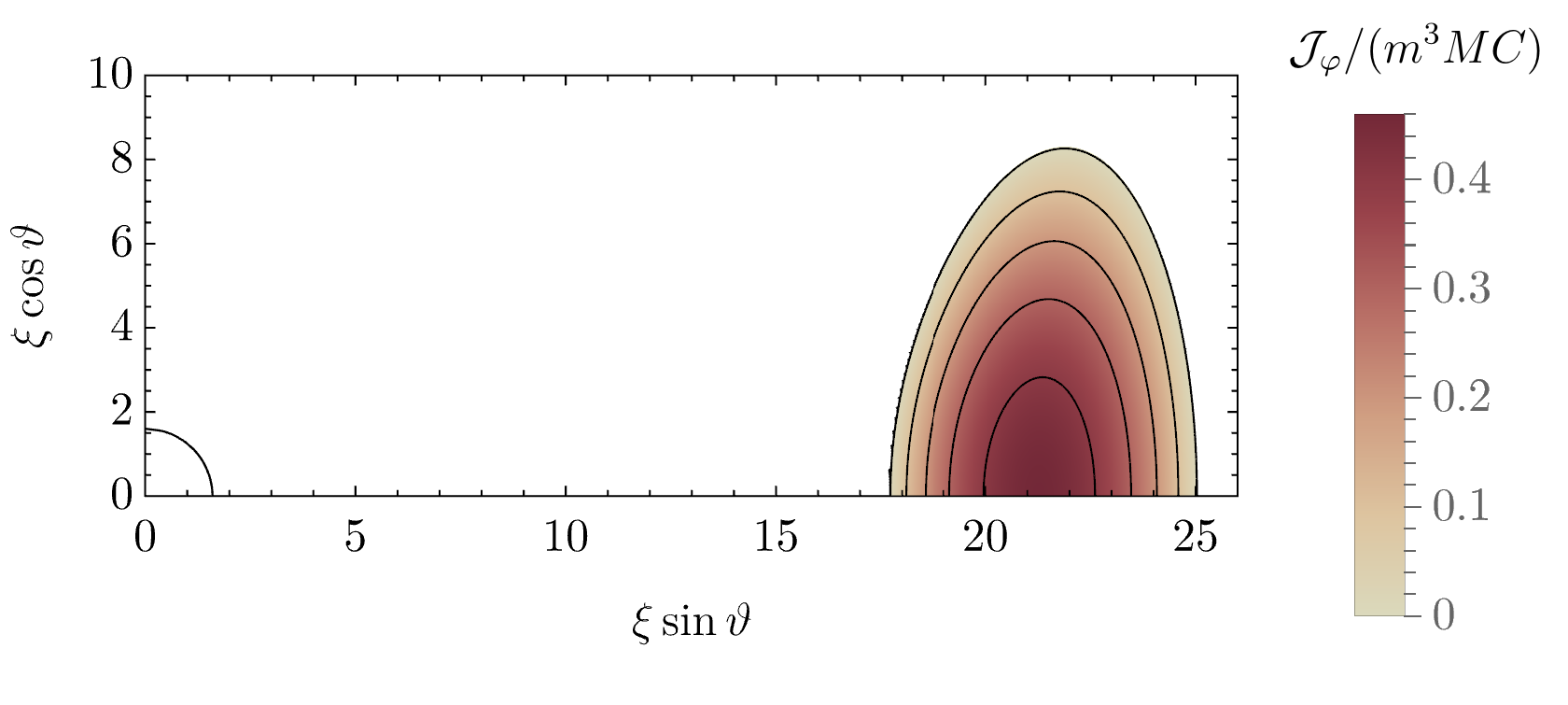}
\caption{Sample thick disk configuration in the Wald magnetosphere for a constant energy (see Eqs.\ \eqref{Jthick8}), with $\alpha = 0.8$, $\beta = 0.01$, $\varepsilon_0 = \sqrt{0.95}$, and $\lambda_\mathrm{min} = 6.5$.}
\label{fig:ThickTorus4}
\end{figure}

The sample graphs of $\mathcal J_t$ and $\mathcal J_\varphi$, computed according to Eqs.\ \eqref{Jthick8}, are shown in Fig.\ \ref{fig:ThickTorus4}. Here again, we assume the Wald magnetosphere with $Q = 0$. Similarly to Figs.\ \ref{fig:ThickTorus1}--\ref{fig:ThickTorus3}, in Fig.\ \ref{fig:ThickTorus4} we also exclude regions close to the black hole with $\lambda_+(\varepsilon_0,\xi,\vartheta) > \lambda_\mathrm{min}$, related with particles plunging into the black hole.

In contrast to constant-angular momentum disks discussed in the previous subsection, for constant-energy disks one should separately ensure that four-momenta of individual particles are future-pointing. This was checked for the examples shown in Figs.\ \ref{fig:ThickTorus4}. We will return to this question, discussing examples of thin disks in Sec.\ \ref{sec:constantE}.

Many other choices of the distribution function $\mathcal F_0$ are, of course, possible. Since this paper ultimately focuses on infinitely thin disks in the equatorial plane, we stop the discussion of thick-disk configurations at this point, leaving the topic for further studies.

\section{Disks in the equatorial plane of the Kerr spacetime}
\label{sec:Thindisks}

\subsection{Disks in the equatorial plane}

In this section, we investigate stationary infinitely thin disks confined to the equatorial plane $\vartheta = \pi/2$. As in the previous section, we first consider a general stationary axially symmetric test electromagnetic field in the Kerr spacetime. In Section \ref{sec:thinWald}, we specialize to the Wald magnetosphere.

To check the consistency of the thin-disk assumption, suppose that a charged particle is located at $\vartheta = \pi/2$ and its polar momentum component vanishes, $\Pi_\vartheta = p_\vartheta = 0$. Hamilton's equations (\ref{hamilton}) give
\begin{subequations}
\begin{eqnarray}
    \frac{\d \vartheta}{\d \tau} & = & p^\vartheta = g^{\vartheta \vartheta} p_\vartheta = 0, \\
    \frac{\d \Pi_\vartheta}{\d \tau} & = & - \frac{1}{2} \frac{\partial g^{\mu\nu}}{\partial \vartheta} p_\mu p_\nu + q \frac{\partial A_\mu}{\partial \vartheta} p^\mu = q \frac{\partial A_\mu}{\partial \vartheta} p^\mu,
\end{eqnarray}
\end{subequations}
where we used the fact that the components $g^{\mu\nu}$ depend on $\vartheta$ only via $\sin^2 \vartheta$ and $\cos^2 \vartheta$, and consequently $\partial g^{\mu \nu} / \partial \vartheta = 0$ at $\vartheta = \pi/2$. We will now require that
\begin{equation}
\label{equatorialconds}
\left. \frac{\partial A_t}{\partial \vartheta} \right|_{\vartheta = \pi/2} = \left. \frac{\partial A_\varphi}{\partial \vartheta} \right|_{\vartheta = \pi/2} = 0.
\end{equation}
Hence, also $\d \Pi_\vartheta / \d \tau = 0$. Thus, a particle with $p_\vartheta = 0$, initially located in the equatorial plane, does not leave this plane. Note that the Wald solution in the Kerr spacetime (\ref{WaldKerr}) satisfies Eq.\ (\ref{equatorialconds}).

A restriction of the distribution function to the equatorial plane can be obtained by setting
\begin{equation}
\label{twodf}
\mathcal F(t,r,\vartheta,\varphi;p_\alpha) = \frac{1}{r} \delta\left(\vartheta - \frac{\pi}{2}\right)  F(t,r,\varphi;p_\alpha), \qquad \mathcal J_\mu(t,r,\vartheta,\varphi) = \frac{1}{r} \delta\left(\vartheta - \frac{\pi}{2}\right) J_\mu(t,r,\varphi).
\end{equation}
The factor $\delta(\vartheta - \pi/2)/r$ can be written as $\delta(\vartheta - \pi/2)/r = \delta(z)$, where $z \coloneqq r \cos \vartheta$. The vectors $\partial_\vartheta$ and $\partial_z$ are both normal to the equatorial plane $\vartheta = \pi/2$. The vector $\partial_z$ is normalized to unity on the equatorial plane. The above definitions of $F$ and $J_\mu$ imply that
\begin{equation}
J_\mu(t,r,\varphi) = \int_{P^+_x} F(t,r,\varphi;p_\alpha) p_\mu \mathrm{dvol}_x(p).
\end{equation}
As a consistency condition, we require that also all momenta $p_\vartheta = 0$ (otherwise the particles would be allowed to leave the equatorial plane). Thus
\begin{equation}
F(t,r,\varphi,p_t,p_r,p_\vartheta,p_\varphi) = r \delta(p_\vartheta) f(t,r,\varphi,p_t,p_r,p_\varphi),
\end{equation}
where again the factor $r \delta(p_\vartheta)$ can be written as $r \delta(p_\vartheta) = \delta(p_z)$ (at the equatorial plane). This follows immediately from the relation $p_z = - p_\vartheta/r$, holding at the equatorial plane.\footnote{Alternatively, one can work with the frame components and write $\delta \left(p^{(\vartheta)} \right)=\delta(\rho p_\vartheta) = \delta(p_\vartheta)/\rho$, to avoid the $z$ coordinate.
} Note that distribution functions $\mathcal F$ and $f$ are conveniently related as
\begin{equation}
\label{dist_flat}
\mathcal F(t,r,\vartheta,\varphi,p_t,p_r,p_\vartheta,p_\varphi) = \delta\left(\vartheta - \frac{\pi}{2}\right)\delta(p_\vartheta)f(t,r,\varphi,p_t,p_r,p_\varphi).
\end{equation}

The particle current surface density can be now expressed in the form
\begin{equation}
\label{jmusurface}
J_\mu(t,r,\varphi) = \int_{P^+_x} r \delta(p_\vartheta) f(t,r,\varphi,p_t,p_r,p_\varphi) p_\mu \frac{\d p_t \d p_r \d p_\vartheta \d p_\varphi}{r^2} = \int_{\bar P^+_x} f(t,r,\varphi,p_t,p_r,p_\varphi) p_\mu \frac{\d p_t \d p_r  \d p_\varphi}{r},
\end{equation}
where
\begin{equation}
    \bar P^+_x \coloneqq \{ (p_t,p_r,p_\varphi) \colon \bar g^{\mu \nu} p_\mu p_\nu < 0, \, p \text{ is future-directed} \},
\end{equation}
and where we have used the fact that at the equatorial plane the volume element \eqref{dvolp} reduces to
\begin{equation}
\mathrm{dvol}_x(p) = \frac{1}{\rho^2 \sin \vartheta} \d p_t \d p_r \d p_\vartheta \d p_\varphi = \frac{1}{r^2} \d p_t \d p_r \d p_\vartheta \d p_\varphi.
\end{equation}
Here
\begin{equation}
\label{gamma3D}
\bar g = \bar g_{tt} \d t^2 + 2 \bar g_{t\varphi} \d t \d\varphi + \bar g_{rr} \d r^2 + \bar g_{\varphi \varphi} \d\varphi^2
\end{equation}
with
\begin{equation}
\bar g_{t t} = -1 + \frac{2M}{r}, \quad
\bar g_{t \varphi} = - \frac{2 M a}{r}, \quad
\bar g_{r r} = \frac{r^2}{\Delta}, \quad
\bar g_{\varphi \varphi} = r^2 + a^2 + \frac{2 M a^2}{r}
\end{equation}
denotes the metric induced at the equatorial plane. Non-zero contravariant components $\bar g^{\mu\nu}$ read
\begin{equation}
    \bar g^{tt} = - \frac{1}{\Delta} \left( r^2 + a^2 + \frac{2 M a^2}{r} \right), \quad \bar g^{t \varphi} = - \frac{2 M a}{\Delta r}, \quad \bar g^{rr} = \frac{\Delta}{r^2}, \quad \bar g^{\varphi \varphi} = \frac{1}{\Delta} \left( 1 - \frac{2M}{r} \right).
\end{equation}

One can now express the Vlasov equation (\ref{vlasovgeneral}) in terms of the distribution function $f$. We have
\begin{equation}
    \frac{\partial \mathcal F}{\partial x^\mu} \frac{\partial H}{\partial \Pi_\mu} -\frac{\partial \mathcal F}{\partial \Pi_\nu} \frac{\partial H}{\partial x^\nu} = \frac{\partial \mathcal F}{\partial \vartheta} \frac{\partial H}{\partial \Pi_\vartheta} -\frac{\partial \mathcal F}{\partial \Pi_\vartheta} \frac{\partial H}{\partial \vartheta} + \delta \left( \vartheta - \frac{\pi}{2} \right) \delta \left( \Pi_\vartheta \right) \sum_{\mu \neq \vartheta, \nu \neq \vartheta}\left(\frac{\partial f}{\partial x^\mu} \frac{\partial H}{\partial \Pi_\mu} -\frac{\partial f}{\partial \Pi_\nu} \frac{\partial H}{\partial x^\nu}\right).
\end{equation}
On the other hand,
\begin{equation}
    \frac{\partial \mathcal F}{\partial \vartheta} \frac{\partial H}{\partial \Pi_\vartheta} -\frac{\partial \mathcal F}{\partial \Pi_\vartheta} \frac{\partial H}{\partial \vartheta} = \delta^\prime \left( \vartheta - \frac{\pi}{2} \right) \delta(\Pi_\vartheta) \frac{\partial H}{\partial \Pi_\vartheta} f - \delta \left( \vartheta - \frac{\pi}{2} \right) \delta^\prime(\Pi_\vartheta) \frac{\partial H}{\partial \vartheta} f.
\end{equation}
It turns out that both terms in the above expression vanish. We have
\begin{equation}
    \delta^\prime \left( \vartheta - \frac{\pi}{2} \right) \delta(\Pi_\vartheta)  \frac{\partial H}{\partial \Pi_\vartheta} f = \delta^\prime \left( \vartheta - \frac{\pi}{2} \right) \delta(\Pi_\vartheta) g^{\vartheta \vartheta} \Pi_\vartheta f = 0
\end{equation}
and
\begin{equation}
     \delta \left( \vartheta - \frac{\pi}{2} \right) \delta^\prime(\Pi_\vartheta)  \frac{\partial H}{\partial \vartheta} f = \delta \left( \vartheta - \frac{\pi}{2} \right) \delta^\prime(\Pi_\vartheta) \left( \frac{1}{2} \frac{\partial g^{\mu \nu}}{\partial \vartheta} p_\mu p_\nu - q \frac{\partial A_\mu}{\partial \vartheta} p^\mu \right) f = 0,
\end{equation}
where the last equality holds because of assumptions (\ref{equatorialconds}). Thus,
\begin{equation}
     \frac{\partial \mathcal F}{\partial x^\mu} \frac{\partial H}{\partial \Pi_\mu} -\frac{\partial \mathcal F}{\partial \Pi_\nu} \frac{\partial H}{\partial x^\nu} = \delta \left( \vartheta - \frac{\pi}{2} \right) \delta \left( \Pi_\vartheta \right) \sum_{\mu \neq \vartheta, \nu \neq \vartheta}\left(\frac{\partial f}{\partial x^\mu} \frac{\partial H}{\partial \Pi_\mu} -\frac{\partial f}{\partial \Pi_\nu} \frac{\partial H}{\partial x^\nu}\right).
\end{equation}
It follows that $\mathcal F$ satisfies the Vlasov equation (\ref{vlasovgeneral}), provided that $f$ satisfies its three dimensional version
\begin{equation}
\label{vlasovreduced}
    \sum_{\mu \neq \vartheta, \nu \neq \vartheta}\left(\frac{\partial f}{\partial x^\mu} \frac{\partial H}{\partial \Pi_\mu} -\frac{\partial f}{\partial \Pi_\nu} \frac{\partial H}{\partial x^\nu}\right) = 0.
\end{equation}

It is clear from the form of Eq.\ (\ref{vlasovreduced}) that any distribution function $f$ depending on $(x^\mu,\Pi_\nu)$ through constants of motion $m$, $E$ and $L$ satisfies the Vlasov equation (\ref{vlasovreduced}). This result can be generalized as follows (cf.\ \cite{Sarbach2014}). Consider a canonical transformation $(x^\mu,\Pi_\nu) \mapsto (Q^\mu,P_\nu)$, where $P_0$, $P_1$, and $P_2$ are the constants of motion
\begin{equation}
    P_0 \coloneqq m, \qquad P_1 \coloneqq E, \qquad P_2 \coloneqq L
\end{equation}
and the coordinates $Q^\mu$ are defined by the generating function
\begin{equation}
S(x^\mu,P_\nu) \coloneqq \int \Pi_\mu dx^\mu = - (t - t_0) E + (\varphi - \varphi_0) L + \int \Pi_r dr
\end{equation}
as
\begin{equation}
    Q^0 \coloneqq \frac{\partial S}{\partial P_0} = \frac{\partial S}{\partial m}, \qquad Q^1 \coloneqq \frac{\partial S}{\partial P_1} = \frac{\partial S}{\partial E}, \qquad \quad Q^2 \coloneqq \frac{\partial S}{\partial P_2} = \frac{\partial S}{\partial L}.
\end{equation}
Since the Poisson bracket is covariant with respect to canonical transformations, one can rewrite Eq.\ (\ref{vlasovreduced}) as
\begin{equation}
    \frac{\partial f}{\partial Q^\mu} \frac{\partial H}{\partial P_\mu} -\frac{\partial f}{\partial P_\nu} \frac{\partial H}{\partial Q^\nu} = - P_0 \frac{\partial f}{\partial Q^0} = 0,
\end{equation}
where we have used the fact that $H = - m^2/2 = -P_0^2/2$. Hence, any distribution function expressed in new coordinates as
\begin{equation}
    f = f(Q^1,Q^2,P_0,P_1,P_2)
\end{equation}
satisfies the Vlasov equation (\ref{vlasovreduced}). Additional restrictions on the distribution function may follow from symmetry assumptions.

A symmetry generated by the Killing vector $X$ defined on the spacetime manifold $(\mathcal M,g)$ can be extended to the cotangent space $T^\ast \mathcal M$ in terms of the so-called lifts of the Killing vector field $X$. For a Killing vector field on $(\mathcal M,g)$ expressed as
\begin{equation}
X_x = X^\mu(x) \left.  \frac{\partial}{\partial x^\mu}  \right|_x,
\end{equation}
we define the lifts of $X$ as vector fields $\hat X$ on $T^\ast \mathcal M$ given by
\begin{equation}
\hat X_{(x,p)} \coloneqq  X^\mu(x) \left.  \frac{\partial}{\partial x^\mu}  \right|_{(x,p)} - p_\alpha \frac{\partial X^\alpha}{\partial x^\mu}(x) \left. \frac{\partial}{\partial p_\mu} \right|_{(x,p)}.
\end{equation}
The lifts defined in this way satisfy Killing equations with respect to the so-called Sasaki metric---a natural metric on $T^\ast \mathcal M$ \cite{Acuna2022}. We will say that a distribution function $\mathcal F$ is symmetric with respect to the Killing vector $X$, if $\mathcal L_{\hat X} \mathcal F = 0$, where $\mathcal L$ denotes the Lie derivative.

The lifts of the two Killing vectors $k = \partial_t$ and $\eta = \partial_\varphi$, admitted by the Kerr spacetime, read
\begin{equation}
\hat k = \frac{\partial}{\partial t}, \qquad \hat \eta = \frac{\partial}{\partial \varphi}.
\end{equation}
They can be expressed, in new coordinates $(Q^\mu,P_\nu)$ (we work on the equatorial plane) as
\begin{subequations}
\begin{eqnarray}
\hat k & = & \frac{\partial}{\partial t} = \frac{\partial Q^\mu}{\partial t} \frac{\partial}{\partial Q^\mu} + \frac{\partial P_\nu}{\partial t} \frac{\partial}{\partial P_\nu} = - \frac{\partial}{\partial Q^1}, \\
\hat \eta & = & \frac{\partial}{\partial \varphi} = \frac{\partial Q^\mu}{\partial \varphi} \frac{\partial}{\partial Q^\mu} + \frac{\partial P_\nu}{\partial \varphi} \frac{\partial}{\partial P_\nu} = \frac{\partial}{\partial Q^2},
\end{eqnarray}
\end{subequations}
where we have used the fact that
\begin{equation}
\frac{\partial Q^1}{\partial t} = -1, \qquad \frac{\partial Q^0}{\partial t} = \frac{\partial Q^2}{\partial t} = 0, \qquad  \frac{\partial Q^2}{\partial \varphi} = 1, \qquad  \frac{\partial Q^0}{\partial \varphi} = \frac{\partial Q^1}{\partial \varphi} = 0.
\end{equation}
Hence, a distribution function of the form
\begin{equation}
    f = f(m,E,L)
\end{equation}
not only satisfies the Vlasov equation, but it is also stationary and axially symmetric.

The particle current density and the energy-momentum tensor are computed as integrals with respect to momenta $(p_t,p_r,p_\varphi)$. Due to the assumed form of $f$, it is convenient to introduce a coordinate change $(p_t,p_r,p_\varphi) \mapsto (m,E,L)$, where
\begin{subequations}
\begin{eqnarray}
    m^2 & = & - 2 H = - \left( \bar g^{tt} p_t^2 + 2 \bar g^{t \varphi} p_t p_\varphi + \bar g^{rr}p_r^2 +  \bar g^{\varphi \varphi} p_\varphi^2  \right), \\
    E & = & - p_t - q A_t, \\
    L & = & p_\varphi + q A_\varphi.
\end{eqnarray}
\end{subequations}
The corresponding Jacobian reads
\begin{equation}
    \frac{\partial (m^2, E, L)}{\partial (p_t, p_r, p_\varphi)} = - 2 \bar g^{rr} p_r.
\end{equation}
Thus
\begin{equation}
    \d p_t \d p_r \d p_\varphi = \frac{\d m^2 \d E \d L}{2 \left| \bar g^{rr} p_r \right|} =  \frac{m \d m \d E \d L}{\left| \bar g^{rr} p_r \right|}.
\end{equation}
On the other hand,
$\bar g^{rr} p_r^2 = \mathcal G(r,m,E,L)$, where we have defined
\begin{equation}
   \mathcal G(r,m,E,L) \coloneqq -m^2 - \bar g^{tt}(E + q A_t)^2 + 2 \bar g^{t \varphi} (E + q A_t)(L - q A_\varphi) - \bar g^{\varphi \varphi}(L - q A_\varphi)^2,
   \label{eq:Gdef}
\end{equation}
and hence
\begin{equation}
    \d p_t \d p_r \d p_\varphi =\frac{m \d m \d E \d L}{\sqrt{\bar g^{rr} \mathcal G(r,m,E,L)}} = \frac{r m \d m \d E \d L}{\sqrt{\Delta \mathcal G(r,m,E,L)}}.
\end{equation}
We emphasize that the two allowed signs of $p_r$ have to be taken into account in the definition of the integration domain. The particle current surface density can now be expressed as
\begin{equation}
\label{jmusurface1}
J_\mu(r) =  \frac{1}{\sqrt{\Delta}} \int_{\bar {\mathcal P}} f(m,E,L) p_\mu \frac{m \d m \d E \d L}{ \sqrt{ \mathcal G(r,m,E,L)}},
\end{equation}
where $\bar{\mathcal P} \subseteq \bar P_x^+$. The electric current $j_\mu$ now reads
\begin{equation}
\label{jmucurrent}
    j_\mu (r,\vartheta) = \frac{q}{r \sqrt{\Delta}} \delta \left( \vartheta - \frac{\pi}{2} \right) \int_{\bar {\mathcal P}} f(m,E,L) p_\mu \frac{m \d m \d E \d L}{\sqrt{\mathcal G(r,m,E,L)}}.
\end{equation}

As in the case of geometrically thick disks, we restrict ourselves to the so-called simple gas of identical particles and set
\begin{equation}
    f(m,E,L) = \delta(m - m_0) f_0(E,L).
\end{equation}
In this case, expressions \eqref{jmusurface1} and \eqref{jmucurrent} read
\begin{equation}
\label{jmusurface2}
J_\mu(r) =  \frac{m_0}{\sqrt{\Delta}} \int_{\bar {\mathcal P}(m_0)} f_0(E,L) p_\mu \frac{\d E \d L}{ \sqrt{ \mathcal G(r,m_0,E,L)}}
\end{equation}
and
\begin{equation}
    j_\mu (r,\vartheta) = \frac{q m_0}{r \sqrt{\Delta}} \delta \left( \vartheta - \frac{\pi}{2} \right) \int_{\bar {\mathcal P}(m_0)} f_0(E,L) p_\mu \frac{\d E \d L}{\sqrt{\mathcal G(r,m_0,E,L)}},
    \label{eq:jEquatorial}
\end{equation}
where
\begin{equation}
\bar{\mathcal P}(m) \subseteq \bar P_x^+(m) \coloneqq \{ (p_t,p_r,p_\varphi) \colon \bar g^{\mu \nu} p_\mu p_\nu = -m^2, \, p \, \text{is future directed}  \}.
\end{equation}

Same expressions can be also obtained from Eq.\ (\ref{Jthick}). This can be seen as follows. Observe that
\begin{equation}
r \delta(p_\vartheta) \frac{m \d E \d L \d\chi}{\sqrt{\Delta}} = r \delta(\rho \zeta \sin \chi) \frac{m \d E \d L \d \chi}{\sqrt{\Delta}}.
\end{equation}
For a test function $\Phi(\chi)$, one obtains
\begin{equation}
\int \Phi(\chi) r \delta(\rho \zeta \sin \chi) \frac{m \d E \d L \d \chi}{\sqrt{\Delta}} = \int [\Phi(0) + \Phi(\pi)] \frac{m \d E \d L}{r \sqrt{\Delta} \sqrt{\mathcal G(r,m,E,L)}},
\end{equation}
where we have used the fact that in the equatorial plane $\mathcal G = \zeta^2$. Note that the values  $\chi = 0$ and $\chi = \pi$ correspond, respectively, to the signs $\epsilon_r = +1$ and $\epsilon_r = -1$ of the radial momentum
\begin{equation}
\label{pr}
    p_r = \epsilon_r \sqrt{\frac{\mathcal G(r,m,E,L)}{\bar g^{rr}}} = \epsilon_r r \sqrt{\frac{\mathcal G(r,m,E,L)}{\Delta}} = \epsilon_r \frac{r \zeta}{\sqrt{\Delta}}.
\end{equation}

Equation \eqref{pr} has a convenient consequence. Choosing the set $\bar {\mathcal P}$ so that it corresponds to particles moving on bound orbits, say between some radii $r_1$ and $r_2$, one obtains\footnote{In principle, the explicit sum over $\epsilon_r = \pm 1$ in Eq.\ \eqref{zeroradialurrent} could be omitted, since the two branches corresponding to $\epsilon_r = \pm$ can be included in the definition of $\bar {\mathcal P}$. We use the summation symbol to emphasize the existence of those two symmetric branches.}
\begin{equation}
\label{zeroradialurrent}
    j_r = \frac{q}{\Delta} \delta \left( \vartheta - \frac{\pi}{2} \right) \sum_{\epsilon_r = \pm 1} \int_{\bar{\mathcal P}} \epsilon_r f(m,E,L) m \d m \d E \d L = 0.
\end{equation}
That means that the resulting current is purely circular.

\subsection{Constant angular momentum disks}
\label{sec:constLgeneral}

In the following, we consider disks characterized by a constant value $L = L_0$ and set $f_0(E,L) = \delta(L - L_0) \tilde f_0(E)$, similar to thick disks discussed in Sec.\ \ref{sec:thickdisks}.

To proceed further, it is convenient to factor $\mathcal G(r,m,E,L)$ as
\begin{eqnarray}
    \mathcal G(r,m,E,L) & = & - \bar g^{tt} \left[ E - \mathcal W_-(r,m,L) \right] \left[ E - \mathcal W_+(r,m,L) \right] \nonumber \\
    & = & \frac{1}{\Delta} \left( r^2 + a^2 + \frac{2 M a^2}{r} \right) \left[ E - \mathcal W_-(r,m,L) \right] \left[ E - \mathcal W_+(r,m,L) \right],
    \label{eq:GfactorizatioE}
\end{eqnarray}
where
\begin{equation}
\label{potW}
    \mathcal W_\pm(r,m,L) \coloneqq -\frac{1}{\bar g^{tt}} \left[  \bar g^{tt} q A_t - \bar g^{t \varphi} (L - q A_\varphi) \pm \sqrt{(L -q A_\varphi)^2\left[(\bar g^{t\varphi})^2-\bar g^{tt}\bar g^{\varphi\varphi}\right] - m^2 \bar g^{tt}} \right].
\end{equation}
Again, we emphasize that the term $(\bar g^{t \varphi})^2 - \bar g^{tt} \bar g^{\varphi \varphi}$ is a geometrical invariant that can be expressed in terms of the Killing vectors according to Eq.\ (\ref{geominv}). In Boyer--Lindquist coordinates $(\bar g^{t \varphi})^2 - \bar g^{tt} \bar g^{\varphi \varphi} = 1/\Delta$. For $r > r_+$, we have $\mathcal W_+ \ge \mathcal W_-$, and consequently $\mathcal G(r,m,E,L) \ge 0$ for $E \le \mathcal W_-$ or $E \ge \mathcal W_+$. As before, only the case with $E \ge \mathcal W_+$ corresponds to future-pointing momenta. We should stress that $\mathcal W_+$ is essentially an effective radial potential associated with equatorial electrogeodesics.

For $f_0(E,L) = \delta( L - L_0) \tilde f_0(E)$, non-vanishing components in Eq.\ (\ref{eq:jEquatorial}) can be written as
\begin{eqnarray}
     \left( \begin{array}{c} j_t (r,\vartheta) \\ j_\varphi (r,\vartheta) \end{array} \right) & = & \frac{q m_0}{r \sqrt{\Delta}} \delta \left( \vartheta - \frac{\pi}{2} \right) \int_{\bar {\mathcal P}(m_0)} \tilde f_0(E) \left( \begin{array}{c} -E - q A_t \\ L_0  - q A_\varphi \end{array} \right) \frac{\d E}{\sqrt{\mathcal G(r,m_0,E,L_0)}} \nonumber \\
     & = & \frac{q m_0^2 \delta \left( \vartheta - \pi/2 \right)}{M^2 \sqrt{\xi \left[\xi^3+\alpha^2 (2 + \xi)\right]}}  \int_{\bar {\mathcal P}(m_0)} \tilde f_0(E) \left( \begin{array}{c}-\varepsilon - \frac{q}{m_0} A_t \\ M\lambda_0  - \frac{q}{m_0} A_\varphi \end{array} \right)  \frac{\d\varepsilon}{\sqrt{ \left[ \varepsilon - w_+(\xi,\lambda_0) \right] \left[ \varepsilon - w_-(\xi,\lambda_0) \right]}},
\end{eqnarray}
where $w_\pm \coloneqq \mathcal W_\pm/m$. If $\tilde f_0(E) = \mathrm{const}$, the integrals with respect to $\varepsilon$ can be evaluated. We construct our disks by integrating over the energies from $w_+(\xi,\lambda_0)$ to some fixed constant value $\varepsilon_\mathrm{max}$. The latter can be, for instance, set to the value of $w_+$ at a local maximum $\xi = \xi_\mathrm{max}$, i.e., $\varepsilon_\mathrm{max} = w_+(\xi_\mathrm{max}, \lambda_0)$. For $\tilde f_0 = \mathrm{const}$, one could use the following explicitly real expressions
\begin{subequations}
\label{integralseps}
\begin{eqnarray}
\int_{w_+}^W \frac{\d\varepsilon}{\sqrt{(\varepsilon - w_-)(\varepsilon - w_+)}} & = & 2 \, \mathrm{arctanh} \sqrt{\frac{W - w_+}{W - w_-}}, \\
\int_{w_+}^W \frac{\varepsilon \d\varepsilon}{\sqrt{(\varepsilon - w_-)(\varepsilon - w_+)}} & = & \sqrt{(W - w_-)(W - w_+)} + (w_- + w_+) \, \mathrm{arctanh} \sqrt{\frac{W - w_+}{W - w_-}},
\end{eqnarray}
\end{subequations}
where $w_- < w_+ < W$ (cf.\ Appendix E in \cite{Mach2026b}). Alternatively, one can also use the form
\begin{equation}
    \mathrm{arcsinh} \sqrt{\frac{W - w_+}{w_+ - w_-}} = \mathrm{arctanh} \sqrt{\frac{W - w_+}{W - w_-}}.
\end{equation}
This yields
\begin{subequations}
\label{jconsthatlintegrated}
\begin{eqnarray}
     j_t (r,\vartheta) & = & \frac{q m_0^2}{M^2 \sqrt{\xi \left[\xi^3+\alpha^2 (2 + \xi)\right]}} \delta \left( \vartheta - \frac{\pi}{2} \right) \tilde f_0 \int_{w_+(\xi, \lambda_0)}^{\varepsilon_\mathrm{max}} \left( -\varepsilon - \frac{q}{m_0} A_t \right) \frac{\d\varepsilon}{\sqrt{ \left[ \varepsilon - w_+(\xi,\lambda_0) \right] \left[ \varepsilon - w_-(\xi,\lambda_0) \right]}}, \nonumber \\
     & = & - \frac{q m_0^2}{M^2 \sqrt{\xi \left[\xi^3+\alpha^2 (2 + \xi)\right]}} \delta \left( \vartheta - \frac{\pi}{2} \right) \tilde f_0 \left\{ \right.  \sqrt{(\varepsilon_\mathrm{max} - w_-(\xi))(\varepsilon_\mathrm{max} - w_+(\xi))} \nonumber \\
     && \left.  + \left(\frac{2 q A_t}{m_0} + w_-(\xi) + w_+(\xi) \right) \,  \mathrm{arctanh} \sqrt{\frac{\varepsilon_\mathrm{max} - w_+(\xi)}{\varepsilon_\mathrm{max} - w_-(\xi)}} \right\}, \\
     j_\varphi (r,\vartheta) & = & \frac{q m_0^2}{M \sqrt{\xi \left[\xi^3+\alpha^2 (2 + \xi)\right]}} \delta \left( \vartheta - \frac{\pi}{2} \right) \tilde f_0 \int_{w_+(\xi, \lambda_0)}^{\varepsilon_\mathrm{max}} \left( \lambda_0 - \frac{q}{M m_0} A_\varphi \right) \frac{\d\varepsilon}{\sqrt{ \left[ \varepsilon - w_+(\xi,\lambda_0) \right] \left[ \varepsilon - w_-(\xi,\lambda_0) \right]}} \nonumber \\
     & = & \frac{2 q m_0^2}{M \sqrt{\xi \left[\xi^3+\alpha^2 (2 + \xi)\right]}} \delta \left( \vartheta - \frac{\pi}{2} \right) \tilde f_0  \left( \lambda_0 - \frac{q}{M m_0} A_\varphi \right) \mathrm{arctanh} \sqrt{\frac{\varepsilon_\mathrm{max} - w_+(\xi)}{\varepsilon_\mathrm{max} - w_-(\xi)}},
\end{eqnarray}
\end{subequations}
where, to shorten the notation, we have omitted the argument $\lambda_0$ in $w_\pm$. We set $j_t = j_\varphi = 0$ outside the region characterized by $\varepsilon_\mathrm{max} > w_+(\xi, \lambda_0)$ and also sufficiently close to the black hole.

We stress that the integrals appearing in the expressions for $j_t$ and $j_\varphi$ can also be performed analytically for the following choices of $\tilde f_0(E)$:
\begin{subequations}
\begin{align}
    \tilde f_0(E) &= \sum a_n E^n\,, &&\text{arbitrary polynomial}\\
    \tilde f_0(E) &= \frac{P(E)}{R(E)}\,, &&\text{rational function of }E\\
    \tilde f_0(E) &= (E-k)^\mu \,, &&\text{for $k$ and $\mu$ real constants}
\end{align}
\end{subequations}
and, naturally, for linear combinations of the last case. Specific examples for the Wald magnetosphere will be given in subsequent sections.

\subsection{Constant energy disks}

We will now consider constant-energy disks with
\begin{equation}
\label{constEf}
f_0(E,L) = \delta(E - E_0) \bar f_0(L).
\end{equation}
In Equation (\ref{eq:GfactorizatioE}) the function $\mathcal{G}$ (defined in Eq.\ \eqref{eq:Gdef}) is factorized with respect to the energy $E$. It is also possible to factorize this function with respect to $L$:
\begin{align}
    \mathcal G(r,m,E,L) & = - \bar g^{\varphi\varphi} \left[ L -  \Lambda_-(r,m,E) \right] \left[ L - \Lambda_+(r,m,E) \right] \nonumber \\
    & = - \frac{1}{\Delta} \left(1 - \frac{2M}{r} \right) \left[ L -  \Lambda_-(r,m,E) \right] \left[ L - \Lambda_+(r,m,E) \right],
    \label{eq:GfactorizatioL}
\end{align}
where
\begin{align}
    \Lambda_\pm (r,m,E) & \coloneqq \frac{1}{\bar g^{\varphi \varphi}} \left\{(E + q A_t) \bar g^{t \varphi} + q A_\varphi \bar g^{\varphi \varphi} \pm \sqrt{(E + q A_t)^2[(\bar g^{t \varphi})^2 - \bar g^{tt} \bar g^{\varphi \varphi}] - \bar g^{\varphi \varphi} m^2} \right\} \nonumber \\
    & = \frac{1}{1-\frac{2M}{r}} \left\{ - \frac{2Ma}{r}(E + q A_t) +\left( 1 - \frac{2M}{r} \right) q A_\varphi \pm \sqrt{\Delta} \sqrt{(E + q A_t)^2 - \left( 1 - \frac{2M}{r} \right) m^2} \ \right\}.
\end{align}
Outside the ergosphere, for $r > 2 M$, we have $ \Lambda_- <  \Lambda_+$.

Assuming the distribution function of the form (\ref{constEf}), and integrating over energies, we get
\begin{eqnarray}
\label{constEj}
     \left( \begin{array}{c} j_t (r,\vartheta)\\ j_\varphi (r,\vartheta) \end{array} \right)  & = & \frac{q m_0}{r \sqrt{\Delta}} \delta \left( \vartheta - \frac{\pi}{2} \right) \int_{\bar {\mathcal P}(m_0)} \bar f_0(L) \left( \begin{array}{c} -E_0 - q A_t \\ L - q A_\varphi \end{array} \right) \frac{\d L}{\sqrt{\mathcal G(r,m_0,E_0,L)}} \\
     & = & \frac{q m_0}{r \sqrt{1 - \frac{2M}{r}}} \delta \left( \vartheta - \frac{\pi}{2} \right) \int_{\bar {\mathcal P}(m_0)} \bar f_0(L) \left( \begin{array}{c} -E_0 - q A_t \\ L - q A_\varphi \end{array} \right) \frac{\d L}{\sqrt{- [L - \Lambda_-(r,m_0,E_0)][L - \Lambda_+(r,m_0,E_0)]}} \nonumber \\
     & = & \frac{q m_0^2}{M \xi \sqrt{1 - \frac{2}{\xi}}} \delta \left( \vartheta - \frac{\pi}{2} \right) \int_{\bar {\mathcal P}(m_0)} \bar f_0(L) \left( \begin{array}{c} -\varepsilon_0 - \frac{q}{m_0} A_t \\ M \lambda - \frac{q}{m_0} A_\varphi \end{array} \right) \frac{\d\lambda}{\sqrt{- [\lambda - \lambda_-(\xi,\varepsilon_0)][\lambda - \lambda_+(\xi,\varepsilon_0)]}}, \nonumber
\end{eqnarray}
where $\lambda_\pm (\xi,\varepsilon) \coloneqq \Lambda_\pm(r,m_0,E)/(m_0 M)$, and $\varepsilon_0 \coloneqq E_0/m_0$.

As before, for $\bar f_0 = \mathrm{const}$, the integrals over $\lambda$ can be computed with the help of the formulas
\begin{subequations}
\begin{eqnarray}
\int_{\lambda_1}^\lambda \frac{\d \lambda^\prime}{\sqrt{(\lambda^\prime - \lambda_1)(\lambda_2 - \lambda^\prime)}} & = & \pi - 2 \arctan \sqrt{ \frac{\lambda_2 - \lambda}{\lambda - \lambda_1}} = 2 \arcsin \sqrt{\frac{\lambda - \lambda_1}{\lambda_2 - \lambda_1}}, \\
    \int_{\lambda_1}^\lambda \frac{ \lambda^\prime \d \lambda^\prime}{\sqrt{(\lambda^\prime - \lambda_1)(\lambda_2 - \lambda^\prime)}} & = & - \sqrt{(\lambda - \lambda_1)(\lambda_2 - \lambda)} + (\lambda_1 + \lambda_2) \arcsin \sqrt{\frac{\lambda - \lambda_1}{\lambda_2 - \lambda_1}},
\end{eqnarray}
\end{subequations}
where $\lambda_1 < \lambda < \lambda_2$. Another convenient form is
\begin{subequations}
\label{eq:lambdaIntegralsAnalyticAll}
\begin{eqnarray}
\label{eq:lambdaIntegralsAnalytic}
    \int_{\lambda}^{\lambda_2} \frac{\d \lambda^\prime}{\sqrt{(\lambda^\prime - \lambda_1)(\lambda_2 - \lambda^\prime)}} & = & 2 \arctan \sqrt{\frac{\lambda_2 - \lambda}{\lambda - \lambda_1}},\\
    \int_{\lambda}^{\lambda_2} \frac{ \lambda^\prime \d \lambda^\prime}{\sqrt{(\lambda^\prime - \lambda_1)(\lambda_2 - \lambda^\prime)}} & = & \sqrt{(\lambda - \lambda_1)(\lambda_2 - \lambda)} + (\lambda_1 + \lambda_2) \arctan \sqrt{\frac{\lambda_2 - \lambda}{\lambda - \lambda_1}}.
\end{eqnarray}
\end{subequations}
The first form gives, by integrating over $\lambda$ from $\lambda_-(\xi,\varepsilon_0)$ to some fixed $\lambda_\mathrm{max}$:
\begin{subequations}
\label{constEthingeneral}
\begin{eqnarray}
    j_t(r,\vartheta) & = & - \frac{2 q m_0^2 \bar f_0}{M\xi \sqrt{1 - \frac{2}{\xi}}} \delta \left( \vartheta - \frac{\pi}{2} \right) \left( \varepsilon_0 + \frac{q}{m_0}A_t \right) \arcsin \sqrt{\frac{\lambda_\mathrm{max} - \lambda_-(\xi,\varepsilon_0)}{\lambda_+(\xi,\varepsilon_0) - \lambda_-(\xi,\varepsilon_0)}}, \\
    j_\varphi(r,\vartheta) & = &\frac{q m_0^2 \bar f_0}{\xi \sqrt{1 - \frac{2}{\xi}}} \delta \left( \vartheta - \frac{\pi}{2} \right) \Bigg\{ - \sqrt{(\lambda_\mathrm{max} - \lambda_-(\xi,\varepsilon_0))(\lambda_+(\xi,\varepsilon_0) - \lambda_\mathrm{max})} \nonumber \\
    && +  \left[ - \frac{2q}{M m_0} A_\varphi + \lambda_-(\xi,\varepsilon_0) + \lambda_+(\xi,\varepsilon_0) \right] \arcsin \sqrt{\frac{\lambda_\mathrm{max} - \lambda_-(\xi,\varepsilon_0)}{\lambda_+(\xi,\varepsilon_0) - \lambda_-(\xi,\varepsilon_0)}}  \Bigg\}
\end{eqnarray}
\end{subequations}
for the range of $\xi$ in which $\lambda_\mathrm{max} \ge \lambda_-(\xi,\varepsilon_0)$. Using the second form, we get, by integrating over $\lambda$ from a fixed value $\lambda_\mathrm{min}$ to $\lambda_+(\xi,\varepsilon_0)$:
\begin{subequations}
\label{constEthingeneral2}
\begin{eqnarray}
    j_t(r,\vartheta) & = & - \frac{2 q m_0^2 \bar f_0}{M\xi \sqrt{1 - \frac{2}{\xi}}} \delta \left( \vartheta - \frac{\pi}{2} \right) \left( \varepsilon_0 + \frac{q}{m_0}A_t \right) \arctan \sqrt{\frac{\lambda_+(\xi,\varepsilon_0) - \lambda_\mathrm{min}}{\lambda_\mathrm{min} - \lambda_-(\xi,\varepsilon_0)}}, \\
    j_\varphi(r,\vartheta) & = &\frac{q m_0^2 \bar f_0}{\xi \sqrt{1 - \frac{2}{\xi}}} \delta \left( \vartheta - \frac{\pi}{2} \right) \Bigg\{ \sqrt{(\lambda_\mathrm{min} - \lambda_-(\xi,\varepsilon_0))(\lambda_+(\xi,\varepsilon_0) - \lambda_\mathrm{min})} \nonumber \\
    && +  \left[ - \frac{2q}{M m_0} A_\varphi + \lambda_-(\xi,\varepsilon_0) + \lambda_+(\xi,\varepsilon_0) \right] \arctan \sqrt{\frac{\lambda_+(\xi,\varepsilon_0) - \lambda_\mathrm{min}}{\lambda_\mathrm{min} - \lambda_-(\xi,\varepsilon_0)}} \Bigg\}
\end{eqnarray}
\end{subequations}
for the range of $\xi$ in which $\lambda_\mathrm{min} \le \lambda_+(\xi,\varepsilon_0)$.

\section{Equatorial disks in the Wald magnetosphere}
\label{sec:thinWald}

At the equatorial plane, Wald's electromagnetic potential \eqref{WaldKerr} with the electric charge $Q = 0$ reduces to
\begin{equation}
\label{WaldEqPlane}
    A_\mu (\vartheta = \pi/2) = \frac{1}{2} B_0 \left( -2 a + \frac{2 M a}{r},0,0, r^2 + a^2 - \frac{2 M a^2}{r} \right),
\end{equation}
or
\begin{equation}
\label{AWalddimless}
    \frac{q}{m}A_t(\vartheta = \pi/2) = - 2 \alpha \beta \left(1 - \frac{1}{\xi} \right), \qquad \frac{q}{m M} A_\varphi(\vartheta = \pi/2) = \beta \left( \xi^2 + \alpha^2 - \frac{2 \alpha^2}{\xi} \right),
\end{equation}
where, to express the components $A_\mu$ in the dimensionless form, we have introduced $\beta \coloneqq q B_0 M/(2m)$, as in Sec.\ \ref{sec:thickdisks}.

In the following subsections, we will discuss examples of thin disks confined to the equatorial plane and immersed in the Wald magnetosphere.

\subsection{Constant $L$}

Substituting Eqs.\ (\ref{AWalddimless}) in the expressions derived in Sec.\ \ref{sec:constLgeneral} for constant $L$ disks, we get
\begin{equation}
\label{potwdimless}
   w_\pm = \frac{2 \alpha(\lambda+\beta\xi\bar{\Delta}) \pm \sqrt{\bar \Delta} \sqrt{w_0}}{\Xi_+} \,, \qquad
   w_0 =\xi\Xi_+ + \left[ -\lambda\xi+\beta\Xi_-\right]^2,
\end{equation}
where, to shorten notation, we have introduced
\begin{equation}
\Xi_+ \coloneqq \xi^3+\alpha^2(2+\xi), \qquad \Xi_- \coloneqq \xi^3+\alpha^2(-2+\xi).
\end{equation}
Note that if $\beta \neq 0$, $w_+ \to +\infty$ for $\xi \to +\infty$. As a consequence, disks made up of particles with a constant $L$ cannot extend to infinity.

\begin{figure}
\centering
\subfloat[Ranges of $\varepsilon$ for $\beta = 1/2$, $\lambda_0 = 20$
\label{fig:constE-ranges}]
{\includegraphics[width=\threefig]{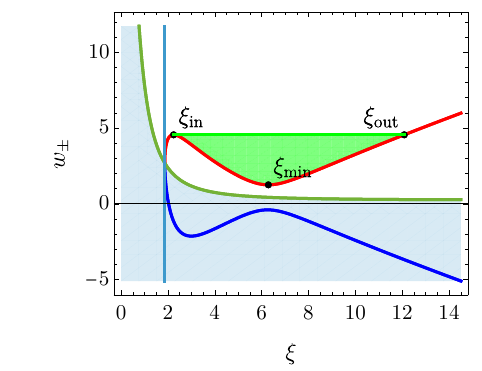}}
\hfill
\subfloat[$\tilde j_t$]
{\includegraphics[width=\threefig]{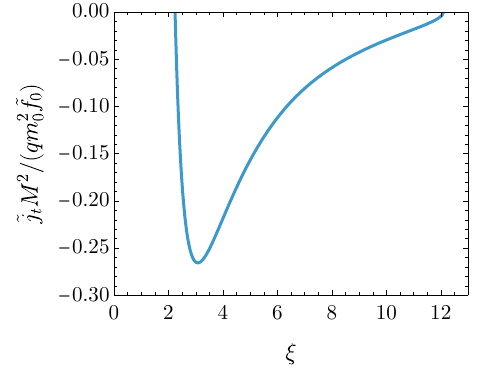}}
\hfill
\subfloat[$\tilde j_\varphi$]
{\includegraphics[width=\threefig]{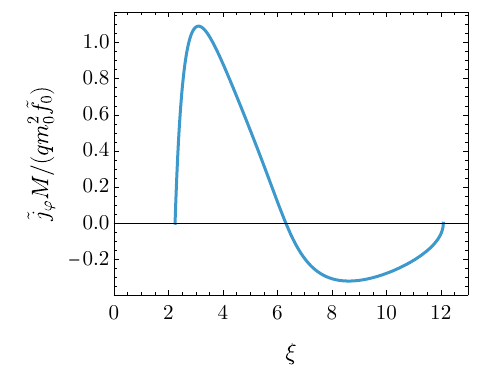}}

\vspace{1ex}

\subfloat[Ranges of $\varepsilon$ for $\beta = -1/2$, $\lambda_0 = 24$]
{\includegraphics[width=\threefig]{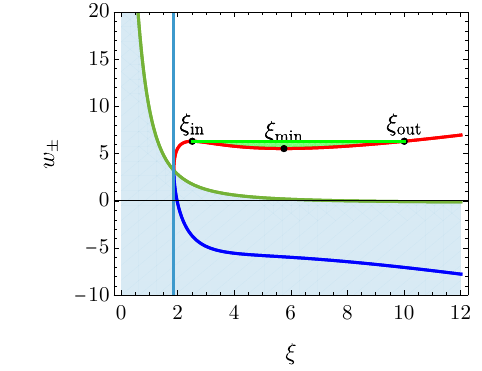}}
\hfill
\subfloat[$\tilde j_t$]
{\includegraphics[width=\threefig]{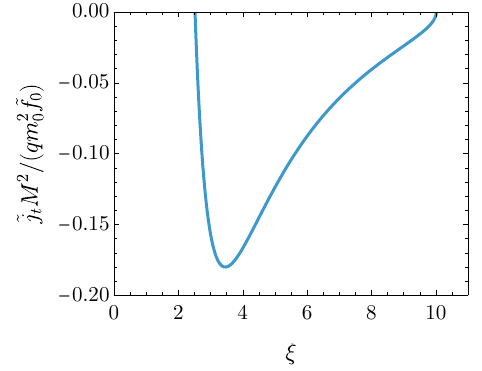}}
\hfill
\subfloat[$\tilde j_\varphi$]
{\includegraphics[width=\threefig]{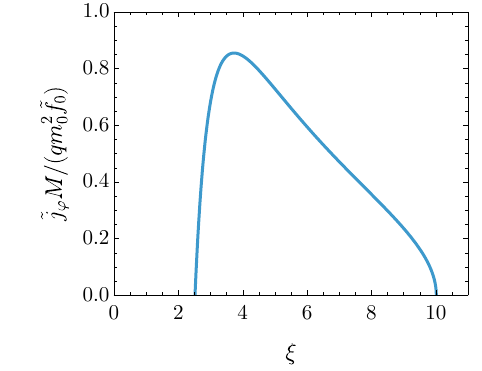}}

\vspace{1ex}

\subfloat[Ranges of $\varepsilon$ for $\beta = -1/2$, $\lambda_0 = -24$]
{\includegraphics[width=\threefig]{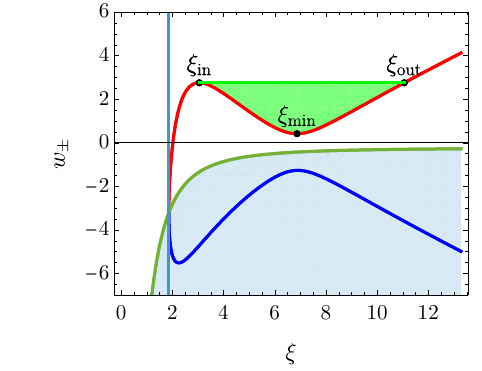}}
\hfill
\subfloat[$\tilde j_t$]
{\includegraphics[width=\threefig]{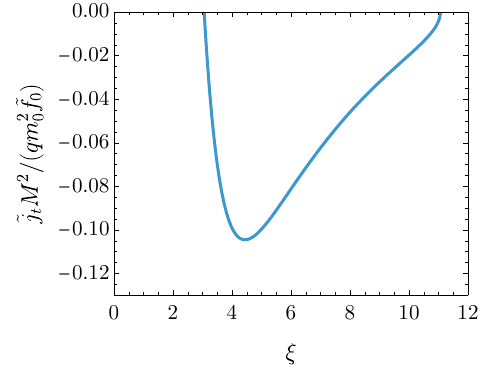}}
\hfill
\subfloat[$\tilde j_\varphi$]
{\includegraphics[width=\threefig]{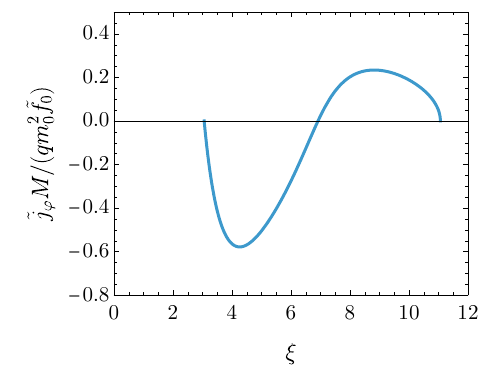}}

\caption{\label{fig:jtjphiconstf} Equatorial models with the fixed angular momentum $\lambda = \lambda_0$. In all figures $\alpha=1/2$. Left panels: The behavior of effective potentials $w_+$ (red) and $w_-$ (blue). Vertical lines mark the locations of the black hole horizon. The blue-shaded region is forbidden---the corresponding momenta are no longer future-directed. The regions marked in green correspond to the currents shown in the middle and right panels. The maximum value of the energy $\varepsilon_\mathrm{max}$ is set to the value of $w_+$ at its local maximum (at $\xi = \xi_\mathrm{in}$). The disks obtained with this choice are, in a sense, maximal---setting $\varepsilon_\mathrm{max}$ to a higher value would include trajectories plunging into the black hole.  Middle and right panels show the components of the electric current for the model with $\tilde f_0 = \mathrm{const}$, computed according to Eqs.\ (\ref{thinconstLwald}), where, in addition, we exclude the region with $\xi < \xi_\mathrm{in}$. We plot normalized values of $\tilde j_t$ and $\tilde j_\varphi$, defined in Eq.\ (\ref{tildej}).
Middle panels: Time components of the electric current. Right panels: Angular components of the electric current. Note that $\tilde j_\varphi$ can change the sign.}
\end{figure}

We choose the value of $\lambda = \lambda_0$, such that $w_+(\xi)$ has a maximum at $\xi = \xi_\mathrm{max}$ and a minimum at $\xi = \xi_\mathrm{min}$, and grows to $+ \infty$ for $\xi > \xi_\mathrm{min}$. This allows for a disk located between $\xi_\mathrm{max}$ and $\xi_\mathrm{out}$, where $\xi_\mathrm{out}$ satisfies $w_+(\xi_\mathrm{max}) = w_+(\xi_\mathrm{out})$. For any $\xi \in (\xi_\mathrm{max},\xi_\mathrm{out})$, we are allowed to select energies $\varepsilon$ in the range $ \varepsilon \in (w_+(\xi),w_+(\xi_\mathrm{max}))$. Configurations obtained in that way are---for a fixed value of $\lambda_0$---maximal. One is also free to select the energies in a smaller range, leading to disk configurations with a smaller spatial extent. In what follows, we choose $\varepsilon \in (w_+(\xi),\varepsilon_\mathrm{max})$, where $\varepsilon_\mathrm{max} \in [w_+(\xi_\mathrm{min}), w_+(\xi_\mathrm{max})]$. Finally, since
\begin{equation}
L_0 - q A_\varphi = m M \left[ \lambda_0 - \beta \left( \xi^2 + \alpha^2 - \frac{2 \alpha^2}{\xi} \right) \right] = mM\left(\lambda_0-\beta\Xi_-/\xi\right),
\end{equation}
we get
\begin{equation}
    \left( \begin{array}{c} j_t (r,\vartheta) \\ j_\varphi (r,\vartheta) \end{array} \right) 
      =  \frac{q m_0^2 \delta \left( \vartheta - \pi/2 \right)}{M^2 \sqrt{\xi \left[\xi^3+\alpha^2 (2 + \xi)\right]}} \int_{w_+(\xi)}^{\varepsilon_\mathrm{max}} \left( \begin{array}{c} - \varepsilon + 2 \beta \alpha \left( 1 - \frac{1}{\xi} \right) \\ M \lambda_0  - M \beta \left( \xi^2 + \alpha^2 - \frac{2 \alpha^2}{\xi} \right)  \end{array} \right) \frac{ \tilde f_0(E) \d \varepsilon}{\sqrt{ \left[ \varepsilon - w_+(\xi, \lambda_0) \right] \left[ \varepsilon - w_-(\xi, \lambda_0) \right]}},
\end{equation}
for $\xi \in (\xi_1,\xi_2)$, where $\xi_\mathrm{min} \le \xi_1 \le \xi_2$, and $w_+(\xi_1) = w_+(\xi_2) = \varepsilon_\mathrm{max}$. For $\tilde f_0(E) = \mathrm{const}$, one obtains
\begin{subequations}
\label{thinconstLwald}
\begin{align}
    j_\varphi(r,\vartheta) &= - \frac{2 m_0^2 q \tilde f_0 \delta(\vartheta - \pi/2)}{M}\frac{\beta \Xi_- - \lambda_0 \xi}{\xi^{3/2}\sqrt{\Xi_+}}\,
    \mathrm{arctanh} \, \sqrt{\frac{ \varepsilon_\mathrm{max} - w_+(\xi)}{\varepsilon_\mathrm{max} - w_-(\xi)}} ,\\
    j_t(r,\vartheta) &= -\frac{2 \alpha}{M \Xi_+}\,j_\varphi(r,\vartheta)  -\frac{m_0^2 q \tilde f_0 \delta(\vartheta - \pi/2)}{M^2}\frac{\sqrt{(\varepsilon_\mathrm{max} - w_+(\xi))(\varepsilon_\mathrm{max} - w_-(\xi))}}{\sqrt{\xi\Xi_+}},
\end{align}
\end{subequations}
where we have abbreviated $w_+(\xi) =  w_+( \xi, \lambda_0 )$, and $w_-(\xi) = w_-( \xi, \lambda_0 )$.

Sample plots of the effective potentials $w_\pm$ as well as the components of the electric current surface density $j_t$ and $j_\varphi$, computed according to Eqs.\ (\ref{thinconstLwald}), are shown in Fig.\ \ref{fig:jtjphiconstf}. To simplify the labels in these plots, we define $\tilde j_t(\xi)$ and $\tilde j_\varphi(\xi)$ by
\begin{equation}
\label{tildej}
    j_t(\xi,\vartheta) \eqqcolon \delta(\vartheta - \pi/2) \tilde j_t(\xi), \qquad  j_\varphi(\xi,\vartheta) \eqqcolon \delta(\vartheta - \pi/2) \tilde j_\varphi(\xi).
\end{equation}

\subsection{Constant $E$}\label{sec:constantE}

Assuming the Wald form of the electromagnetic potential (\ref{AWalddimless}), we obtain 
\begin{align}
    \label{eq:lambdasDefs}
    \lambda_\pm (\xi,\varepsilon) &=\frac{\beta \xi\bar{\Delta}-2\alpha \varepsilon \pm \sqrt{\bar{\Delta}}\sqrt{\lambda_0}}{\xi-2} \,, & 
    \lambda_0 &= \left[\varepsilon\xi-2\alpha\beta(\xi-1)\right]^2-(\xi-2)\xi.
\end{align}
and
\begin{equation}
     \left( \begin{array}{c} j_t (r,\vartheta)\\ j_\varphi (r,\vartheta) \end{array} \right)  =  \frac{q m_0^2 \delta \left( \vartheta - \pi/2 \right)}{M \xi \sqrt{1 - \frac{2}{\xi}}}  \int_{\tilde P^+_x} \bar f_0(L) \left( \begin{array}{c} -\varepsilon_0 + 2 \alpha \beta \left(1 - \frac{1}{\xi} \right) \\ M \left[\lambda - \beta \left(\alpha^2 - \frac{2 \alpha^2}{\xi} + \xi^2 \right)\right] \end{array} \right) \frac{\d\lambda}{\sqrt{[\lambda - \lambda_-(\xi,\varepsilon_0)][\lambda_+(\xi,\varepsilon_0) - \lambda]}}.
\end{equation}
In this case, choosing suitable integration limits turns out to be more involved than for constant angular momentum disks discussed in the previous section. The obvious requirement is to demand that $\lambda_-<\lambda<\lambda_+$ for all $\xi > 2$, i.e., for disks located outside the ergosphere.

We can now discuss the apparent singularity of $\lambda_\pm$ at $\xi = 2$. Let us define an auxiliary parameter $\Gamma \coloneqq \alpha(\varepsilon - \alpha\beta)$. The behavior of $\lambda_\pm$ at $\xi = 2$ splits into the following cases:
\begin{enumerate}
    \item $\Gamma > 0$: $\lambda_+$ is continuous at $\xi = 2$ and $\lambda_-\to \pm\infty$ as $\xi \to 2^{\mp}$, respectively;
    \item $\Gamma < 0$: $\lambda_-$ is continuous at $\xi = 2$ and $\lambda_+\to \pm\infty$ as $\xi \to 2^{\pm}$, respectively.
\end{enumerate}
This can be seen as follows. One has, at $\xi = 2$,
\begin{equation}
    \beta \xi \bar \Delta - 2 \alpha \varepsilon \pm \sqrt{\bar \Delta} \sqrt{\lambda_0} = 2(-\Gamma \pm \sqrt{\Gamma^2}) = 2 (-\Gamma \pm |\Gamma|).
\end{equation}
This result implies that for $\Gamma > 0$, $\lambda_- \to \pm \infty$ for $\xi \to 2^\mp$ and, conversely, for $\Gamma < 0$, $\lambda_+ \to \pm \infty$ for $\xi \to 2^\pm$. The statement about continuity of $\lambda_\pm$ at $\xi = 2$ follows from the following representation:
\begin{equation}
    \lambda_\pm = \frac{(\beta \xi\bar{\Delta}-2\alpha \varepsilon)^2 -  \bar{\Delta}\lambda_0}{(\xi-2) \left( \beta \xi\bar{\Delta}-2\alpha \varepsilon \mp \sqrt{\bar{\Delta}}\sqrt{\lambda_0} \right)} = \frac{\mathcal P(\xi,\alpha,\beta,\varepsilon)}{ \beta \xi\bar{\Delta}-2\alpha \varepsilon \mp \sqrt{\bar{\Delta}}\sqrt{\lambda_0} },
\end{equation}
where $\mathcal P(\xi,\alpha,\beta,\varepsilon)$ is a polynomial with respect to $\xi$ (as well as $\alpha$, $\beta$, and $\varepsilon$), and we have assumed that $\bar \Delta \ge 0$ and $\lambda_0 \ge 0$. Note that $\lambda_\pm$ are real only if $\lambda_0(\xi) \geq 0$, which is a quadratic polynomial in $\xi$. The discriminant of this polynomial is $\delta := 1 + 4\beta\Gamma$. There are cases for which $\delta < 0$ and $\lambda_\pm$ are real for all $\xi \geq \xi_+$, but there also exist examples with $\delta > 0$, for which the sign of $\lambda_0$ can change (see, e.g.\ Figs.\ \ref{fig:specialTypeDisks}). Provided that $\lambda_\pm$ are real, they behave asymptotically as $\lambda_\pm \to \pm\infty$  for $\xi\to +\infty$ when $\sgn{\beta} = \pm 1$, respectively.

\begin{figure}
\centering
\subfloat[Ranges of $\lambda$ for $\beta = 1/2$, $\varepsilon_0 = 20$ \label{fig:constEBetaPositiveRanges}]{\includegraphics[width=\threefig]{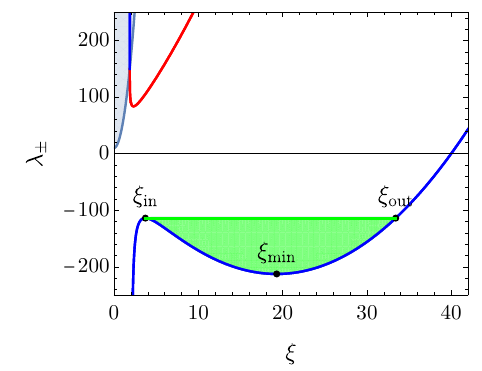}}
\hfill
\subfloat[$\tilde j_t$, $\bar f_0(\lambda) = \mathrm{const}$]{\includegraphics[width=\threefig]{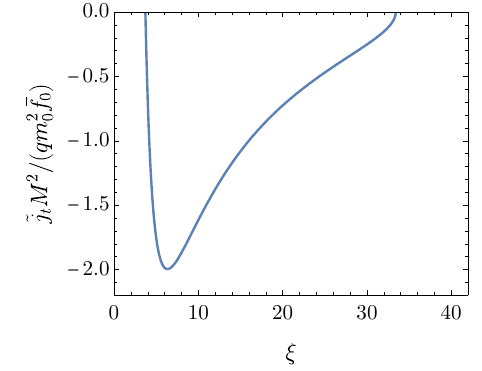}}
\hfill
\subfloat[$\tilde j_\varphi$, $\bar f_0(\lambda) = \mathrm{const}$]{\includegraphics[width=\threefig]{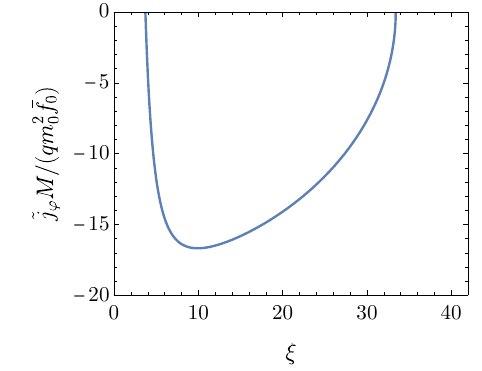}}

\vspace{1ex}
    
\subfloat[Ranges of $\lambda$ for $\beta = 1/2$, $\varepsilon_0 = 20$]{\includegraphics[width=\threefig]{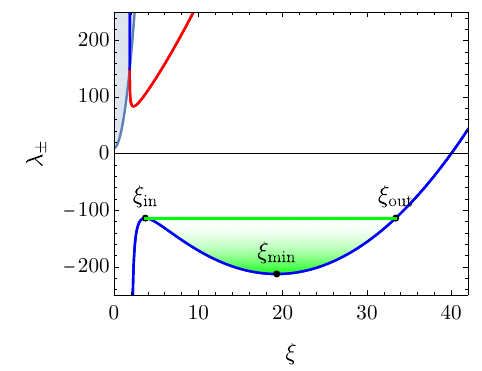}}
\hfill
\subfloat[$\tilde j_t$, $\bar f_0(\lambda) = f_2(\lambda)$]{\includegraphics[width=\threefig]{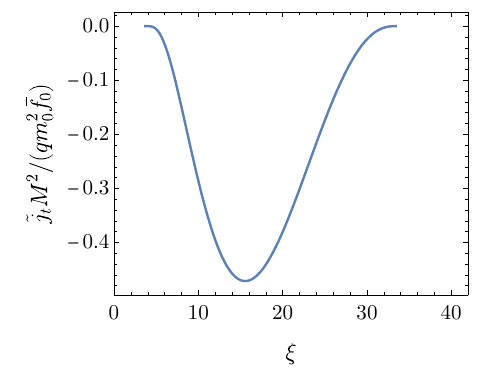}}
\hfill
\subfloat[$\tilde j_\varphi$, $\bar f_0(\lambda) = f_2(\lambda)$]{\includegraphics[width=\threefig]{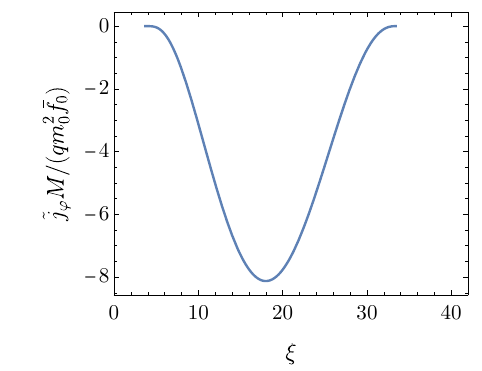}}

\vspace{1ex}

\subfloat[Ranges of $\lambda$ for $\beta = -1/2$, $\varepsilon_0 =20$ \label{fig:constEBetaNegativeRanges}]{\includegraphics[width=\threefig]{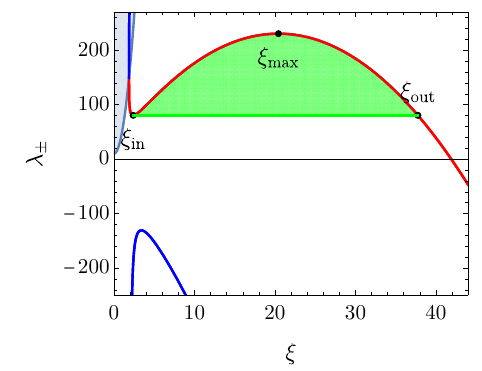}}
\hfill
\subfloat[$\tilde j_t$, $\bar f_0(\lambda) = \mathrm{const}$]{\includegraphics[width=\threefig]{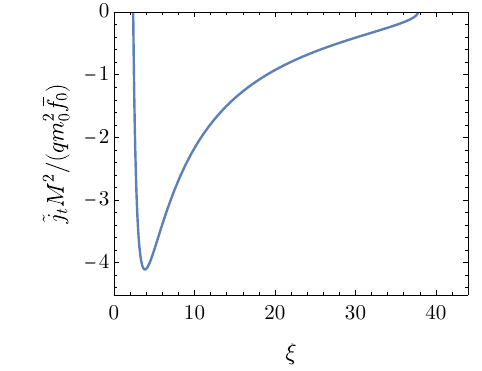}}
\hfill
\subfloat[$\tilde j_\varphi$, $\bar f_0(\lambda) = \mathrm{const}$]{\includegraphics[width=\threefig]{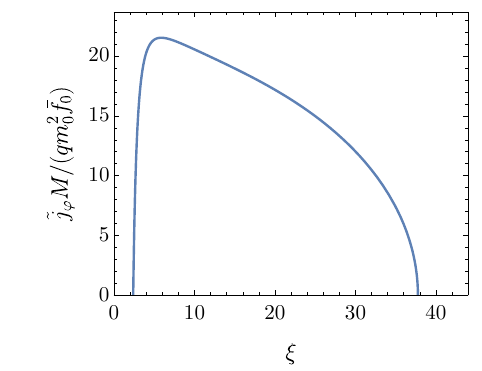}}

\caption{Equatorial models with the fixed energy $\varepsilon = \varepsilon_0$. In all panels $\alpha = 1/2$. Left panels: Plots of $\lambda_-$ (blue) and $\lambda_+$ (red). Middle panels: Time components of the current surface density $\tilde j_t$. Right panels: Azimuthal component of the current surface density $\tilde j_\varphi$.  In left panels, the blue shaded region in top-left corners is forbidden---the corresponding momenta are no longer future-directed. In the top row, in Fig.\ \ref{fig:constEBetaPositiveRanges} the local maximum and minimum of $\lambda_-$ occur at $\xi_{\mathrm{in}}$ and $\xi_\mathrm{min}$, respectively. The value of $\lambda_\mathrm{max}$ is chosen so that $\lambda_\mathrm{max} = \lambda_-(\xi_\mathrm{in})$. This also defines the outer edge of the disk at $\xi_\mathrm{out} > \xi_\mathrm{in}$, such that $\lambda_-(\xi_\mathrm{out}) = \lambda_\mathrm{max}$. The currents shown in middle and right panels of the top row are obtained assuming $\bar f_0(\lambda) = \mathrm{const}$ and integrating over  $\lambda \in [ \lambda_-(\xi), \lambda_\mathrm{max}]$ for $\xi\in [\xi_\mathrm{in},\,\xi_\mathrm{out}]$ (the corresponding region is shown in green in Fig.\ \ref{fig:constEBetaPositiveRanges}). The second row shows a similar model with  the distribution function $\bar f_0(\lambda) = f_2(\lambda) := \bar{f}_0 [(\lambda_\mathrm{max} - \lambda)/(\lambda_\mathrm{max} - \lambda_-(\xi_\mathrm{min}))]^2$, where, with a slight abuse of notation, the symbol $\bar f_0$ without the argument $\lambda$ is used to denote a scaling constant. The third row shows an analogous model, corresponding to a co-rotating disk, instead of counter-rotating ones. In this case, $\bar f_0(\lambda) = \mathrm{const}$.}
\label{fig:Econst}
\end{figure}

\begin{figure}
\centering
\subfloat[Ranges of $\lambda$ for $\alpha = 1/2$, $\varepsilon_0 = 2$, $\beta = 3/2$]{\includegraphics[width=\threefig]{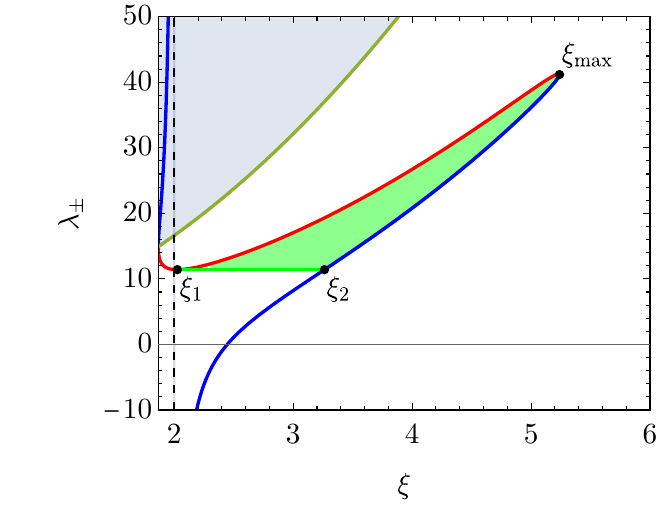}}
\hfill
\subfloat[$\tilde j_t$]{\includegraphics[width=\threefig]{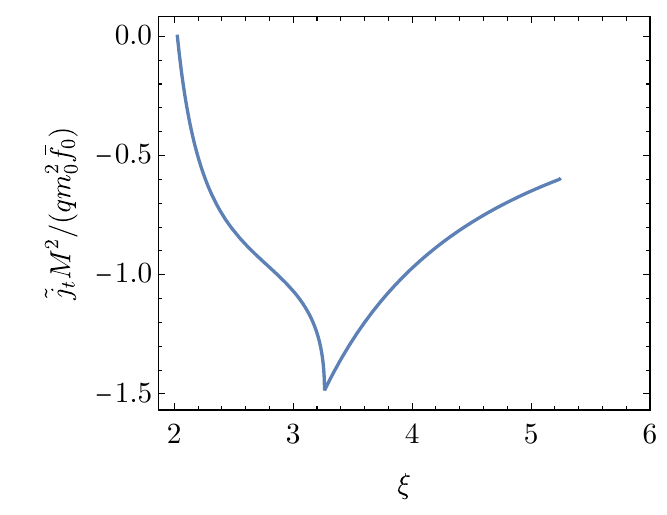}}
\hfill
\subfloat[$\tilde j_\varphi$]{\includegraphics[width=\threefig]{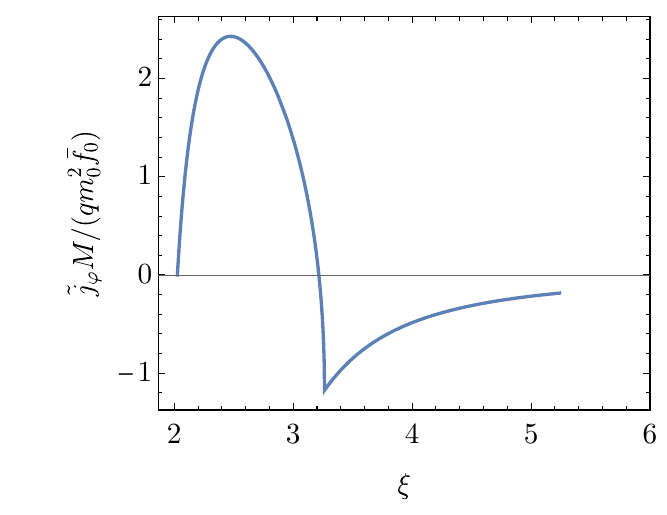}}

\vspace{1ex}

\subfloat[Ranges of $\lambda$ for $\alpha = .95$, $\varepsilon_0 = 6$, $\beta = 5$]{\includegraphics[width=\threefig]{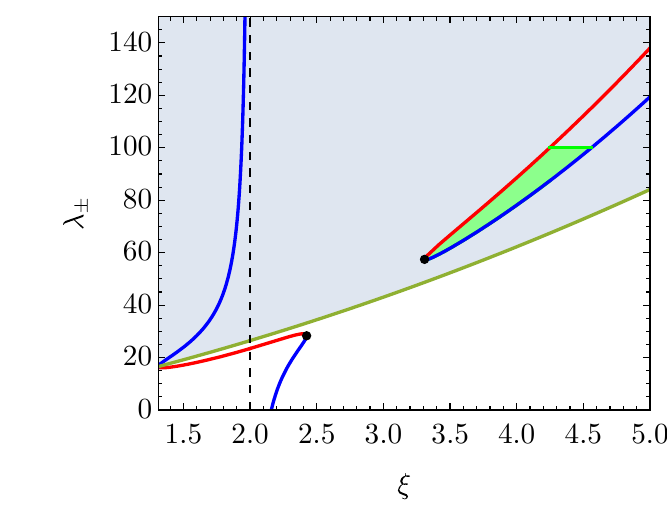}}
\hfill
\subfloat[$\tilde j_t$]{\includegraphics[width=\threefig]{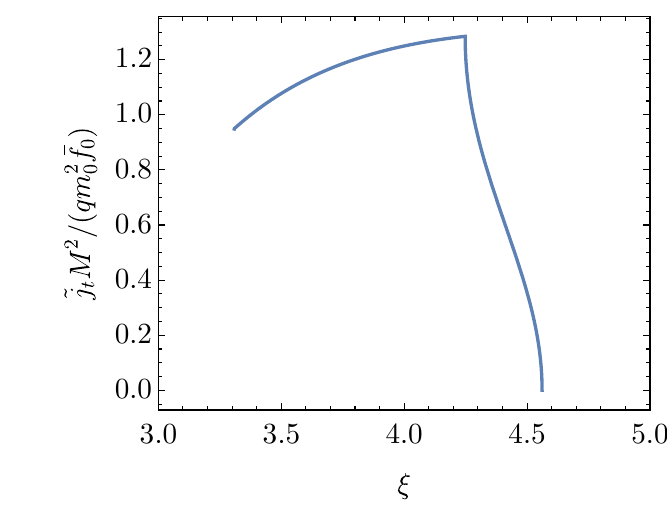}}
\hfill
\subfloat[$\tilde j_\varphi$]{\includegraphics[width=\threefig]{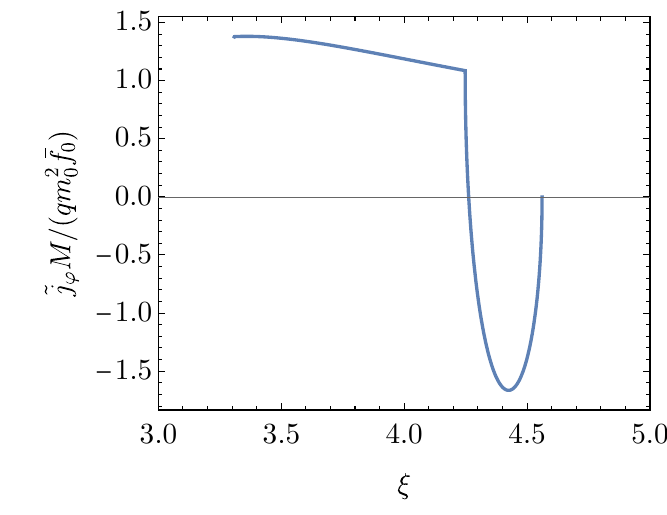}}

\caption{Same as in Fig.\ \ref{fig:Econst}. We plot two special cases of monoenergetic disks with $\bar f_0 = \mathrm{const}$. Upper row: The branches of $\lambda_+$ and $\lambda_-$ merge at $\xi = \xi_\mathrm{max} \approx 5.24$. Integration with respect to $\lambda$ is performed for $\lambda$ between $\lambda_\pm$ and $\lambda \geq \lambda_1 \approx 11.41$ which is the value at $\xi = \xi_1 \approx 2.03$ (the minimum of $\lambda_+(\xi)$). Lower row: real solutions $\lambda_\pm$ reappear for $\xi$ large enough. The radii at which $\lambda_\pm$ coincide are $\xi \approx 2.4$ and $\xi = \xi_{\mathrm{min}} \approx 3.31$. Integration is performed for $\lambda$ between $\lambda_\pm$ and $\lambda \leq \lambda_1 = 100$. One can check that all trajectories contributing to (b--c) are future-directed, meanwhile the disk (e--f) consists of purely past-directed trajectories, hence it is nonphysical. In this case $j_t$ is positive.}
\label{fig:specialTypeDisks}
\end{figure}

For $\xi_+ < \xi < 2$ (within the ergosphere), we have $\lambda_+ \le \lambda_-$ for $\lambda_0(\xi) \ge 0$. Consequently, $\mathcal G$ remains positive for $\lambda < \lambda_+$ or $\lambda > \lambda_-$. Sample graphs of potentials $\lambda_\pm(\xi)$ are shown in Fig.\ \ref{fig:Econst}.

The criterion for the future-oriented momentum, $- N^\mu p_\mu > 0$, reads, in dimensionless variables,
\begin{equation}
    -N^\mu p_\mu = \frac{m}{\bar \Delta \xi} \left\{ - 2 \alpha^3 \beta \xi + \varepsilon \xi^3+\alpha^2 \varepsilon(2+\xi)-2 \alpha \left[ \lambda + \beta (-2 + \xi) \xi^2 \right] \right\}  > 0.
\end{equation}
It can also be expressed in a slightly simpler form by considering the vector
\begin{equation}
    X = (r^2 + a^2) \partial_t + a \partial_\varphi.
\end{equation}
It is easy to check that $g(X,X) = - \Delta \rho^2$, so that $X$ is timelike for $r > r_+$. For $r > r_+$, it is also future directed, since $g(X,N) = -(r^2 + a^2) < 0$. As such, $X$ can also be used to test the time orientation. The criterion for the future-oriented momentum at the equatorial plane can now be written as
\begin{equation}
    - X^\mu p_\mu = M^2 m \left[ -\alpha(\beta\bar{\Delta}+\lambda)+\varepsilon\left(\alpha^2+\xi^2\right) \right] > 0.
\label{eq:futuredirecteddimless}
\end{equation}

\begin{figure}
\centering
\subfloat[Regions of $\lambda$ for $\alpha = 1/2$, $\varepsilon_0 = 3$, $\beta = -1/2$]{\includegraphics[width=\threefig]{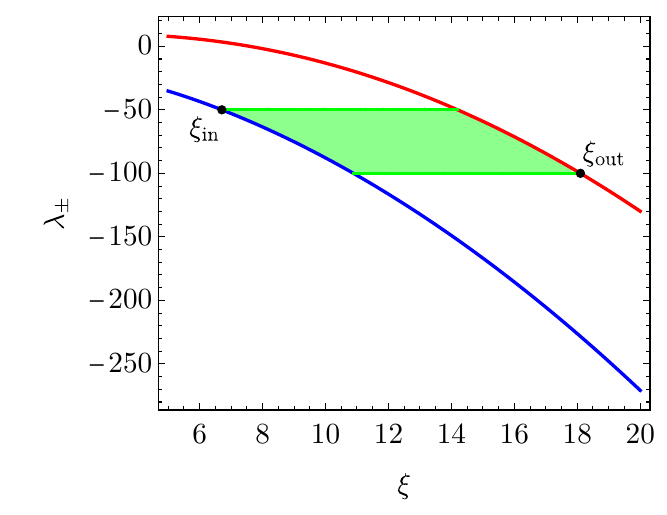}}
\hfill
\subfloat[$\tilde j_t$]{\includegraphics[width=\threefig]{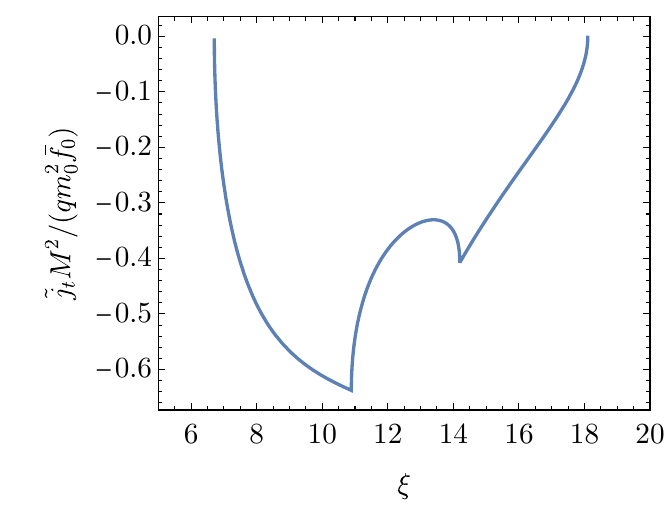}}
\hfill
\subfloat[$\tilde j_\varphi$]{\includegraphics[width=\threefig]{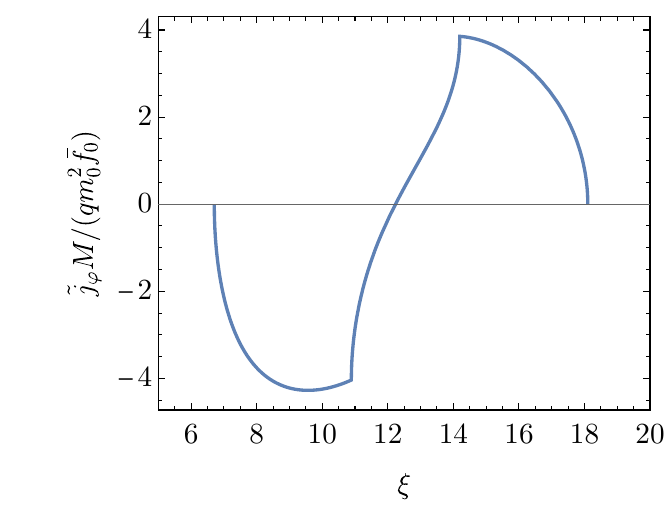}}

\vspace{1ex}

\subfloat[Regions of $\lambda$ for $\alpha = .95$, $\varepsilon_0 = 10$, $\beta = 5$]{\includegraphics[width=\threefig]{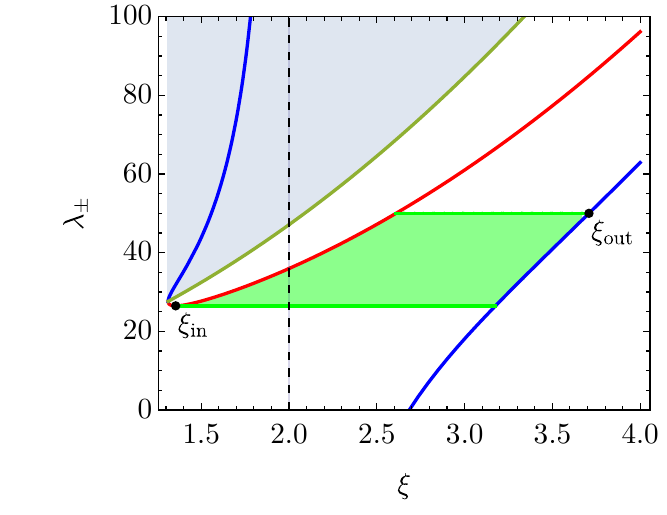}}
\hfill
\subfloat[$\tilde j_t$]{\includegraphics[width=\threefig]{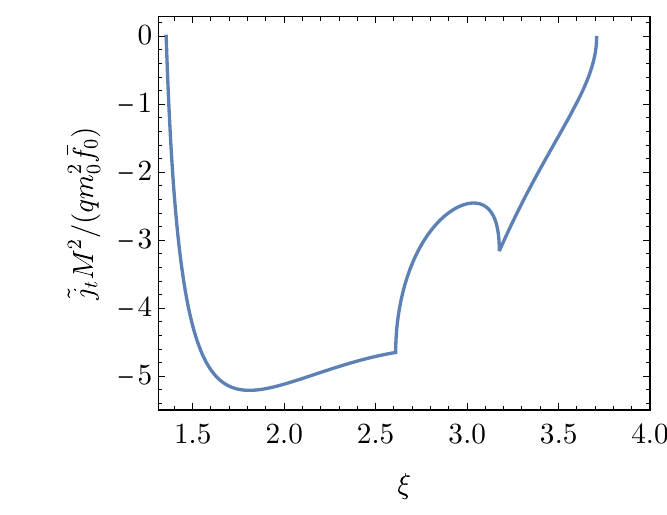}}
\hfill
\subfloat[$\tilde j_\varphi$]{\includegraphics[width=\threefig]{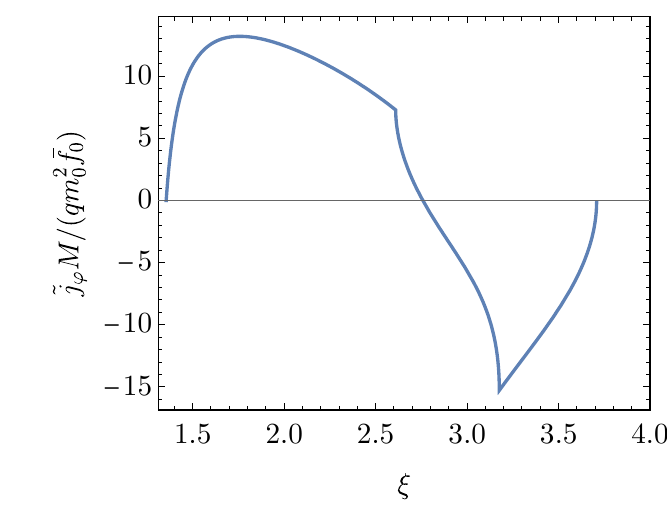}}

\vspace{1ex}

\subfloat[Regions of $\lambda$ for $\alpha = .9$, $\varepsilon_0 = 3$, $\beta = -1/2$]{\includegraphics[width=\threefig]{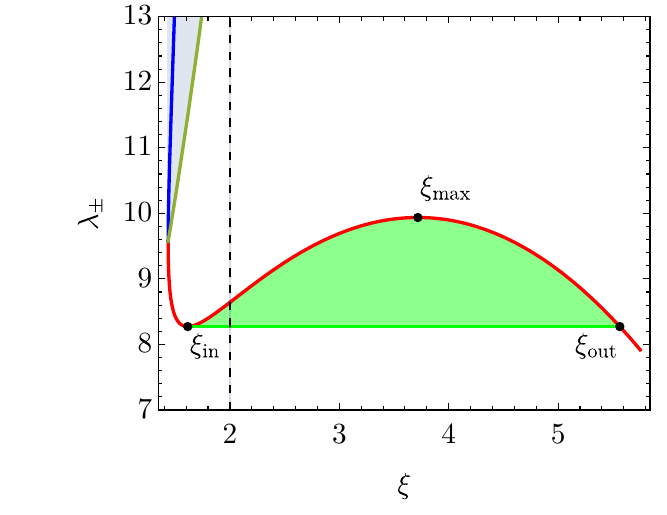}}
\hfill
\subfloat[$\tilde j_t$]{\includegraphics[width=\threefig]{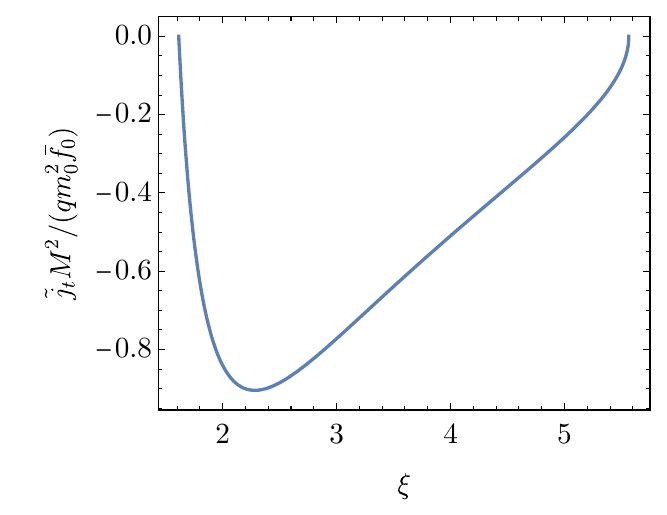}}
\hfill
\subfloat[$\tilde j_\varphi$]{\includegraphics[width=\threefig]{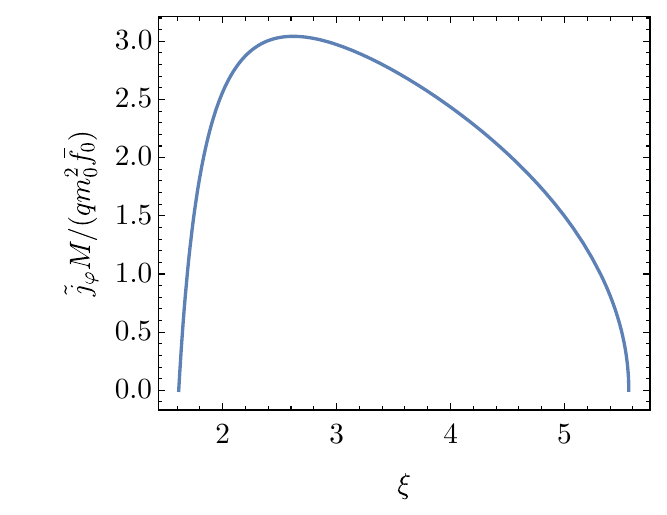}}
        
\caption{Same as in Fig.\ \ref{fig:Econst}. In all cases $\bar f_0 = \mathrm{const}$. Regions of integration with respect to $\lambda$ are marked in green in left panels. Models shown in the second and third rows correspond to disks entering the ergosphere $\xi \leq 2$.}
\label{fig:altDiskIntegrationLims}
\end{figure}

Explicit solutions corresponding to $\bar f_0(L) = \mathrm{const}$ can be obtained as follows. Since
\begin{subequations}
\begin{eqnarray}
\varepsilon_0 + \frac{q}{m_0} A_t & = & \frac{\varepsilon_0 \xi - 2 \alpha \beta (\xi - 1)}{\xi}, \\
    - \frac{2 q}{M m_0} A_\varphi + \lambda_+(\xi,\varepsilon_0) + \lambda_-(\xi,\varepsilon_0) & = & - \frac{4 \alpha [\varepsilon_0 \xi - 2 \alpha \beta (\xi - 1)]}{\xi (\xi - 2)},
\end{eqnarray}
\end{subequations}
we get, from Eqs.\ (\ref{constEthingeneral}), i.e., by integrating over $\lambda$ from $\lambda_-(\xi,\varepsilon_0)$ to $\lambda_\mathrm{max}$:
\begin{subequations}
\label{thinnconstEwald1}
\begin{align}
    j_t (r,\vartheta) &= \frac{2m_0^2 q \bar f_0 \delta(\vartheta - \pi/2)}{M}\frac{2 \alpha \beta (\xi - 1) - \varepsilon_0 \xi}{\xi^{3/2}\sqrt{\xi-2}}\,\arcsin\sqrt{\frac{\lambda_\mathrm{\max} - \lambda_- (\xi,\varepsilon_0)}{\lambda_+(\xi,\varepsilon_0)-\lambda_-(\xi,\varepsilon_0)}}\,,\\
    j_\varphi(r,\vartheta) &= \frac{2\alpha M}{\xi-2}\,j_t(r,\vartheta) - m_0^2 q \bar f_0 \delta \left( \vartheta - \frac{\pi}{2} \right) \frac{\sqrt{(\lambda_+(\xi,\varepsilon_0)-\lambda_\mathrm{max})(\lambda_\mathrm{\max}-\lambda_-(\xi,\varepsilon_0))}}{\sqrt{\xi(\xi-2)}}.
\end{align}
\end{subequations}
Similarly, Eqs.\ (\ref{constEthingeneral2}) yield, by integrating over $\lambda$ from $\lambda_\mathrm{min}$ to $\lambda_+(\xi,\varepsilon_0)$,
\begin{subequations}
\label{thinnconstEwald2}
\begin{eqnarray}
    j_t(r,\vartheta) & = & - \frac{2 q m_0^2 \bar f_0 \delta \left( \vartheta - \pi/2 \right)}{M}  \frac{\varepsilon_0 \xi - 2 \alpha \beta (\xi - 1)}{\xi^{3/2} \sqrt{\xi - 2}} \arctan \sqrt{\frac{\lambda_+(\xi,\varepsilon_0) - \lambda_\mathrm{min}}{\lambda_\mathrm{min} - \lambda_-(\xi,\varepsilon_0)}}, \\
    j_\varphi(r,\vartheta) & = & \frac{2 \alpha M}{\xi - 2} j_t(r,\vartheta) + \frac{q m_0^2 \bar f_0 \delta \left( \vartheta - \pi/2 \right)}{ \sqrt{\xi(\xi - 2)}}  \sqrt{(\lambda_\mathrm{min} - \lambda_-(\xi,\varepsilon_0))(\lambda_+(\xi,\varepsilon_0) - \lambda_\mathrm{min})}.
\end{eqnarray}
\end{subequations}
Examples of solutions given by Eqs.\ (\ref{thinnconstEwald1}) and (\ref{thinnconstEwald2}) are shown in Fig.\ \ref{fig:Econst}, \ref{fig:specialTypeDisks}, and \ref{fig:altDiskIntegrationLims}, together with the corresponding behavior of potentials $\lambda_\pm$.

We emphasize that for constant-energy models, the condition \eqref{eq:futuredirecteddimless} has to be checked separately. Figure \ref{fig:specialTypeDisks} shows examples of (unphysical) disk models containing particles with past-oriented momenta, for which $\lambda_- < \lambda < \lambda_+$.

\begin{figure}
    \centering
    \includegraphics[width=0.5\linewidth]{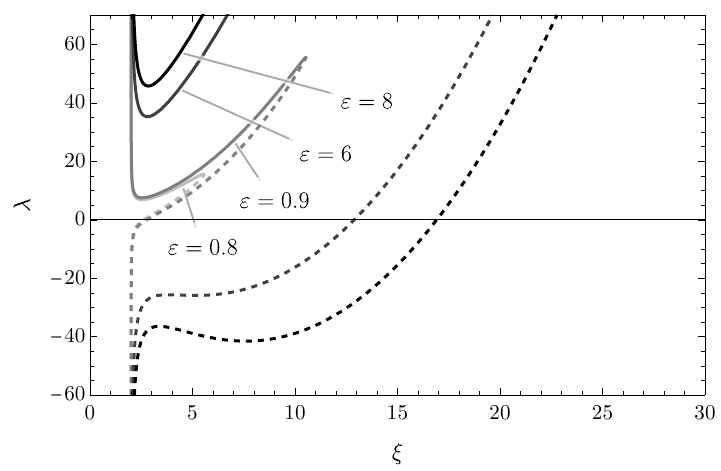}
    \caption{Critical values $\lambda_+(\xi,\varepsilon)$ (solid lines) and $\lambda_-(\xi,\varepsilon)$ (dashed lines) for $\alpha = 0$ and $\beta = 1/2$. The location $\xi$ at which $\lambda_+(\xi,\varepsilon) = \lambda_-(\xi,\varepsilon)$  tends to $+\infty$ for $\varepsilon \nearrow 1$.}
    \label{fig:critical-e}
\end{figure}

By checking the condition
\begin{equation}
\label{inf1}
\lambda_+(\xi,\varepsilon)-\lambda_-(\xi,\varepsilon)=0
\end{equation}
one can realize that there is a possibility to construct infinitely extended disks, composed of particles moving on bound orbits. This can be seen as follows. Except for $\xi = \xi_\pm$, condition \eqref{inf1} implies that $\lambda_0 = 0$. For $\varepsilon \neq \pm 1 + 2 \alpha \beta$, this is a quadratic equation with respect to $\xi$, with the solutions of the form
\begin{equation}
\label{inf2}
\xi=\frac{-1 - 2\alpha\beta(\varepsilon-2\alpha\beta) \pm \sqrt{1+4\alpha\beta(\varepsilon-\alpha\beta)}}{(\varepsilon-2\alpha\beta)^2-1}.
\end{equation}
This expression can diverge for $\varepsilon \to \pm 1 + 2 \alpha \beta$. As the simplest example, consider the case with $\alpha = 0$. In this case, Eq.\ \eqref{inf2} yields two roots $\xi = 0$ and $\xi = 2/(1 - \varepsilon^2)$. The latter tends to $+ \infty$ for $\varepsilon \nearrow 1$. This behavior is illustrated in Fig.\ \ref{fig:critical-e}. Also note that, in this case, both $\lambda_\pm$ tend to  infinity, as $\xi \to \infty$.

%%%%%%%%%%%

\subsection{Three dimensional behavior of $\mathcal G$}
\label{subsec:3DG}

\begin{figure}
\centering
\subfloat[Crossection $\varepsilon=10$ (cf.\  Fig.~\ref{fig:constEBetaPositiveRanges}).]{\includegraphics[width=0.9\twofig]{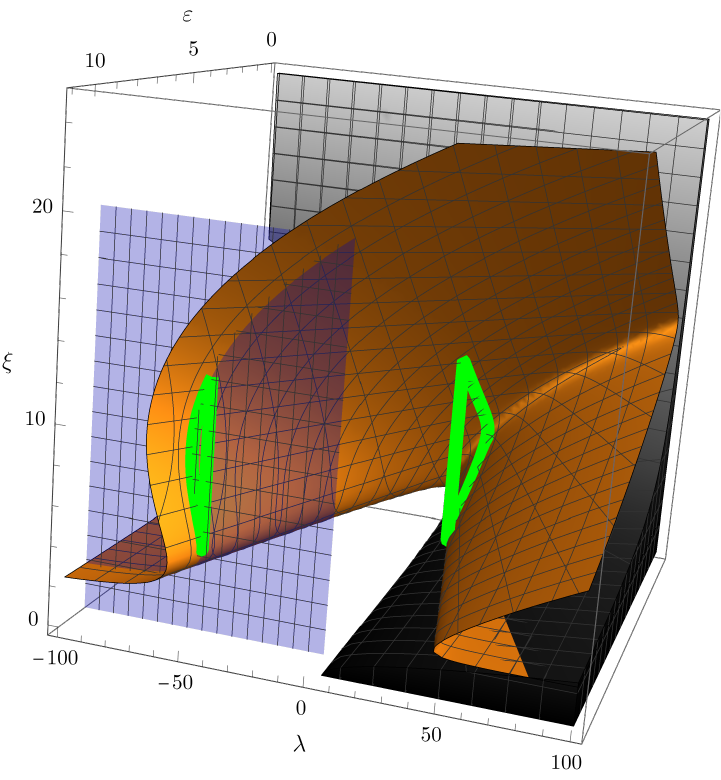}}
\hspace{1em}
\subfloat[Crossection $\lambda=20$ (cf.\ Fig.~\ref{fig:constE-ranges}).]{\includegraphics[width=0.9\twofig]{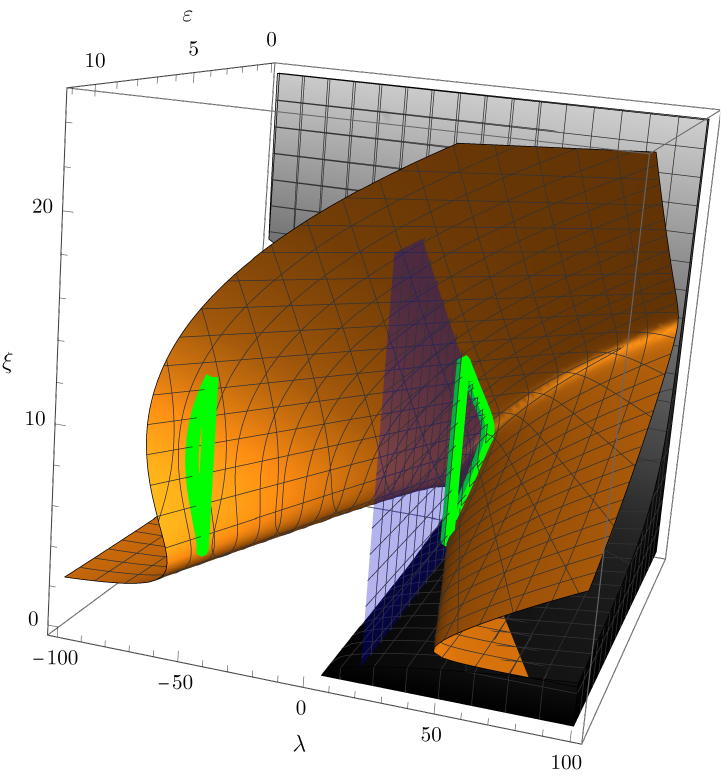}}

\caption{3D picture of zeros of $\hat{\mathcal{G}}$ with $\alpha=1/2$ and $\beta=1/2$ and integration regions for constant $\varepsilon$ or $\lambda$ (depicted by blue opaque planes). The region corresponding to four-momenta pointing to the past is depicted in black/gray. When comparing the subfigure (b) to Fig.~\ref{fig:constE-ranges} note that in the 3D picture the negative part of the $\varepsilon$ axis is missing.}
\label{fig:3D}
\end{figure}

It is also instructive to show the general behavior of the function $\mathcal G$ in the specific case of the Wald magnetosphere with $Q  = 0$. In this case, the dimension-less variant of $\mathcal{G}$, defined as $\hat{\mathcal{G}} = \mathcal G/m^2$, can be written as
\begin{align}
\label{Ghat}
    \hat{\mathcal{G}} (\xi,\varepsilon,\lambda) &= -1-\frac{\left[-2\alpha\varepsilon+\alpha^2\beta\xi+(\xi-2)(\xi^2\beta-\lambda)\right]^2}{\bar{\Delta}(\xi-2)\xi}+\frac{\left[-2\alpha\beta(\xi-1)+\varepsilon\xi\right]^2}{(\xi-2)\xi}.
\end{align}
For a fixed $\xi$, the condition $\hat {\mathcal G}(\xi,\varepsilon,\lambda) = 0$ represents an implicitly defined quadratic curve in the $(\varepsilon,\,\lambda)$ plane. Its discriminant reads $4/\bar \Delta$; it is independent of $\alpha$ and $\beta$ and is positive for $\xi > \xi_+$. Consequently, for $\xi > \xi_+$, the condition $\hat{\mathcal G}(\xi,\varepsilon,\lambda) = 0$ defines a hyperbola in the $(\varepsilon,\,\lambda)$ plane. Sample 3D plots of zeros of $\hat{\mathcal G}(\xi,\varepsilon,\lambda)$ are shown in Fig.~\ref{fig:3D}.

Note that the functions $w_\pm(\xi,\lambda)$ and $\lambda_\pm(\xi,\varepsilon)$ can be understood as implicitly defined implicitly by the relations $\hat{\mathcal{G}}(\xi,w_\pm,\lambda)=\hat{\mathcal{G}}(\xi,\varepsilon,\lambda_\pm)=0$. A trivial consequence of the implicit function theorem states that the extrema of $w_+(\xi,\lambda)$ with respect to $\xi$ coincide with the extrema of $\lambda_\pm(\xi,\varepsilon)$.

%Added by TG

\section{Equatorial Vlasov disks in the Wald magnetosphere: Schwarzschild limit}
\label{sec:SchwarzschildLimit}

In this section, we consider stationary disks confined to the equatorial plane, characterized by more general distribution functions $f_0(E,L)$. We exclude particle trajectories plunging into the black hole. To provide a characterization of the corresponding phase-space regions, we have to find, for each value of the radius $\xi$, the allowed range of energy $E$, and, for every value of the energy $E$ in that range, the values of $L$ that lead to bound orbits.

The general expressions for the effective potentials $\mathcal W_\pm$ and $\Lambda_\pm$ (or $w_\pm$ and $\lambda_\pm$ in dimensionless form) obtained in the preceding sections are difficult to analyze in full generality. The situation simplifies in the Schwarzschild limit $a \to 0$, in which Eq.\ (\ref{potW}) yields
\begin{equation}\label{eq:effPotWald}
            w_+(\xi) = \frac{1}{\xi^2}\left[- \xi^2 \frac{q A_t}{m} + \sqrt{\bar\Delta}\sqrt{\xi^2 + \left( \lambda - \frac{q A_\varphi}{m M} \right)^2}\right].
        \end{equation}
In this case, the Wald solution can also be written in a simpler form. For $a = 0$ and $Q = 0$, we obtain from Eq.\ (\ref{WaldEqPlane}):
\begin{equation}\label{eq:WaldMagSpec}
A_t = 0, \quad A_\varphi = \frac{1}{2}B_0r^2.
\end{equation}
Consequently, Eq.\ (\ref{potwdimless}) simplifies to
\begin{equation}
\label{eq:effPotWaldSchw}
w_+(\xi) = \sqrt{\frac{(\xi-2)[\xi^2 + (\lambda - \beta\xi^2)^2]}{\xi^3}}.
\end{equation}
It is also convenient to note that for $\alpha = 0$, Eq.\ \eqref{Ghat} reduces to the form
\begin{equation}
\label{Ghatazero}
    \hat{\mathcal{G}} (\xi,\varepsilon,\lambda) = -1 - \frac{(\xi^2 \beta - \lambda)^2}{\xi^2} + \frac{\varepsilon^2 \xi}{\xi - 2}.
\end{equation}

In this case, construction of disks consisting of particles with varying energy and momentum becomes viable. We choose the energy cut-off $\varepsilon_{\mathrm{max}}$ for which bound trajectories can be found and integrate over all lower energies $\varepsilon \leq \varepsilon_{\mathrm{max}}$ and angular momenta $\lambda$ for which the inner radial turning point also exists, i.e., the particles do not plunge into the black hole.

For $\alpha = 0$ the currents (\ref{eq:jEquatorial}) can be written as
\begin{eqnarray}
             \left( \begin{array}{c} j_t (r, \vartheta) \\ j_\varphi (r, \vartheta) \end{array} \right) & = & \frac{q m_0^2\delta\left(\vartheta - \pi/2 \right)}{M \sqrt{\xi(\xi-2)}} \int_{\varepsilon_{\mathrm{min}}(\xi)}^{\varepsilon_{\mathrm{max}}(\xi)}\int_{\lambda \in \mathbf{\Lambda}} f_0(E, L) \left( \begin{array}{c}-\varepsilon - \frac{q}{m_0} A_t \\ M\lambda  - \frac{q}{m_0} A_\varphi \end{array} \right)  \frac{d\lambda d\varepsilon}{\sqrt{- [\lambda - \lambda_-(\xi,\varepsilon)][\lambda - \lambda_+(\xi,\varepsilon)]}} \nonumber \\
             & = & \frac{q m_0^2\delta\left( \vartheta - \pi/2 \right)}{M \sqrt{\xi(\xi-2)}} \int_{\varepsilon_{\mathrm{min}}(\xi)}^{\varepsilon_{\mathrm{max}}(\xi)}\int_{\lambda \in \mathbf{\Lambda}} f_0(E, L) \left( \begin{array}{c}-\varepsilon\\ M(\lambda - \beta\xi^2) \end{array} \right)  \frac{d\lambda d\varepsilon}{\sqrt{- [\lambda - \lambda_-(\xi,\varepsilon)][\lambda - \lambda_+(\xi,\varepsilon)]}}\,,
\label{eq:schwWaldCurrents}
\end{eqnarray}
in the appropriately determined bounds $\mathbf{\Lambda}$, which we will now discuss qualitatively. A more detailed analysis of the effective potential (\ref{eq:effPotWaldSchw}), determination of the inner and outer radii of the disk, and derivation of the integration bounds can be found in Appendix \ref{app:schwarzschilBounds}.
In explicit examples given in this section, these bounds are determined numerically.

\begin{figure}
    \centering
    \subfloat[The plots of $\lambda_\pm(\xi,\varepsilon)$ for different values of $\varepsilon$ and $\beta=0$. The existence of bounded orbits starts for $\varepsilon\geq2\sqrt{2}/3\approx.94$, where the ISCO starts to exist. The $\lambda_\mathrm{min,max}^{\circlearrowleft,\circlearrowright}$ are plotted for $\varepsilon=0.99$.]{\includegraphics[keepaspectratio,width=\twofig]{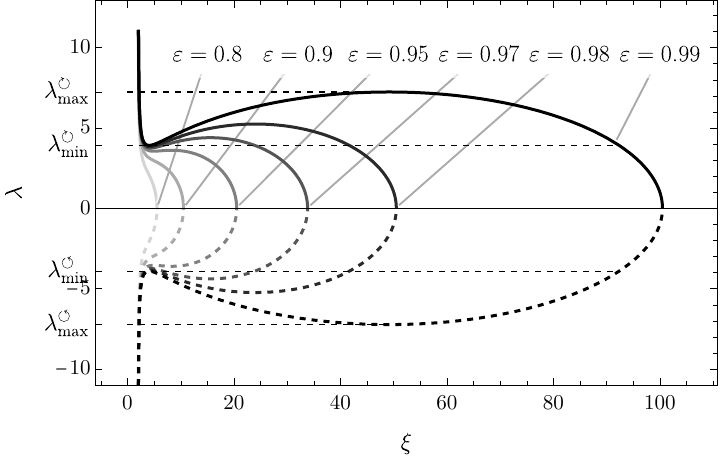}\label{subfig:epsilon}}\hfill
    \subfloat[The plots of $\lambda_\pm(\xi,\varepsilon=0.75)$ for different values of $\beta$.]{\includegraphics[keepaspectratio,width=\twofig]{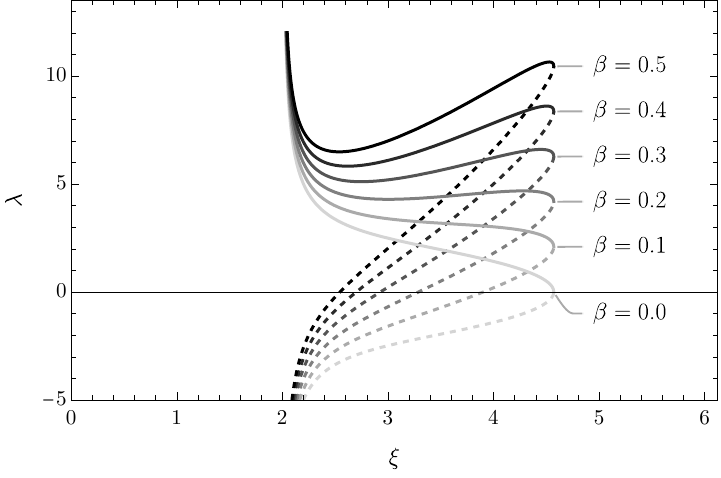}\label{subfig:beta}}
    \caption{The plots of $\lambda_+(\xi,\varepsilon)$ (full line) and $\lambda_-(\xi,\varepsilon)$ (dashed) for different configurations. The particle can exist in the interval $[\lambda_-(\xi,\varepsilon),\lambda_+(\xi,\varepsilon)]$ and the bounded orbits exist in union of intervals $\left[\lambda^\circlearrowleft_{\mathrm{max}}(\varepsilon),\lambda^\circlearrowleft_{\mathrm{min}}(\varepsilon) \right] \cup \left[ \lambda^\circlearrowright_{\mathrm{min}}(\varepsilon),\lambda^\circlearrowright_{\mathrm{max}}(\varepsilon) \right]$.}
    \label{fig:Schw-eb}
\end{figure}

In the first step, we set the upper bound on the allowed energy $\varepsilon_\mathrm{max}$ and search for the values of $\lambda$ corresponding to the 
Innermost Stable Circular Orbit (ISCO), i.e., we solve the system of equations
\begin{equation}
\frac{\partial w_+(\xi,\lambda)}{\partial\xi} =0 \,, \qquad
\frac{\partial^2 w_+(\xi,\lambda)}{\partial\xi^2} =0\,,
\label{eq:ISCO}
\end{equation}
for $\xi_{\mathrm{ISCO}}$ and $\lambda_{\mathrm{ISCO}}$. The corresponding value of $w_+$ is denoted by $\varepsilon_{\mathrm{ISCO}}$. Bound orbits can only exist for the energies $\varepsilon \ge \varepsilon_\mathrm{ISCO}$, and consequently for $\varepsilon_\mathrm{max} \ge \varepsilon_\mathrm{ISCO}$. As a reference, one can think of the Schwarzschild spacetime with no magnetic field. In this case, the solution to Eqs.~(\ref{eq:ISCO}) is given by $\xi_\mathrm{ISCO} = 6$, $\lambda_\mathrm{ISCO} = \pm 2\sqrt{3}$, $\varepsilon_\mathrm{ISCO} = 2\sqrt{2}/3$. For energies $\varepsilon \ge \varepsilon_\mathrm{ISCO}$, bound orbits exist for angular momenta $\lambda$ in a union of intervals $\left[\lambda^\circlearrowleft_{\mathrm{max}}(\varepsilon),\lambda^\circlearrowleft_{\mathrm{min}}(\varepsilon) \right] \cup \left[ \lambda^\circlearrowright_{\mathrm{min}}(\varepsilon),\lambda^\circlearrowright_{\mathrm{max}}(\varepsilon) \right]$. Here, the superscripts $\circlearrowright$ and $\circlearrowleft$ refer to co- and counter-rotating particles, respectively (see Fig.\ (\ref{subfig:epsilon}) for visualization). On the Schwarzschild background with no magnetic field, these two channels are symmetric with respect to the transformation $\lambda\rightarrow-\lambda$, and a distribution function $f_0(\varepsilon,\lambda)$ even in $\lambda$ leads to disks with a non-zero charge density $j_t$ and a vanishing current component $j_\varphi$. In other words, the co- and counter-rotating streams of charged particles balance each other.

\begin{figure}
    \centering
    %%%%%%%%%%%
    \subfloat[3D contour of $\hat{\mathcal{G}}(\xi,\varepsilon,\lambda)=0$.]{\includegraphics[keepaspectratio,width=\twofig]{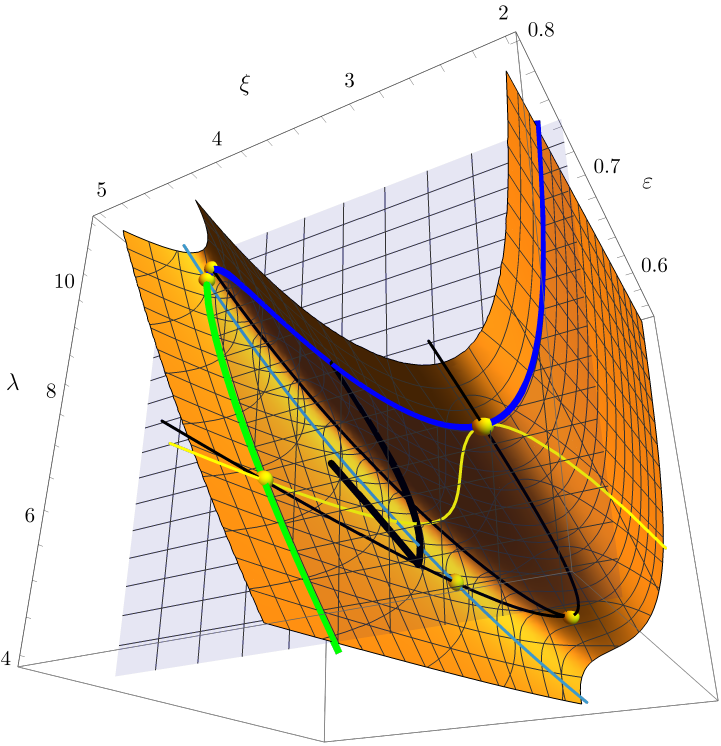}}\hfill
    %%%%%%%%%%%
    \subfloat[Orthographic projection of (a) suppressing $\varepsilon$.]{\includegraphics[keepaspectratio,width=\twofig]{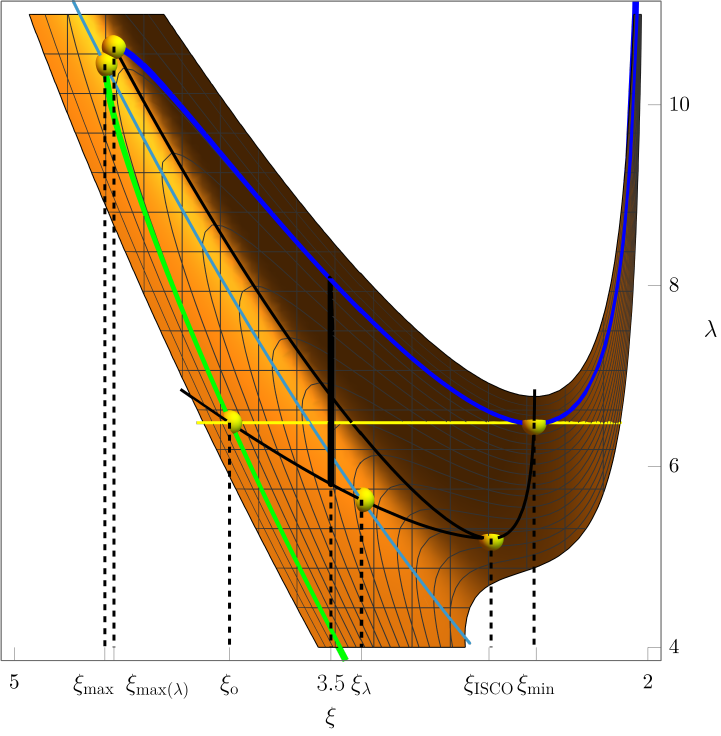}}\\
    %%%%%%%%%%%
    \subfloat[$\varepsilon-\lambda$ integration domain for $\xi = 3.5$ (as in SubFigs.\ (a) and (b)). The blue line corresponds to $\varepsilon=0.7$.]{\includegraphics[keepaspectratio,width=\twofig]{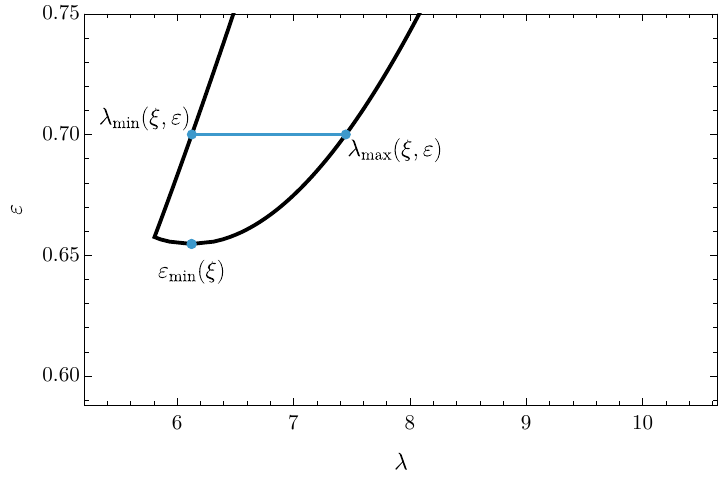}\label{subfig:3DdomainAc}}\hfill
    %%%%%%%%%%%
    \subfloat[$\varepsilon-\lambda$ integration domain for various fixed $\xi\in {\left[\xi_{\mathrm{min}},\xi_{\mathrm{max}}\right]}$. For $\xi=\xi_{\mathrm{max}}$ and $\xi=\xi_{\mathrm{min}}$ these regions shrink to zero-measure sets and therefore are not depicted.]{\includegraphics[keepaspectratio,width=\twofig]{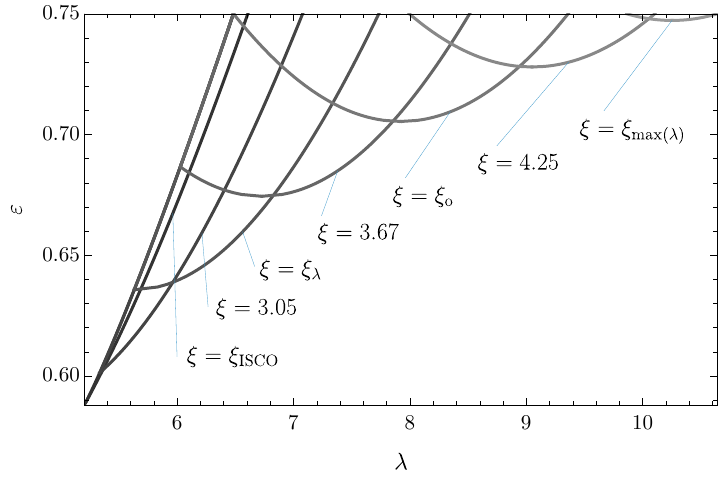}\label{subfig:3DdomainAd}}
    %%%%%%%%%%%
    \caption{\label{fig:3DdomainA} 3D domain of integration for $\alpha=0$, $\beta=1/2$, $\varepsilon_{\mathrm{max}}=3/4$.
    In SubFigs.\ (a) and (b): 3D contour of $\hat{\mathcal{G}}(\xi,\varepsilon,\lambda)=0$; in (b) the orthographic projection of (a) is shown. The transparent blue flat surface in (a) depicts the $\varepsilon=\varepsilon_{\mathrm{max}}$ cut-off level. Blue and green lines correspond to $\lambda_+(\xi,\varepsilon_{\mathrm{max}})$ and $\lambda_-(\xi,\varepsilon_{\mathrm{max}})$, respectively. The black line connecting $\xi_{\mathrm{ISCO}}$ to $\xi_{\mathrm{min}}$ is the maximum of the effective potential $w_+(\xi,\lambda)$ (with respect to the $\xi$ coordinate) and this defines the relation $w_{\mathrm{cu}}(\lambda)=\varepsilon(\lambda)$ so that $\d w_+(\xi,\lambda)/\d\xi=0$ and $\hat{\mathcal{G}}(\xi,\varepsilon(\lambda),\lambda)=0$. We define $\xi_{\mathrm{max}(\lambda)}$ as a point where $\lambda_+(\xi,\varepsilon_{\mathrm{max}})$ attains a local maximum with respect to $\xi$. The line connecting $\xi_{\mathrm{ISCO}}$ to $\xi_{\mathrm{max}(\lambda)}$ corresponds to the minima of the effective potential $w_+(\xi,\lambda)$ with respect to the $\xi$ coordinate. The remaining black line connecting $\xi_{\mathrm{ISCO}}$ to $\xi_{\mathrm{o}}$ is defined implicitly by the solution of $\hat{\mathcal{G}}(\xi,\varepsilon = w_{\mathrm{cu}}(\lambda),\lambda)=0$ for a given $\xi$. The yellow line denotes an effective potential $w_+(\xi,\lambda)$ for a particular value of $\lambda$, and the pale blue line represents points where $\lambda_+(\xi,\varepsilon)=\lambda_-(\xi,\varepsilon)$ which, for $\alpha=0$, is given as $( \varepsilon,\lambda,\xi)=\left(\sqrt{(\xi-2)/\xi},\beta\xi^2,\xi \right)$. The point in which this line crosses the $\varepsilon-\lambda$ region is denoted by $\xi_\lambda$. For $\xi>\xi_\lambda$ it can happen, depending upon the energy, that the integration is performed over $[\lambda_-(\xi,\varepsilon),\lambda_+(\xi,\varepsilon)]$. The thick black line (a deformed wedge) in SubFig.\ (a) and its detail in SubFig.\ (c) visualizes the domain of integration which, for $\alpha = 0$, is bounded by the hyperbola in the $(\varepsilon,\lambda)$ plane (see Eq.\ \eqref{Ghatazero}), possibly cut by the relation $\varepsilon = w_{\mathrm{cu}}(\lambda)$. The points on the hyperbola are given by $\lambda_\pm(\xi,\varepsilon)$. In SubFig.\ (d) the integration domain is shown for different values of $\xi$ (the larger the $\xi$ the lighter shade of gray is used).}
\end{figure}

A non-vanishing external electromagnetic field introduces an asymmetry, even in the Schwarzschild background. The change in allowed regions is visualized in Fig.~(\ref{subfig:beta}). Unfortunately, when the electromagnetic field is present, one is no longer able to provide exact analytic expressions to describe the disks. Moreover, the corresponding parameter space is large enough to make the general discussion almost impossible or, at least, incomprehensible. We will in detail discuss the case of $\alpha=0,\,\beta=1/2,\,\varepsilon_{\mathrm{max}}=3/4$ for which we construct the disk composed of particles with all possible values of $\lambda > 0$ and $\varepsilon$ limited only by $\varepsilon<\varepsilon_\mathrm{max}$. The relevant values of $\varepsilon$ and $\lambda$ are visualized in Fig.~(\ref{fig:3DdomainA}).

For $\alpha = 0$, $\beta = 1/2$, and $\varepsilon_\mathrm{max} = 3/4$, the following boundaries in the phase space play a role in the characterization of the regions available for the motion of particles on bound orbits.
\begin{description}
    \item[ISCO] The parameters  $\xi_{\mathrm{ISCO}}$, $\varepsilon_{\mathrm{ISCO}}$, and  $\lambda_{\mathrm{ISCO}}$ characterizing the ISCO can be obtained by solving Eqs.\ (\ref{eq:ISCO}). In general, it is possible to obtain two sets of parameters, corresponding to co- and counter-rotating orbits. In the case $\alpha=0,\,\beta=1/2$ we get
    \begin{itemize}
    \item $\xi^\circlearrowright_\mathrm{ISCO}\approx 2.743,\,\varepsilon^\circlearrowright_\mathrm{ISCO}\approx 0.587,\,\lambda^\circlearrowright_\mathrm{ISCO}\approx 5.203$ 
    \item $\xi^\circlearrowleft_\mathrm{ISCO}\approx 4.322\,,\varepsilon^\circlearrowleft_\mathrm{ISCO}\approx 5.683,\,\lambda^\circlearrowleft_\mathrm{ISCO}\approx -23.894$
    \end{itemize}
    \item[Edge of the potential well] The location of the local maximum of $w_+(\xi,\lambda)$ defines the edge of the potential well. It corresponds to the circular unstable orbit and can be obtained by solving the equations $\hat{\mathcal{G}}(\xi,\varepsilon,\lambda)=0$ and $\d w_+(\xi,\lambda)/\d\xi=0$ for $\xi<\xi_\mathrm{ISCO}$. The solution can be parameterized as $\varepsilon = w_{\mathrm{cu}}(\xi)$ and $\lambda = \lambda_{\mathrm{cu}}(\xi)$; by eliminating $\xi$ we obtain the relation $\varepsilon = w_{\mathrm{cu}}(\lambda)$.
    \item[Inner radius] For the case with $\alpha = 0$, $\beta = 1/2$, and $\varepsilon_\mathrm{max} = 3/4$, the inner edge of the disk $\xi_\mathrm{min}$ is given by the location of the edge of the potential well for $\varepsilon = \varepsilon_\mathrm{max}$, i.e., $w_\mathrm{cu}(\xi_\mathrm{min}) = \varepsilon_\mathrm{max}$. In general, it is also possible that the location $\xi$ of the maximum of $w_+(\xi,\lambda)$ moves towards the black hole with a decreasing $\lambda > 0$ (cf.\ Appendix \ref{app:schwarzschilBounds}, Fig.\ (\ref{fig:lambdaSame})). In this case, the inner edge of the disk would correspond to $w_\mathrm{cu}(\xi) = \varepsilon$, for some $\varepsilon < \varepsilon_\mathrm{max}$. In Appendix \ref{app:schwarzschilBounds}, this location is denoted as $\xi_\mathrm{tp}$.
    \item[$\varepsilon-\lambda$ boundary] An extension of the edge of the potential well for $\xi>\xi_\mathrm{ISCO}$ is given by the solution of
    \begin{equation}
    \label{extensioneq}
    \hat{\mathcal{G}}(\xi,\varepsilon=w_{\mathrm{cu}}(\lambda),\lambda)=0.
    \end{equation}
    To find this extension, for a given $\xi$, one has to numerically solve  Eq.\ \eqref{extensioneq} with respect to $\lambda=\lambda_\mathrm{p}(\xi)$. The value of $\xi$ for which $w_{\mathrm{cu}}(\lambda_{\mathrm{p}})=\varepsilon_\mathrm{max}$ is denoted by $\xi_\mathrm{o}$ (and also as $\xi_2(\varepsilon_\mathrm{max})$ in Appendix \ref{app:schwarzschilBounds}). 
    \item[Outer radius] To find the outer radius of the disk, we compare $\xi_{\mathrm{o}}$ with a location where $\lambda_+(\xi,\varepsilon_{\max})=\lambda_-(\xi,\varepsilon_{\max})$, and the greater of these two values is the outermost occupied radius $\xi_{\max}$.
\end{description}

Performing the integration in Eq.\ \eqref{eq:schwWaldCurrents} requires finding the minimum energy $\varepsilon_{\mathrm{min}}(\xi)$ which, for a given $\xi$, defines the ranges of the outer integral $\varepsilon \in \left[ \varepsilon_\mathrm{min}(\xi),\varepsilon_\mathrm{max}\right]$ (vertical axis in Figs.\ \ref{subfig:3DdomainAc} and \ref{subfig:3DdomainAd}). The $\varepsilon_{\mathrm{min}}(\xi)$ can be either the local minimum of the hyperbola $\hat{\mathcal G}(\xi,\varepsilon,\lambda) = 0$ in variables $\varepsilon$ -- $\lambda$ (see Eq.\ \ref{Ghatazero}) or the $\varepsilon-\lambda$ boundary discussed around Eq.~(\ref{extensioneq}).

\begin{figure}
\centering
\subfloat[$\tilde j_t$]{\includegraphics[width=\threefig]{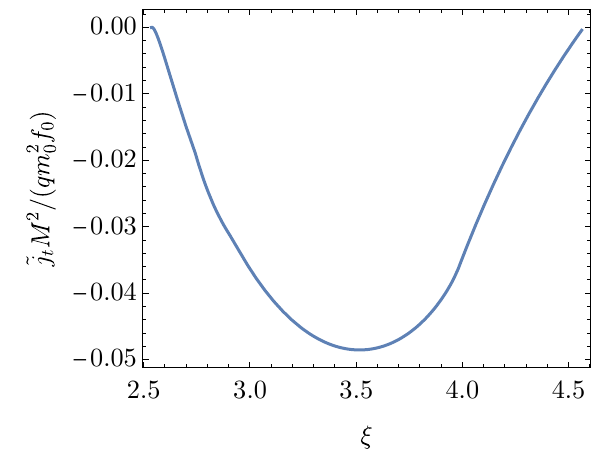}}
\hfill
\subfloat[$\tilde j_t$]{\includegraphics[width=\threefig]{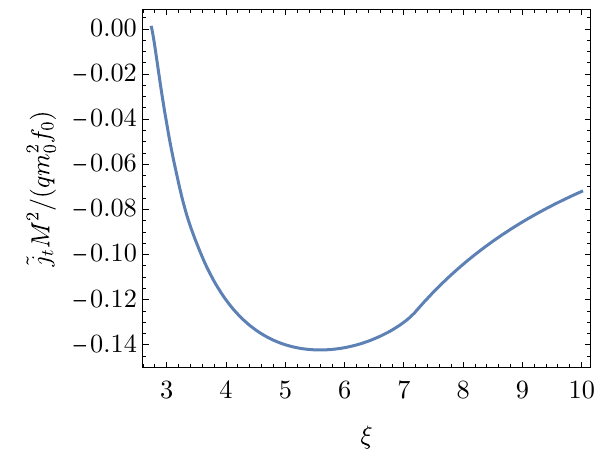}}
\hfill
\subfloat[$\tilde j_t$]{\includegraphics[width=\threefig]{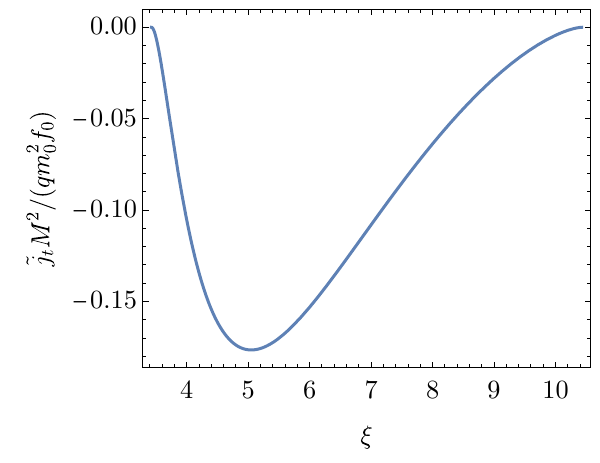}}

\vspace{1ex}

\subfloat[$\tilde j_\varphi$, $\beta = 1/2$, $\varepsilon_{\mathrm{max}} = 3/4$]{\includegraphics[width=\threefig]{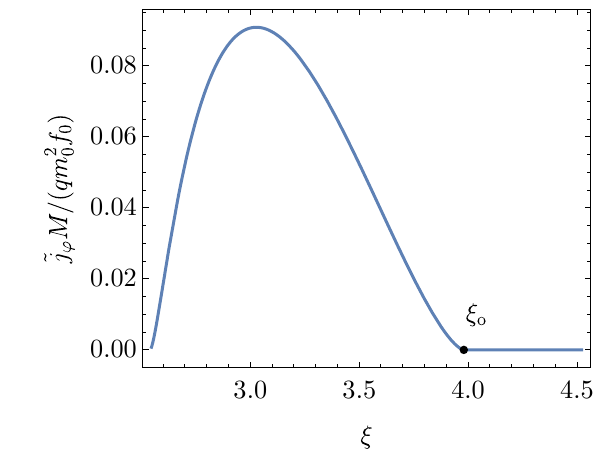}}
\hfill
\subfloat[$\tilde j_\varphi$, $\beta = 1/4$, $\varepsilon_{\mathrm{max}} = 1.1$]{\includegraphics[width=\threefig]{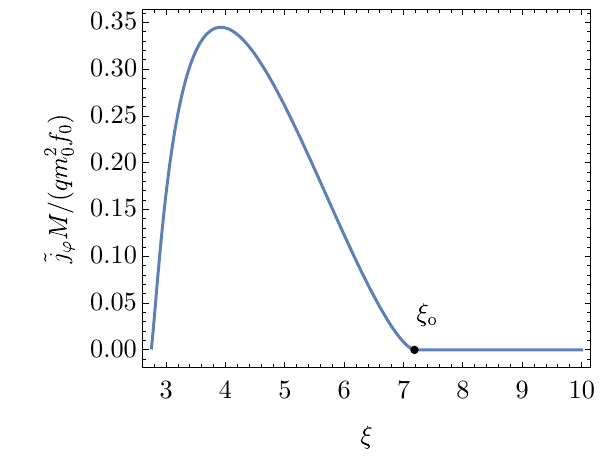}}
\hfill
\subfloat[$\tilde j_\varphi$, $\beta = -1/4$, $\varepsilon_{\mathrm{max}} = 4$]{
\includegraphics[width=\threefig]{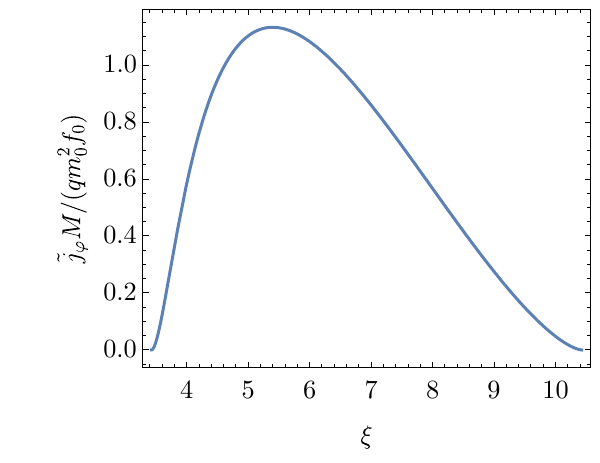}}

\caption{Current components $\tilde j_t$ (upper row) and $\tilde j_\varphi$ (lower row) in the Wald magnetosphere for $\alpha = 0$ and $f_0 = \mathrm{const}$. The region of integration is determined by all trajectories that do not plunge into the black hole, whose energy is lower than $\varepsilon_{\mathrm{max}}$. The standard situation is shown in Figs.\ (c) and (f) (third column) for $\beta<0$. In this case, the disk is finite and the current components vanish at its boundaries $\xi_{1,2}(\varepsilon_{\mathrm{max}})$ (determined in Appendix \ref{app:schwarzschilBounds}). For $\beta > 0$ (first two columns), the situation is more intricate: There always exists a point (labeled on the graphs as $\xi_\mathrm{o} = \xi_2(\varepsilon_{\mathrm{max}})$) beyond which $\tilde j_\varphi$ vanishes identically even though $\tilde j_t$ is non-zero. Furthermore, one can show that for $\varepsilon_{\mathrm{max}} \geq 1$ there are bounded orbits reaching arbitrarily large distances $\xi$ with energy smaller than $\varepsilon_{\mathrm{max}}$. This is illustrated in panel (b), where $\tilde j_t(\xi)$ is non-zero for arbitrarily large $\xi$. Nonetheless, for the same parameters, in panel (e), $\tilde j_\varphi(\xi)$ must vanish for $\xi \geq \xi_\mathrm{o}$. For $\varepsilon_{\mathrm{max}} < 1$ the disk is necessarily finite. This case is depicted in panel (a).}
\label{fig:schwarzschildDisks}
\end{figure}

The inner integral is performed  over $\lambda \in \left[ \lambda_\mathrm{min}(\xi,\varepsilon),\lambda_\mathrm{max}(\xi,\varepsilon) \right]$. In the examples discussed in this section---the case with $\beta = 1/2$ and $\varepsilon_\mathrm{max} = 3/4$ and also two other examples shown in Fig.\ \ref{fig:schwarzschildDisks}---the correct upper boundary is $\lambda_\mathrm{max}(\xi,\varepsilon)=\lambda_+(\xi,\varepsilon)$, and then for $\varepsilon>w_\mathrm{cu}$ the $\lambda_{\mathrm{min}}$ is defined by the $\varepsilon-\lambda$ boundary (to be determined numerically). For $\varepsilon<w_\mathrm{cu}(\xi)$ the integration is done over $\lambda \in \left[ \lambda_-(\xi,\varepsilon),\lambda_+(\xi,\varepsilon) \right]$ and this integral can be,  depending on the distribution function $f_0(\varepsilon,\lambda)$, performed analytically. For example, for the function $f_0(\varepsilon,\lambda)=f_0(\varepsilon)$ the expression for $j_\varphi$ given by Eq.~(\ref{eq:schwWaldCurrents}) vanishes. Once $\lambda_\mathrm{min}$ jumps to the cut-off branch $\lambda$, the integral must be evaluated numerically.

In Figure \ref{fig:schwarzschildDisks} we present examples of currents with integration bounds $\lambda \in [\lambda_{\mathrm{min}}(\xi,\varepsilon), \lambda_{\mathrm{max}}(\xi, \varepsilon)]$ and $\varepsilon \in [\varepsilon_{\mathrm{min}}(\xi), \varepsilon_{\mathrm{max}}]$ discussed above and also, in more detail, in Appendix \ref{app:schwarzschilBounds}. Apart from the case with $\alpha = 0$, $\beta = 1/2$ and $\varepsilon_\mathrm{max} = 3/4$, we also provide two examples with $\alpha = 0$, $\beta = 1/4$, $\varepsilon_\mathrm{max} = 1.1$ and $\alpha = 0$, $\beta = - 1/4$, $\varepsilon_\mathrm{max} = 4$. In all examples, we set $f_0 = \mathrm{const}$. For this choice, one can analytically perform the integration over $\lambda$, by  noticing that (cf.\ Eq.~(\ref{eq:lambdaIntegralsAnalyticAll}))
\begin{subequations} 
\label{eq:analyticIntegrals}
\begin{eqnarray}
\int\frac{(\lambda - \beta\xi^2)\dd\lambda}{\sqrt{-\xi(\xi-2)(\lambda - \lambda_-)(\lambda - \lambda_+)}} & = & \sqrt{\frac{(\lambda - \lambda_-)(\lambda - \lambda_+)}{-\xi(\xi-2)}} + \mathrm{const}, \label{eq:analyticIntegralsa} \\
\int\frac{-\varepsilon\dd\lambda}{\sqrt{-\xi(\xi-2)(\lambda - \lambda_-)(\lambda - \lambda_+)}} & = & \frac{2\varepsilon}{\sqrt{\xi(\xi-2)}}\arctan{\sqrt{-\frac{\lambda - \lambda_+}{\lambda-\lambda_-}}} + \mathrm{const}, \label{eq:analyticIntegralsb}
\end{eqnarray}
\end{subequations}
where we used the identity $\lambda_+ + \lambda_- = 2\beta\xi^2$. Integration over $\varepsilon$ cannot be performed analytically, as the functions $\lambda_{\mathrm{min}}(\xi, \varepsilon)$ in the integral are calculated numerically. 

Furthermore, notice that for $\xi \geq \xi_2(\varepsilon)$ (defined in Appendix \ref{app:schwarzschilBounds}), $\lambda_{\mathrm{max}}(\xi, \varepsilon)$ and $\lambda_{\mathrm{min}}(\xi, \varepsilon)$  coincide with $\lambda_{\pm}(\xi, \varepsilon)$. We immediately conclude that the first expression on the right-hand side of (\ref{eq:analyticIntegralsa}) vanishes at $\lambda = \lambda_{\pm}$, while the corresponding expression in Eq.\ (\ref{eq:analyticIntegralsb}) does not. Therefore, for $\xi \geq \xi_2(\varepsilon_{\mathrm{max}})$ we necessarily must have $j_\varphi(\xi) = 0$, even though $j_t(\xi)$ might be non-zero. In the region $\xi \geq \xi_2(\varepsilon_{\mathrm{max}})$, we obtain:
\begin{equation}
    j_t(\xi) = - \frac{\pi q m_0^2 f_0 \delta(\vartheta - \pi/2)}{2 M \sqrt{\xi(\xi - 2)}} \left[ \varepsilon_\mathrm{max}^2 - \varepsilon^2_\mathrm{min}(\xi) \right].
\end{equation}
This results in the possibility of disks for which $j_\varphi(\xi)$ vanishes entirely beyond some finite $\xi = \xi_2$, but $j_t(\xi)$ remains non-zero for $\xi$ arbitrarily large. The plots shown in the first two columns of Fig.\ \ref{fig:schwarzschildDisks} illustrate the subtle cases with $j_t (\xi)\neq 0$ at the locations for which $j_\varphi(\xi) = 0$. 

\section{Conclusions}
\label{sec:Conclusions}

We have constructed a family of stationary charged disks composed of a collisionless simple gas of identical particles orbiting a Kerr black hole immersed in an external electromagnetic field. We considered models of geometrically thick disks, but also infinitely thin ones, confined to the equatorial plane. Both of these families can be described in a single framework by imposing appropriate assumptions on the distribution function. Since in our case the electrogeodesics confined to the equatorial plane are integrable, thin disk models allow for a slightly more detailed analysis. In both cases, specific numerical examples were given assuming the electromagnetic field in the form of the Wald magnetosphere.

We focused on disks composed of particles with fixed energy or fixed angular momentum. In those cases, it is possible to construct analytic solutions, without a detailed control of the behavior of the corresponding effective potentials. To a reasonable degree, the effective potentials can be controlled in the Schwarzschild limit of $\alpha \to 0$ for the Wald magnetosphere. In this case, it is possible to construct thin disks composed of variable energies and angular momenta.

The astrophysical plasma is composed of several kinds of different particles of different masses and charges. Since we neglected collisions between the particles, the self-gravity of the gas, and its backreaction on the electromagnetic field, solutions corresponding to several different particle species can be obtained by superimposing the solutions obtained in this work. However, relaxing any of these assumptions would require taking into account all species of particles at the same time.

We decided to present our results mostly in terms of the electric current generated by the charged gas, as a step toward a consistent coupling of the kinetic disk models analyzed in this paper with solutions to the Maxwell equations. This is also the main reason for focusing on the analysis that can be applied to general stationary and axially symmetric electromagnetic fields and treating the Wald magnetosphere as a particular example only. The remaining component---solutions of the Maxwell equations for circular currents in the Kerr spacetime---can be found in \cite{Kofron2022}.

\begin{acknowledgments}
P.\ D.\ acknowledges support from the Faculty of Physics, Astronomy, and Applied Computer Science under the Strategic Program Excellence Initiative at Jagiellonian University.
D.\ K.\ acknowledges the use of \emph{Wolfram Mathematica}$^{\copyright}$ and the \emph{xAct} package \cite{xAct} in the version 1.3.0.
\end{acknowledgments}

\appendix

\section{Volume element on the mass hyperboloid}
\label{appvolume}

In this appendix, we derive the volume element induced on the future mass hyperboloid $P_x^+(m)$. This derivation is adapted from \cite{Acuna2022}. To maintain contact with the original calculation presented in \cite{Acuna2022}, we work with oriented volume elements.

It is shown in \cite{Acuna2022} that the unit vector normal to $P_x^+(m)$
is given by
\begin{equation}
N = \frac{1}{m} p_\mu \frac{\partial}{\partial p_\mu}.
\end{equation}
The volume form induced on $P_x^+(m)$ can be computed as
\begin{equation}
\eta_{P_x^+(m)} = i_N \eta_{T_x^\ast \mathcal M}, \end{equation}
where $i_N$ denotes the interior derivative with respect to the normal vector $N$, and
\begin{equation}
\eta_{T_x^\ast \mathcal M} = - \sqrt{- \mathrm{det} \, (g^{\alpha \beta})} d p_0 \wedge \dots \wedge dp_3.
\end{equation}
Here, the choice of the sign (related with the chosen orientation in $T^\ast \mathcal M$) is compatible with $p^0 > 0$ and $p_0 < 0$.
Hence,
\begin{eqnarray}
    \eta_{P_x^+(m)} & = & i_N \eta_{T_x^\ast \mathcal M} = - \sqrt{- \mathrm{det} \, (g^{\alpha \beta})} \left[ (i_N dp_0) dp_1 \wedge dp_2 \wedge dp_3 - (i_N dp_1) dp_0 \wedge dp_2 \wedge dp_3 \right. \nonumber \\
    && \left. + (i_N dp_2) dp_0 \wedge dp_1 \wedge dp_3 - (i_N dp_3) dp_0 \wedge dp1 \wedge dp_2 \right].
\end{eqnarray}
It is now convenient to use the fact that $i_N dp_\alpha = dp_\alpha (N) = p_\alpha/m$. This yields
\begin{eqnarray}
    \eta_{P_x^+(m)} & = & - \frac{1}{m} \sqrt{- \mathrm{det} \, (g^{\alpha \beta})} \left[ p_0 dp_1 \wedge dp_2 \wedge dp_3 - p_1 dp_0 \wedge dp_2 \wedge dp_3 \right. \nonumber \\
    && \left. + p_2 dp_0 \wedge dp_1 \wedge dp_3 - p_3 dp_0 \wedge dp1 \wedge dp_2 \right].
\end{eqnarray}
Since, $p^\alpha dp_\alpha = 0$, one has $dp_0 = - p^a dp_a/p^0$. Thus,
\begin{eqnarray}
    \eta_{P_x^+(m)} & = & - \frac{1}{m p^0} \sqrt{- \mathrm{det} \, (g^{\alpha \beta})} \left[ p^0 p_0 dp_1 \wedge dp_2 \wedge dp_3 + p^a p_a dp_1 \wedge dp_2 \wedge dp_3 \right] \nonumber \\
    &= & \frac{m \sqrt{- \mathrm{det} \, (g^{\alpha \beta})}}{p^0} dp_1 \wedge dp_2 \wedge dp_3.
\end{eqnarray}

\section{Conservation of the particle current density and the energy-momentum tensor}
\label{appconservation}

In this appendix, we show that  $\nabla_\mu \mathcal J^\mu = 0$ and $\nabla_\mu \mathcal T\indices{^\mu_\nu} = F_{\nu \alpha} j^\alpha$. This is also a known result. We provide it here mainly for completeness. A straightforward calculation gives (for simplicity, we omit the domain of integration with respect to momenta $P_x^+$)
\begin{eqnarray}
    \nabla_\mu \mathcal J^\mu & = & \frac{1}{\sqrt{- \mathrm{det} g_{\alpha \beta}}} \frac{\partial}{\partial x^\mu} \left( \sqrt{- \mathrm{det} g_{\alpha \beta}} \mathcal J^\mu  \right) = \frac{1}{\sqrt{- \mathrm{det} g_{\alpha \beta}}} \frac{\partial}{\partial x^\mu} \left[ \sqrt{- \mathrm{det} g_{\alpha \beta}} \int \mathcal F (x,p) p^\mu \sqrt{- \mathrm{det} g^{\alpha \beta}} d^4 p \right] \nonumber \\
    & = & \frac{1}{\sqrt{- \mathrm{det} g_{\alpha \beta}}} \frac{\partial}{\partial x^\mu} \int \mathcal F (x,p) p^\mu d^4 p = \frac{1}{\sqrt{- \mathrm{det} g_{\alpha \beta}}}  \int \left[ \frac{\partial \mathcal F (x,p)}{\partial x^\mu} g^{\mu \nu} p_\nu + \mathcal F(x,p) \frac{\partial g^{\mu \nu}}{\partial x^\mu} p_\nu \right] d^4 p.
\end{eqnarray}
Substituting Eq.\ (\ref{vlasovem}) into the above expression, we obtain
\begin{equation}
\label{B2}
\nabla_\mu \mathcal J^\mu = \frac{1}{\sqrt{- \mathrm{det} g_{\alpha \beta}}}  \int \left[ \frac{1}{2} \frac{\partial g^{\alpha \beta}}{\partial x^\mu} p_\alpha p_\beta \frac{\partial \mathcal F (x,p)}{\partial p_\mu} + \mathcal F(x,p) \frac{\partial g^{\mu \nu}}{\partial x^\mu} p_\nu - q \frac{\partial \mathcal F(x,p)}{\partial p_\mu} g^{\alpha \beta} p_\beta F_{\mu \alpha}\right] d^4 p.
\end{equation}
The first term in the integrand can be written as
\begin{equation}
\frac{1}{2} \frac{\partial g^{\alpha \beta}}{\partial x^\mu} p_\alpha p_\beta \frac{\partial \mathcal F (x,p)}{\partial p_\mu} = \frac{\partial}{\partial p_\mu} \left[  \frac{1}{2} \frac{\partial g^{\alpha \beta}}{\partial x^\mu} p_\alpha p_\beta \mathcal F(x,p) \right] - \frac{\partial g^{\mu \beta}}{\partial x^\mu} p_\beta \mathcal F(x,p). \end{equation}
On the other hand, the last term in the integrand in Eq.\ \eqref{B2} reads
\begin{equation}
- q \frac{\partial \mathcal F(x,p)}{\partial p_\mu} g^{\alpha \beta} p_\beta F_{\mu \alpha} = - \frac{\partial}{\partial p_\mu} \left[ q g^{\alpha \beta} p_\beta F_{\mu \alpha} \mathcal F(x,p) \right] - q g^{\alpha \mu} F_{\mu \alpha} \mathcal F(x,p) = - \frac{\partial}{\partial p_\mu} \left[ q g^{\alpha \beta} p_\beta F_{\mu \alpha} \mathcal F(x,p) \right],
\end{equation}
where in the last step we used the antisymmetry of $F_{\mu\nu}$. Therefore,
\begin{equation}
\nabla_\mu \mathcal J^\mu = \frac{1}{\sqrt{- \mathrm{det} g_{\alpha \beta}}} \int \frac{\partial}{\partial p_\mu} \left[  \frac{1}{2} \frac{\partial g^{\alpha \beta}}{\partial x^\mu} p_\alpha p_\beta \mathcal F(x,p) - q g^{\alpha \beta} p_\beta F_{\mu \alpha} \mathcal F(x,p) \right] d^4p.
\end{equation}
Writing the above integral as a surface term and assuming that it vanishes asymptotically, we get $\nabla_\mu \mathcal J^\mu = 0$.

The divergence of the Vlasov energy momentum tensor (\ref{vlasovemtensor}) can be computed as follows. A general symmetric tensor $T^{\mu\nu}$ satisfies
\begin{equation}
\nabla_\mu T\indices{^\mu_\nu} = \frac{1}{\sqrt{- \mathrm{det} g_{\alpha \beta}}} \partial_\mu \left( \sqrt{- \mathrm{det} g_{\alpha \beta}} T\indices{^\mu_\nu} \right) - \frac{1}{2} (\partial_\nu g_{\rho \sigma}) T^{\rho \sigma}.
\end{equation}
This gives
\begin{eqnarray}
    \nabla_\mu \mathcal T\indices{^\mu_\nu} & = & \frac{1}{\sqrt{- \mathrm{det} g_{\alpha \beta}}} \frac{\partial}{\partial x^\mu} \int \mathcal F(x,p) g^{\mu \alpha} p_\alpha p_\nu d^4 p - \frac{1}{2} (\partial_\nu g_{\rho \sigma}) \mathcal T^{\rho \sigma} \nonumber \\
    & = & \frac{1}{\sqrt{- \mathrm{det} g_{\alpha \beta}}} \int \left[ \frac{\partial \mathcal F(x,p)}{\partial x^\mu} g^{\mu \alpha} p_\alpha p_\nu + \mathcal F(x,p) \frac{\partial g^{\mu \alpha}}{\partial x^\mu} p_\alpha p_\nu \right] d^4 p - \frac{1}{2} (\partial_\nu g_{\rho \sigma}) \mathcal T^{\rho \sigma} \nonumber \\
    & = & \frac{1}{\sqrt{- \mathrm{det} g_{\alpha \beta}}} \int \left[ \frac{1}{2} \frac{\partial g^{\alpha \beta}}{\partial x^\mu} p_\alpha p_\beta p_\nu \frac{\partial \mathcal F(x,p)}{\partial p_\mu} - q \frac{\partial \mathcal F(x,p)}{\partial p_\mu} g^{\alpha \beta} p_\beta p_\nu F_{\mu \alpha} + \mathcal F(x,p) \frac{\partial g^{\mu \alpha}}{\partial x^\mu} p_\alpha p_\nu \right] d^4 p \nonumber \\
    & & - \frac{1}{2} (\partial_\nu g_{\rho \sigma}) \mathcal T^{\rho \sigma},
\end{eqnarray}
where in the last step we used the Vlasov equation (\ref{vlasovem}). A straightforward calculation gives
\begin{eqnarray}
   \lefteqn{\frac{1}{2} \frac{\partial g^{\alpha \beta}}{\partial x^\mu} p_\alpha p_\beta p_\nu \frac{\partial \mathcal F(x,p)}{\partial p_\mu} + \mathcal F(x,p) \frac{\partial g^{\mu \alpha}}{\partial x^\mu} p_\alpha p_\nu} \nonumber \\
    && = \frac{\partial}{\partial p_\mu} \left[ \frac{1}{2} \frac{\partial g^{\alpha \beta}}{\partial x^\mu} p_\alpha p_\beta p_\nu \mathcal F(x,p) \right] - \frac{1}{2} \frac{\partial g^{\alpha \beta}}{\partial x^\nu} p_\alpha p_\beta \mathcal F(x,p).
\end{eqnarray}
Similarly,
\begin{equation}
- q \frac{\partial \mathcal F(x,p)}{\partial p_\mu} g^{\alpha \beta} p_\beta p_\nu F_{\mu \alpha} = - \frac{\partial}{\partial p_\mu} \left[ q \mathcal F (x,p) g^{\alpha \beta} p_\beta p_\nu F_{\mu \alpha} \right] + q \mathcal F(x,p) F_{\nu \alpha} g^{\alpha \beta} p_\beta,
\end{equation}
where we have again used the antisymmetry of $F_{\mu \nu}$. Thus,
\begin{eqnarray}
    \frac{\partial}{\partial x^\mu} \int \mathcal F(x,p) g^{\mu \alpha} p_\alpha p_\nu d^4 p & = & \int \frac{\partial}{\partial p_\mu} \left[ \frac{1}{2} \frac{\partial g^{\alpha \beta}}{\partial x^\mu} p_\alpha p_\beta p_\nu \mathcal F(x,p) - q \mathcal F (x,p) g^{\alpha \beta} p_\beta p_\nu F_{\mu \alpha} \right] d^4 p \nonumber \\
    & & + \int q \mathcal F (x,p) F_{\nu \alpha} g^{\alpha \beta} p_\beta d^4 p - \frac{1}{2} \int \frac{\partial g^{\alpha \beta}}{\partial x^\nu} p_\alpha p_\beta \mathcal F (x,p) d^4 p.
\end{eqnarray}
Expressing the first of the above integrals as a surface term and assuming suitable fall-off conditions of $\mathcal F(x,p)$, so that this term vanishes, we get
\begin{equation}
\frac{\partial}{\partial x^\mu} \int \mathcal F(x,p) g^{\mu \alpha} p_\alpha p_\nu d^4 p = \int q \mathcal F (x,p) F_{\nu \alpha} g^{\alpha \beta} p_\beta d^4 p - \frac{1}{2} \int \frac{\partial g^{\alpha \beta}}{\partial x^\nu} p_\alpha p_\beta \mathcal F (x,p) d^4 p. \end{equation}
This gives
\begin{equation} \nabla_\mu \mathcal T\indices{^\mu_\nu} = q F_{\nu \alpha} g^{\alpha \beta} \mathcal J_\beta - \frac{1}{2} \frac{\partial g^{\alpha \beta}}{\partial x^\nu} \mathcal T_{\alpha \beta} - \frac{1}{2} \frac{\partial g_{\rho \sigma}}{\partial x^\nu} \mathcal T^{\rho \sigma} = q F_{\nu \alpha} g^{\alpha \beta} \mathcal J_\beta, \end{equation}
where we have used the fact that for a generic tensor $T_{\mu \nu}$,
\begin{equation}
(\partial_\nu g^{\alpha \beta}) T_{\beta \alpha} = - (\partial_\nu g_{\beta \mu}) T^{\mu \beta}. \end{equation}
Taking into account Eq.\ (\ref{elcurrent}), we obtain
\begin{equation}
\nabla_\mu \mathcal T\indices{^\mu_\nu} = F_{\nu \alpha} j^\alpha. \end{equation}

\section{Comparison with the phase space of the Kerr metric}
\label{app:kerr}

In this appendix, we compare the formalism used to construct thick disk models in Sec.\ \ref{sec:thickdisks} with the description of the phase space associated with the Kerr metric in the absence of the electromagnetic field. In this case, the geodesic motion is integrable, and four independent constants of motion can be used as convenient phase-space coordinates (in addition to four coordinates specifying the position in the spacetime). Moreover, the phase-space regions available for the motion of the particles of the gas can be determined by the radial and polar effective potentials, appearing in the description of Kerr geodesics. In this respect, it is natural to ask how the variables $E$, $L$, and $\chi$ used in Sec.~\ref{sec:thickdisks} are restricted when applied to the description of the phase-space in the case with the vanishing electromagnetic field. This question applies, in particular, to the variable $\chi$, on which no restricting conditions were imposed in Sec.~\ref{sec:thickdisks}.

The total integrability of geodesic motion in the Kerr spacetime is a consequence of hidden symmetries: the Kerr spacetime admits not only two Killing vectors $k$ and $\eta$, but also a  Killing tensor
\begin{equation}
    K^{\mu \nu} = 2 \rho^2 l^{( \mu} n^{\nu )} + r^2 g^{\mu \nu},
\end{equation}
where the vectors $n$ and $l$ are defined by
\begin{equation}
    l^\mu = \left( \frac{r^2 + a^2}{\Delta},1,0,\frac{a}{\Delta} \right), \qquad n^\mu = \left( \frac{r^2 + a^2}{2 \rho^2}, - \frac{\Delta}{2 \rho^2},0,\frac{a}{2 \rho^2} \right).
\end{equation}
This means that we have 4 constants of motion
\begin{align}
    p^\mu p_\mu &= -m^2\,, & 
    p^\mu K_{\mu \nu} p^\nu &= K=\kappa m^2M^2\,,\\
    p^\mu k_\mu &= -E \,, & 
    p^\mu \eta_\mu &= L\,,
\end{align}
where $\kappa$ is the dimensionless version of the Carter constant $K$. In terms of these constants, we can express the covector $p$ as
\begin{align}\label{eq:momentumNoField}
    p &= 
    -E \, \mathrm d t
    +\epsilon_r \frac{1}{\Delta} \sqrt{R}\, \mathrm d r
    +\epsilon_\vartheta \sqrt{\Theta} \, \mathrm d \vartheta
    +L \, \mathrm d \varphi,
\end{align}
where the radial and angular potentials (together with their dimension-less alternatives) are defined as follows
\begin{align}
    R &= \left[E(r^2+a^2)-aL\right]^2-(K+m^2r^2)\Delta\,,& 
    \bar{R} &= \left[\varepsilon(\alpha^2+\xi^2)-\alpha\lambda\right]^2-\bar{\Delta}(\kappa+\xi^2)\,,\\
    \Theta &= K-m^2a^2\cos^2\vartheta-\left(\frac{L}{\sin\vartheta}-aE\sin\vartheta\right)^2\,, &
    \bar{\Theta} &= \kappa-\alpha^2\cos^2\vartheta-\left(\frac{\lambda}{\sin\vartheta}-\alpha\varepsilon\sin\vartheta\right)^2,
    \label{ThetaDef}
\end{align}
where $R = m^2 M^4 \bar R$, and $\Theta = m^2 M^2 \bar \Theta$. The volume element $\mathrm{dvol}_x^{(m)} (p)$ can be expressed in terms of $\varepsilon$, $\lambda$, and $\kappa$ as
\begin{align}\label{eq:intMeasureNoField}
    \mathrm{dvol}_x^{(m)} (p) &= \frac{m}{2} \frac{1}{\sin \vartheta} \frac{\mathrm{d} E \, \mathrm{d} L\, \mathrm{d}K}{\sqrt{R} \sqrt{\Theta}} = \frac{m^3}{2}\,\frac{1}{\sin\vartheta}\frac{\mathrm{d}\varepsilon\,\mathrm{d}\lambda\,\mathrm{d}\kappa}{\sqrt{\bar{R}}\sqrt{\bar{\Theta}}}.
\end{align}
This expression can be compared with Eq.\ (\ref{eq:intMeasureThick1}). To this end, first note that although each of the momentum components $p^{(r)}$ and $p^{(\vartheta)}$ depends on the Carter constant $K$, the expression $\zeta^2 = \left( p^{(r)} \right)^2 + \left( p^{(\vartheta)} \right)^2$ is independent of~$K$. Writing
$\zeta^2 = \left( \Delta p_r^2 +  p_\vartheta^2 \right)/\rho^2$
and substituting the expressions for $p_r$ and $p_\vartheta$ from Eq.\ (\ref{eq:momentumNoField}), we get
\begin{equation}
    \zeta^2 = \frac{1}{\Delta \rho^2} \left[ E(r^2 + a^2) - a L \right]^2 - m^2 - \frac{1}{\rho^2} \left( \frac{L}{\sin \vartheta} - a E \sin \vartheta \right)^2.
\end{equation}
Moreover, the tetrad component
\begin{equation}
    p^{(\varphi)} = \frac{\rho L}{\sin \vartheta \sqrt{A}}
\end{equation}
is also independent of $K$. Consequently, the only variable in the volume element (\ref{eq:intMeasureThick1}) or (\ref{Pvolume}) that depends on~$K$ is~$\chi$. Note that $\tan \chi = p^{(\vartheta)}/p^{(r)}$, where
\begin{equation}
\label{chiEq}
\frac{p^{(\vartheta)}}{p^{(r)}} = \frac{\epsilon_r \epsilon_\vartheta \sqrt{\Delta} \sqrt{\Theta}}{\sqrt{R}}.
\end{equation}
A straightforward calculation yields
\begin{equation}
    \frac{d \chi}{d K} = \frac{\epsilon_r \epsilon_\vartheta \sqrt{\Delta}}{2 \sqrt{R} \sqrt{\Theta}}.
\end{equation}
Substituting this result in Eq.\ (\ref{Pvolume}), one obtains Eq.\ (\ref{eq:intMeasureNoField}), as expected.

The phase space region available for the motion of the particles is restricted by the relations $R \ge 0$ and $\Theta \ge 0$, which can also be interpreted as conditions for the admissible values of $K$. Equation (\ref{chiEq}) suggests that no restrictions are imposed on the allowed values of $\chi \in [0,2\pi)$. For $K$ changing in the range corresponding to the limits of $\Theta = 0$ and $R = 0$, the ratio $p^{(\theta)}/p^{(r)}$ changes from 0 to $\pm \infty$, depending on the signs $\epsilon_r$ and $\epsilon_\vartheta$. However, a formal proof, similar to the one in Appendix A of \cite{Mach2026b}, should still be provided.

We should also emphasize the role of the signs $\epsilon_r$ and $\epsilon_\vartheta$. Except for the turning points, the map $(p_r,p_\vartheta,p_\varphi) \to (E,L,K)$ is a four-fold cover, with sheets labeled with four different combinations of signs $\epsilon_r$ and $\epsilon_\vartheta$. This is clearly visible when computing the integrals over momentum space. Consider, for instance, 
\begin{align}
    \mathcal J_\mu &= \int_{\mathcal P(m)} p_\mu \mathcal F_0(x;p_1,p_2,p_3) \mathrm{dvol}_x^{(m)} (p)
\end{align}
and set $\mathcal F_0 = C \delta(E - E_0)\delta(L - L_0)$. This gives
\begin{equation}
    \left( \begin{array}{c} \mathcal J_t \\ \mathcal J_\varphi \end{array} \right) =  \frac{m C}{2 \sin \vartheta}\left ( \begin{array}{c} -E_0 \\ L_0 \end{array} \right)  \int_{\mathcal P(m)} \frac{ \mathrm{d} K}{\sqrt{R}\sqrt{\Theta}}.
\end{equation}
The primitive function reads
\begin{align}
\int\frac{\mathrm{d} K}{{\sqrt{R}\sqrt{\Theta}}}  &= -\frac{2}{\sqrt{\Delta}} \arctan\left(\frac{\sqrt{R}}{\sqrt{\Delta}\sqrt{\Theta}}\right)\,,
\end{align}
which needs to be evaluated assuming proper boundaries. A natural possibility is to take $K_\Theta < K < K_R$, where $\Theta(K_{\Theta}) = 0$ and $R(K_R) = 0$. This yields
\begin{equation}
\label{CarterIntegration}
    \left( \begin{array}{c} \mathcal J_t \\ \mathcal J_\varphi \end{array} \right) =  4 \times \frac{\pi m C}{2 \sin \vartheta \sqrt{\Delta}}\left ( \begin{array}{c} -E_0 \\ L_0 \end{array} \right),
\end{equation}
where the multiplicative factor of 4 has been introduced to take into account the possible values of signs $\epsilon_r$ and $\epsilon_\vartheta$, as discussed above. This result agrees with Eqs.\ (\ref{Jthick3}) or (\ref{Jthick8}).

\section{}
\label{app:schwarzschilBounds}

In the following, we discuss the properties of $w_+(\xi)$ in the Schwarzschild limit, given by (\ref{eq:effPotWaldSchw}), with the objective of constructing disks composed of particles of varying energy $\varepsilon$ and angular momentum $\lambda$. Without loss of generality, we only consider $\lambda \geq 0$. Alternatively, one could also take into account $\beta \ge 0$ and both positive and negative values of $\lambda$. Equation (\ref{eq:effPotWaldSchw}) implies that $w_+ \to 0$ as $\xi \to 2^+$. As $\xi \to +\infty$, $w_+ \to +\infty$ for $\beta \neq 0$ and $w_+ \to 1$ for $\beta = 0$. The derivative of $w_+(\xi)$ with respect to $\lambda$ reads
\begin{equation}
\label{eq:behaviorWithLambdaWald}
            \frac{\partial w_+(\xi)}{\partial \lambda} = \sqrt{\frac{1-2/\xi}{1 + \xi^2\left(\lambda/\xi^2 - \beta\right)^2}}\left(\frac{\lambda}{\xi^2} - \beta\right).
\end{equation}
This implies that for $\beta > 0$, $w_+(\xi)$ may decrease with $\lambda$. This happens for $\xi \geq \xi_{\mathrm{sc}} \coloneqq  \sqrt{\lambda/\beta}$. Otherwise, the effective potential (\ref{eq:effPotWaldSchw}) is always an increasing function of $\lambda$ at a fixed point $\xi$.

The locations of the extrema of $w_+$ can be described by solutions to $\d  w_+/\d\xi = 0$. Provided that
\begin{equation}
\label{undersq}
\xi[1  +\beta^2\xi(\xi-2)^2]-3 > 0,
\end{equation}
a local extremum occurs at $\xi$ for $\lambda = \bar\lambda_\pm(\xi)$ (not to be confused with $\lambda_\pm$ from Sec.\ \ref{sec:constantE}), where
\begin{equation}\label{eq:extremaWald}
\bar\lambda_\pm(\xi) \coloneqq \frac{-\beta\xi^2 \mp \xi\sqrt{\xi[1  +\beta^2\xi(\xi-2)^2]-3}}{\xi-3}.
\end{equation}
Condition (\ref{undersq}) is equivalent to       \begin{equation}
\label{eq:extremumEx}
\xi > 3 \quad \text{or} \quad (2 < \xi \leq 3 \quad \text{and} \quad |\beta| \geq \mathcal{F}(\xi)), \quad \text{where} \quad \mathcal{F}(\xi) \coloneqq \frac{\sqrt{3-\xi}}{\xi(\xi - 2)}.
\end{equation}
Notice that $\mathcal{F}(\xi)$ is a decreasing function reaching $0$ at $\xi = 3$ and diverging to $+\infty$ as $\xi \to 2^+$. Therefore, there always exists a solution to $|\beta| = \mathcal{F}(\xi)$, denoted by $\xi = \xi_\mathrm{tp}$ and coinciding with a turning point, i.e., the smallest radius an orbit with radii bounded from below can reach. It corresponds to $\bar \lambda_+ =  \bar \lambda_- = -\beta \xi^2/(\xi - 3)$, which is positive for $\beta > 0$ and negative for $\beta < 0$. We emphasize that the particles that do not plunge into the black hole can only be found for $\xi \ge \xi_\mathrm{tp}$. Note that since for $\beta \neq 0$ one has $w_+ \to +\infty$ as $\xi \to + \infty$, there are no unbound orbits that can reach infinity.

The signs of $\bar\lambda_{\pm}$ depend only on the radius $\xi$ and the sign of $\beta$, as summarized in Tab.\ \ref{tab:lambdaSign}. Since we restrict ourselves to $\lambda \ge 0$, we also choose $\bar \lambda_\pm \ge 0$. The values $\lambda = \bar \lambda_\pm$ satisfying this condition correspond to the extrema of $w_+(\xi)$: a minimum occurring for $\lambda = \lambda_\mathrm{cs}$, associated with a stable circular orbit, and a maximum occurring for $\lambda = \lambda_\mathrm{cu}$, associated with an unstable circular orbit.

\begin{table}[h!]
\centering
\begin{tabular}{c|c|c}
                $\xi$ & $\beta > 0$ & $\beta < 0$ \\
                \hline
                $\xi_{\mathrm{tp}} \leq \xi < 3$ & $\bar\lambda_+ >0,\bar\lambda_->0$ & $\bar\lambda_+ < 0,\bar\lambda_- < 0$ \\
                \hline
                $\xi > 3$ & $\bar\lambda_+ < 0,\bar\lambda_- > 0$ & $\bar\lambda_+ < 0,\bar\lambda_- > 0$
\end{tabular}
\caption{The sign of the solution (\ref{eq:extremaWald}) in the respective region. In our convention, we consider only $\lambda > 0$.}
\label{tab:lambdaSign}
\end{table}

\begin{figure}
\centering
\subfloat[$\bar\lambda_-(\xi)$, $\beta = -1/2$]{\includegraphics[width=0.9\twofig]{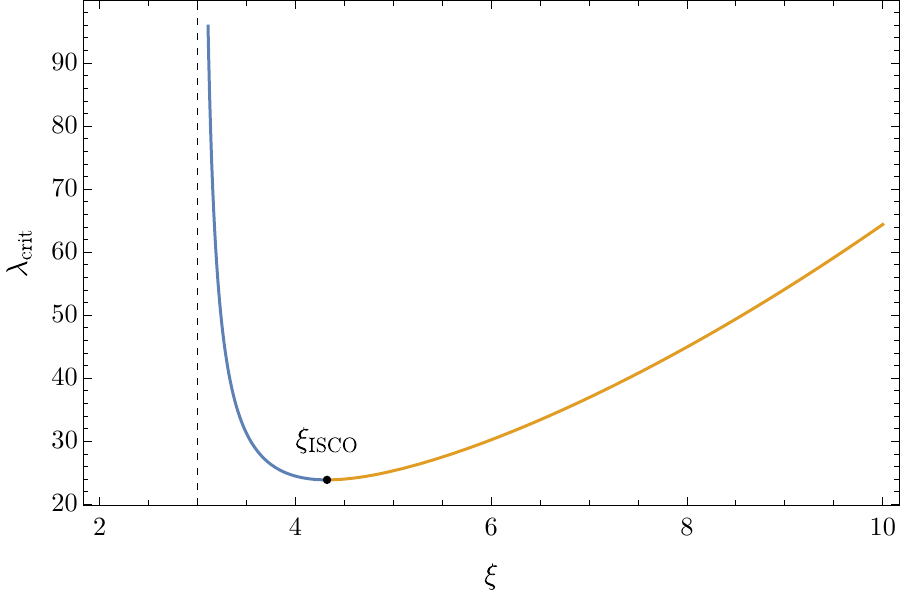}}
\hspace{1em}
\subfloat[$w_{\mathrm{cu},\mathrm{cs}}(\xi)$, $\beta = -1/2$]{\includegraphics[width=0.9\twofig]{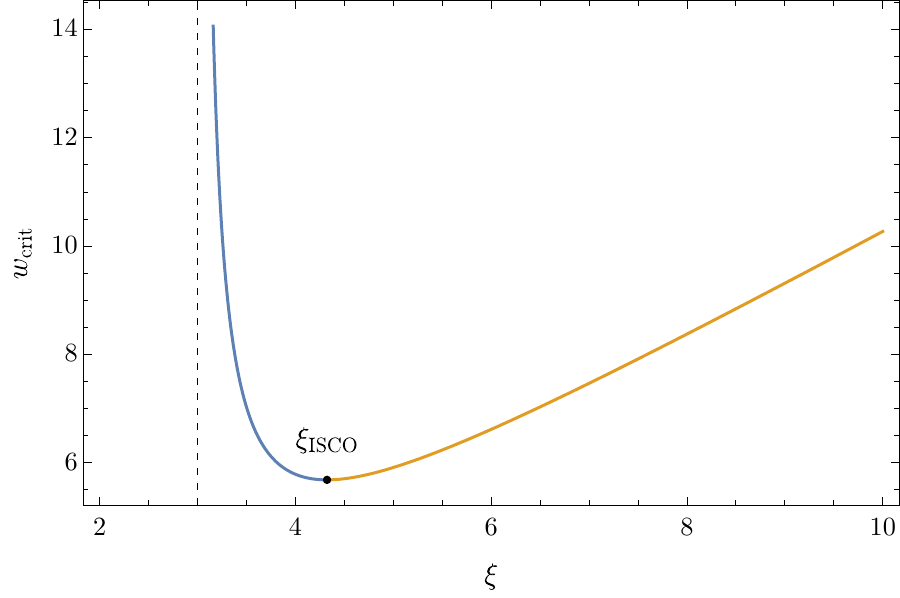}}            

\caption{Left panel: Critical values $\bar\lambda_{-}(\xi)$ for $\beta < 0$. Blue and orange lines represent $\lambda_{\mathrm{cu}}(\xi)$ and $\lambda_{\mathrm{cs}}(\xi)$, respectively. The branch $\lambda_\mathrm{cu}(\xi)$ diverges to $+\infty$ at $\xi_\mathrm{tp} = 3$. The branch $\lambda_\mathrm{cs}(\xi)$ tends to $+\infty$ for $\xi \to +\infty$. Both branches meet at the common minimum at $\xi_\mathrm{ISCO}.$ Right panel: $w_{\mathrm{cu}}(\xi)$ (blue) and $w_{\mathrm{cs}}(\xi)$ (orange) for $\beta < 0$. The branches $\lambda_\mathrm{cu,cs}(\xi)$ and $w_\mathrm{cu,cs}(\xi)$ exhibit the same asymptotic behavior.}
\label{fig:lambdaOpp}
\end{figure}

\begin{figure}
\centering
\subfloat[The effective potential $w_+(\xi)$ for $\lambda = 10$, $\beta = 1/2$]{\includegraphics[width=0.9\twofig]{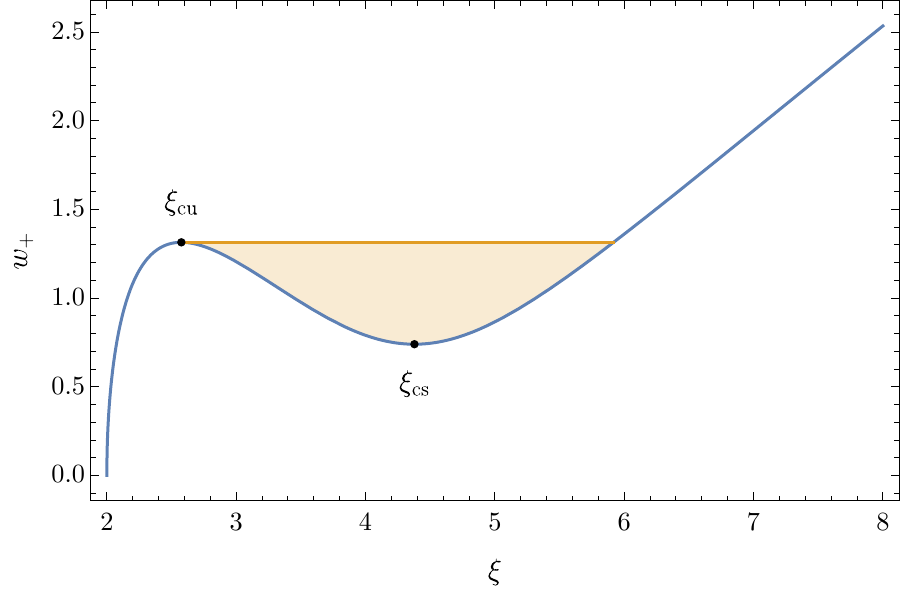}}
\hspace{1em}
\subfloat[The effective potential $w_+(\xi)$ for $\lambda = \lambda_{\mathrm{ISCO}} \approx 5.2$, $\beta = 1/2$]{\includegraphics[width=0.9\twofig]{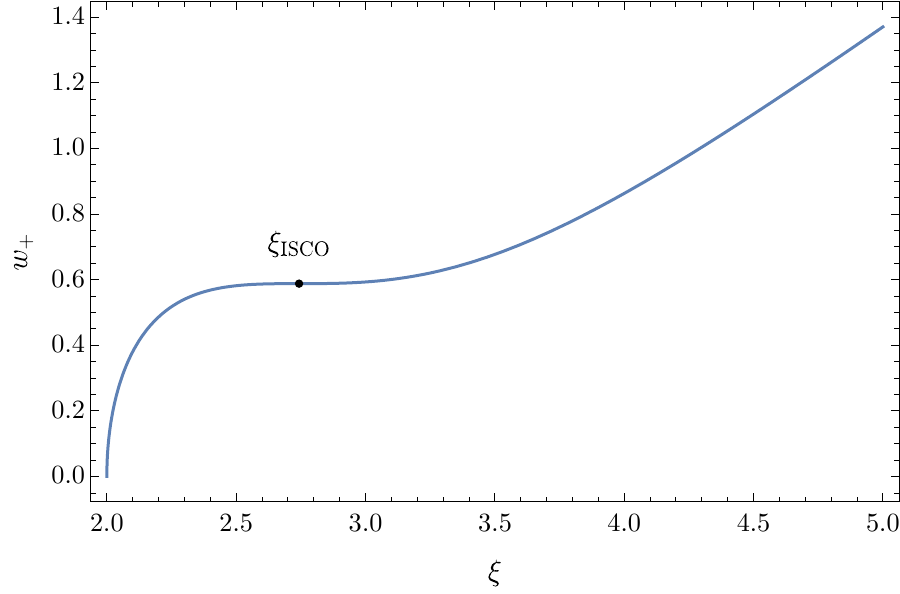}}
                
\caption{Sample plots of the effective potential $w_+(\xi)$ defined in Eq.\ (\ref{eq:effPotWaldSchw}) for $\alpha = 0$. Left panel: $\lambda > \lambda_{\mathrm{ISCO}}$. In this case, there is a maximum at $\xi_\mathrm{cu}$ and a minimum at $\xi_\mathrm{cs}$. The shaded region corresponds to energies $\varepsilon$ of orbits with radii bounded from below and above. Right panel: $\lambda = \lambda_{\mathrm{ISCO}}$. In this case, an inflection point occurs at $\xi_\mathrm{ISCO}$.}
\label{fig:wExample}
\end{figure}

The further analysis can be split into two cases: $\beta > 0$ and $\beta < 0$, respectively. When $\beta < 0$ (and $\lambda \ge 0$), there is no extremum for $\xi \leq 3$. In this case, the radius of the smallest circular orbit approaches $\xi_{\mathrm{tp}} = 3$. Otherwise, it corresponds to the solution to $|\beta| = \mathcal{F}(\xi_{\mathrm{tp}})$, as discussed above. For $\xi > 3$ there exists only one extremum of $w_+(\xi)$, corresponding to $\lambda = \bar\lambda_-(\xi)$, at each point $\xi$. Moreover, $w_+(\xi)$ is always an increasing smooth function of $\lambda$ for fixed $\xi$. We conclude that $w_{\mathrm{cs}}$ (the potential at the circular stable orbit) and $w_{\mathrm{cu}}$ (the potential at the circular unstable orbit) are also increasing functions of $\lambda$.\footnote{This fact follows as a direct consequence of the chain rule. Consider $w_+ = w_+(\xi,\lambda)$ understood as a function of two arguments. Let $\xi_\mathrm{cs} = \xi_\mathrm{cs}(\lambda)$ denote the location of the local minimum of $w_+$ with respect to $\xi$, so that $\partial w_+(\xi_\mathrm{cs} (\lambda),\lambda)/\partial \xi = 0$, and let $w_\mathrm{cs}(\lambda) = w_+(\xi_\mathrm{cs}(\lambda),\lambda)$ be the value of $w_+$ at its local minimum. Since $w_\mathrm{cs}(\lambda)$ is differentiable, one has
\[ \frac{\d w_\mathrm{cs}(\lambda)}{\d \lambda} = \frac{\partial w_+}{\partial \xi} (\xi_\mathrm{cs}(\lambda),\lambda) \frac{\d \xi_\mathrm{cs}}{\d \lambda} + \frac{\partial w_+}{\partial \lambda} (\xi_\mathrm{cs}(\lambda),\lambda) = \frac{\partial w_+}{\partial \lambda} (\xi_\mathrm{cs}(\lambda),\lambda) \ge 0. \]
An analogous reasoning holds for the local maximum.} Furthermore, let us examine the roots of $\mathrm d w_+/\mathrm d \xi = 0$. This condition translates to a polynomial equation of order $5$ in $\xi$: $\tilde P(\xi) = 0$. We are interested in the solutions above the horizon, i.e., for $\xi > 2$, hence we substitute $\xi = y + 2$ to obtain $P(y) = 0$, where $P(y) = \sum_{k=0}^5P_ky^k$. We list the coefficients:
\begin{eqnarray*}
    P_5 = \beta^2, \; P_4 = 9\beta^2, \; P_3 = 32\beta^2, \; P_2 = 1 + 56 \beta^2 - 2 \beta\lambda, \; P_1 = 4 - (\lambda - 4\beta)(\lambda + 12\beta), \; P_0 = 4 + (\lambda - 4\beta)^2.
\end{eqnarray*}
The structure of signs in this sequence is $\{+,+,+, \pm, \pm,+\}$, hence at most $2$ changes of signs can occur: $\{+,+,+,+,-,+\}$, $\{+,+,+,-,+,+\}$, $\{+,+,+,-,-,+\}$. Otherwise, we have no sign change: $\{+,+,+,+,+,+\}$. By Descartes' rule of signs, the polynomial $P(y)$ has either zero or two positive roots, counted with multiplicity. Consequently, the effective potential $w_+(\xi)$ has at most two distinct extrema for $\xi > 2$, with the possibility of a degenerate stationary point corresponding to the ISCO. In this case, the critical value of $\lambda$ for a circular stable $\lambda_{\mathrm{cs}}$ and unstable $\lambda_{\mathrm{cu}}$ are given by a single continuous function $\bar\lambda_{-}(\xi)$ (as long as it remains positive). Hence, the equation $\lambda = \bar\lambda_{-}(\xi)$ must always have exactly $0$ or $2$ solutions: $\xi_{\mathrm{cs}}$ (circular stable orbit) and $\xi_{\mathrm{cu}}$ (circular unstable orbit), with a possible degeneracy at $\xi_{\mathrm{ISCO}} = \xi_{\mathrm{cu}} = \xi_{\mathrm{cs}}$ for $\lambda = \lambda_{\mathrm{ISCO}}$ (innermost stable circular orbit or ISCO, see Fig.\ \ref{fig:wExample}). Moreover, $\bar\lambda_{-}(\xi)$ diverges to $+\infty$ as $\xi \to 3^+$ or $\xi \to +\infty$. Therefore, $\bar\lambda_{-}(\xi)$ is decreasing for $\xi \in (3, \xi_{\mathrm{ISCO}}]$ (corresponding to $\lambda_{\mathrm{cu}}$) and increasing on $\xi \in [\xi_{\mathrm{ISCO}}, +\infty)$ (corresponding to $\lambda_{\mathrm{cs}}$). This behavior is depicted in Fig.~\ref{fig:lambdaOpp}.

\begin{figure}
\centering
\subfloat[$\lambda_{\mathrm{cu},\mathrm{cs}}(\xi)$, $\beta = 1/2$]{\includegraphics[width=0.9\twofig]{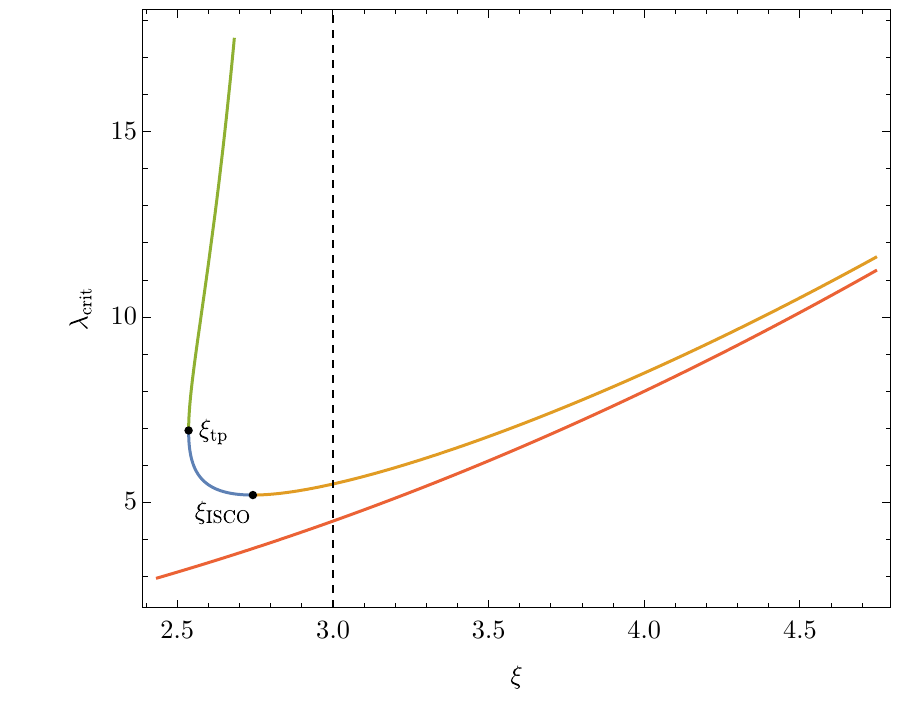}}
\hspace{1em}
\subfloat[$w_{\mathrm{cu},\mathrm{cs}}(\xi)$, $\beta = 1/2$]{\includegraphics[width=0.9\twofig]{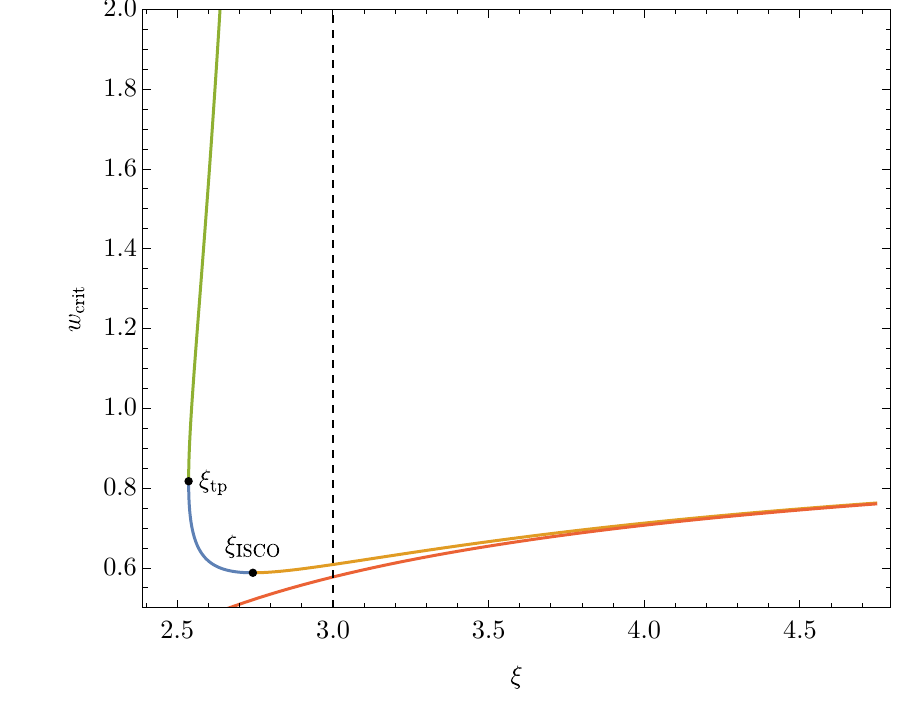}}

\caption{Left panel: Critical values  $\bar\lambda_{\pm}(\xi)$ for $\beta > 0$. ``Lower'' and ``upper'' branches of $\lambda_{\mathrm{cu}}(\xi)$ are shown in blue and green, respectively. The orange line represents the ``lower'' branch $\lambda_{\mathrm{cs}}(\xi)$. The ``upper'' branch $\lambda_{\mathrm{cu}}(\xi)$ diverges as $\xi \to 3^-$. The ``upper'' and ``lower'' branches of $\lambda_\mathrm{cu}(\xi)$ meet at $\xi_\mathrm{tp}$. The ``lower'' branches of $\lambda_\mathrm{cu}(\xi)$ and $\lambda_\mathrm{cs}(\xi)$ join at $\xi_\mathrm{ISCO}$. The red line represents $\lambda_{\mathrm{sc}}(\xi)$. Right panel: The analogous behavior of $w_\mathrm{cu,cs}(\xi)$.}
\label{fig:lambdaSame}
\end{figure}

The case $\beta > 0$ is less obvious, as the critical angular momentum $\bar\lambda_{\pm}(\xi)$ is a multi-valued function for $\xi_{\mathrm{tp}} \leq \xi < 3$, i.e., both branches $\bar \lambda_\pm(\xi)$ remain positive (see Tab.\ \ref{tab:lambdaSign}). In particular, $\bar\lambda_+(\xi)$ is relevant only for $\xi_{\mathrm{tp}} \leq \xi < 3$ and $\bar\lambda_-(\xi)$ for $\xi \geq \xi_{\mathrm{tp}}$. Hence, the latter controls the behavior for large $\xi$. By Equation (\ref{eq:behaviorWithLambdaWald}), the effective potential $w_+(\xi)$ no longer increases with $\lambda$ at a fixed point $\xi$ for $\xi \geq \xi_{\mathrm{sc}} = \sqrt{\lambda/\beta}$. Let $\lambda_{\mathrm{sc}}(\xi) := \beta\xi^2$ be the angular momentum corresponding to this change in monotonicity at a given $\xi$. Notice that $\bar\lambda_- - \lambda_{\mathrm{sc}} \to 0$ as $\xi \to +\infty$.  Moreover, one can check that $\bar\lambda_- > \lambda_{\mathrm{sc}}$,  provided that $\bar\lambda_-,\lambda_{\mathrm{sc}} > 0$. This coincides with $\lambda_{\mathrm{cs}} > \lambda_{\mathrm{sc}}$, therefore $\xi_{\mathrm{sc}} > \xi_{\mathrm{cs}}$, and $w_+(\xi)$ is always an increasing function of $\lambda$ in the neighborhood of $\lambda_{\mathrm{cs},\mathrm{cu}}(\xi)$; therefore $w_{\mathrm{cs},\mathrm{cu}}$ also increases with $\lambda$. By the previous discussion, we still expect $0$ or $2$ extrema of $w_+(\xi)$ (degenerating to one when $\xi_{\mathrm{cs}} = \xi_{\mathrm{cu}} = \xi_{\mathrm{ISCO}}$) although now we have two branches $\bar \lambda_\pm(\xi)$ present in the region $\xi < 3$. Hence, there must exist exactly one ``turning'' point from which they branch out. This turning point corresponds to $\xi_{\mathrm{tp}}$. Notice that $\bar\lambda_+ \to +\infty$ as $\xi \to 3^-$. Hence, we can recognize $\bar \lambda_+(\xi)$ as the ``upper'' $\lambda_{\mathrm{cu}}$ branch. On the other hand, $\bar\lambda_-(\xi)$ remains continuous on $\xi_{\mathrm{tp}} \leq \xi$, hence it can be identified as the joint ``lower'' $\lambda_{\mathrm{cu},\mathrm{cs}}$ branch. By reasoning similar to the previous case, we conclude that the ``upper'' branch increases with $\xi$ and the ``lower'' branch first decreases for $\xi_{\mathrm{tp}} \leq \xi \leq \xi_{\mathrm{ISCO}}$ (corresponding to $\lambda_{\mathrm{cu}}$) and then increases for $\xi_{\mathrm{ISCO}} \leq \xi$ (corresponding to $\lambda_{\mathrm{cs}}$). This is depicted in Fig.\ \ref{fig:lambdaSame}.

We are now ready to discuss the options available in constructing disks composed of particles with varying $\varepsilon$ and $\lambda$. Let us first consider the available range of $\lambda$ corresponding to a given (fixed) energy $\varepsilon = \varepsilon_0$. We choose $\varepsilon_\mathrm{ISCO} < \varepsilon_0 < \varepsilon_\mathrm{max}$, where $\varepsilon_{\mathrm{ISCO}} := w_+(\xi_{\mathrm{ISCO}})|_{\lambda=\lambda_{\mathrm{ISCO}}}$ is the energy corresponding to the ISCO, and $\varepsilon_\mathrm{max}$ is some upper bound. Denote by $\xi_1(\varepsilon_0)$ the (unique) solution of $w_+(\xi)|_{\lambda = \lambda_{\mathrm{cu}}(\xi)} = \varepsilon_0$ with respect to $\xi$. It corresponds to the inner boundary of a ``monoenergetic'' disk composed of particles with the energy $\varepsilon_0$. Trajectories characterized by $\lambda < \lambda_\mathrm{cu}(\xi_1(\varepsilon_0))$ would plunge into the black hole. Trajectories with $\lambda > \lambda_\mathrm{cu}(\xi_1(\varepsilon_0))$ are allowed, but they encounter a radial turning point for $\xi > \xi_1(\varepsilon_0)$. Let $\xi_2(\varepsilon_0)$ denote the other solution of $w_+(\xi)|_{\lambda=\lambda_{\mathrm{cu}}(\xi_1)} = \varepsilon_0$. If $\beta < 0$, the radius $\xi_2(\varepsilon_0)$ coincides with the outer radius of the ``monoenergetic'' disk of energy $\varepsilon_0$. In this case, all trajectories with $\lambda > \lambda_\mathrm{cu}(\xi_1(\varepsilon_0))$ encounter two turning points in the range $(\xi_1(\varepsilon_0),\xi_2(\varepsilon_0))$.

The situation for $\beta > 0$ is more involved, as the coordinate of the outer radial turning point can increase with $\lambda$. In particular, if $\xi_\mathrm{sc}(\lambda_\mathrm{cu}(\xi_1)) < \xi_2(\varepsilon_0)$, the outer boundary of the disk would be reached by particles characterized by the angular momentum larger than $\lambda_\mathrm{cu}(\xi_1)$. Note that the value of $w_+$ at $\xi = \xi_\mathrm{sc} = \sqrt{\lambda/\beta}$ is given by
\begin{equation}
\label{esc}
   \varepsilon_\mathrm{sc}(\xi) := w_+(\xi)|_{\lambda = \xi^2 \beta} = \sqrt{1 - \frac{2}{\xi}},
\end{equation}
and, remarkably, it does not depend on $\beta$. The solution of the equation $w_+(\xi)|_{\lambda = \xi^2 \beta} = \varepsilon_0$ reads
\begin{equation}
\label{wsc}
    \xi = \xi_\mathrm{sc}(\varepsilon_0) := \frac{2}{1 - \varepsilon_0^2}.
\end{equation}
Also note that
\begin{equation}
    \left. \frac{\partial w_+(\xi)}{\partial \xi} \right|_{\lambda = \xi^2 \beta} = \frac{1}{\xi^2} \sqrt{\frac{\xi}{\xi - 2}} = \frac{\d \varepsilon_\mathrm{sc}(\xi)}{\d\xi}.
\end{equation}
Consequently, the graphs of $w_+(\xi)$ and $\varepsilon_\mathrm{sc}(\xi)$ are tangent to each other, at $\xi = \xi_\mathrm{sc}$.
As long as $0 < \varepsilon_0 < 1$, $\xi_\mathrm{sc}(\varepsilon_0)$ coincides with the outer radius of a ``monoenergetic'' disk of the energy $\varepsilon_0$. On the other hand, for $\xi > 2$ the right-hand side of Eq.\ (\ref{esc}) is a strictly growing function of $\xi$ and tends to $1$ for $\xi \to \infty$. Accordingly, for a fixed energy $\varepsilon_0 > 1$ there exist bound trajectories reaching arbitrarily large radii. Hence, we define $\xi_{\mathrm{sc}}(\varepsilon_0) = \infty$ for $\varepsilon_0 \geq 1$. As a consequence, if we pick the cut-off energy of particles $\varepsilon_{\mathrm{max}} \geq 1$, we expect the disk to be infinite and finite, otherwise.

We conclude that for the fixed energy $\varepsilon=\varepsilon_0$, admissible values of $\lambda$ at a given point $\xi$ are given by $\lambda \in [\lambda_{\mathrm{min}}(\xi,\varepsilon_0), \lambda_{\mathrm{max}}(\xi, \varepsilon_0)]$, where:
\begin{enumerate}
            \item $\lambda_{\mathrm{max}}(\xi, \varepsilon_0)$ is the larger $\lambda$ for which $w_+(\xi) = \varepsilon_0$ (in fact, equal to $\lambda_+(\xi, \varepsilon_0)$),
            \item $\lambda_{\mathrm{min}}(\xi, \varepsilon_0) := \lambda_{\mathrm{cu}}(\xi_1(\varepsilon_0))$ for $\beta < 0$, $\xi_1(\varepsilon_0) \leq \xi < \xi_2(\varepsilon_0)$, and
            \begin{equation*}
                \lambda_{\mathrm{min}}(\xi, \varepsilon_0) := \left\{\begin{matrix}
                    \lambda_{\mathrm{cu}}(\xi_1(\varepsilon_0)), & \text{if} & \xi_1(\varepsilon_0) \leq \xi < \xi_2(\varepsilon_0),\\
                    \text{smaller sol. to $w_+(\xi) = \varepsilon_0$ (in fact, $\lambda_-(\xi, \varepsilon_0)$)}, & \text{if} & \xi_2(\varepsilon_0) \leq \xi \leq \xi_{\mathrm{sc}}(\varepsilon_0)
                \end{matrix}\right.
            \end{equation*}
            for $\beta > 0$.
\end{enumerate}
We emphasize that these bounds were obtained by rejecting the trajectories that plunge into the black hole.

We will now specify the limits of the allowed energy $\varepsilon$. For $2 \leq \xi < \xi_{\mathrm{ISCO}}$, the minimal energy of the orbits that do not plunge into the black hole is given by $w_{\mathrm{cu}}(\xi)$. For the points located at $\xi \geq \xi_{\mathrm{ISCO}}$ and $\beta < 0$ one still has to ensure that there is a centrifugal barrier at some radius $\xi'$ smaller than $\xi$. This means that $w_+(\xi)|_{\lambda = \lambda_{\mathrm{cu}}(\xi')} = w_{\mathrm{cu}}(\xi')$ for some $\xi' < \xi_{\mathrm{ISCO}}$. The minimum energy corresponds to the smallest value $w_\mathrm{cu}(\xi')$ that satisfies this condition. We shall denote $\tilde w_{\mathrm{cu}}(\xi) := w_{\mathrm{cu}}(\xi')$ and the maximal outer radius as $\xi = \xi_{\mathrm{o}}$, satisfying $\varepsilon_{\mathrm{max}} = \tilde w_{\mathrm{cu}}(\xi_{\mathrm{o}})$. On the other hand, for $\beta > 0$, $w_+(\xi)$ may decrease with increasing $\lambda$. Consequently, $\varepsilon_{\mathrm{sc}}(\xi)$ (the minimal energy allowed at $\xi$) may become smaller than $\tilde w_{\mathrm{cu}}(\xi)$. We denote the radius at which this occurs as $\xi = \tilde\xi$: $\varepsilon_{\mathrm{sc}}(\tilde\xi) = \tilde w_{\mathrm{cu}}(\tilde\xi)$. For $\xi > \tilde \xi$ the minimum energy allowed is given by $\varepsilon_\mathrm{sc}(\xi)$. Furthermore, for $\varepsilon_{\mathrm{max}} > w_{\mathrm{cu}}(\xi_\mathrm{tp})$, i.e., when one allows for trajectories with the energy higher than the energy of the orbit reaching the turning point at $\xi_\mathrm{tp}$ (cf.\ Fig.\ \ref{fig:lambdaSame}), one also needs to modify the upper bound of integration with respect to the energy. This is because for $\xi < \xi_{1}(\varepsilon_{\mathrm{max}})$, as $\lambda$ increases, the location of the maximum of $w_+(\xi)$ eventually starts to move away from the black hole and crosses $\xi$ before the value of $w_+(\xi)$ at the maximum reaches $\varepsilon_{\mathrm{max}}$. Consequently, the trajectories reaching $\xi$ with the energies $w^{\mathrm{(upper)}}_{\mathrm{cu}}(\xi) < \varepsilon \leq \varepsilon_{\mathrm{max}}$ plunge into black hole and should be excluded (we now choose specifically the ``upper'' branch of $w_{\mathrm{cu}}(\xi)$, which we denote as $w^{\mathrm{(upper)}}_{\mathrm{cu}}(\xi)$). Note that if $\varepsilon_{\mathrm{max}} > w_{\mathrm{cu}}(\xi_{\mathrm{tp}})$, the inner radius of the disk will shift to $\xi_{\mathrm{tp}}$, instead of $\xi_1(\varepsilon_{\mathrm{max}})$. In conclusion, the allowed energies satisfy $\varepsilon\in[\varepsilon_{\min}(\xi),\varepsilon_{\max}(\xi)]$, where, for $\beta>0$,
\begin{equation}
\varepsilon_{\min}(\xi)=
\begin{cases}
w_{\mathrm{cu}}(\xi),
&\xi < \xi_{\mathrm{ISCO}}, \\
\tilde w_{\mathrm{cu}}(\xi),
&\xi_{\mathrm{ISCO}} \leq \xi < \tilde\xi, \\
\varepsilon_{\mathrm{sc}}(\xi),
& \xi \geq \tilde\xi,
\end{cases}
\end{equation}
and, for \(\beta<0\),
\begin{equation}
\varepsilon_{\min}(\xi)=
\begin{cases}
w_{\mathrm{cu}}(\xi),
&\xi<\xi_{\mathrm{ISCO}},\\
\tilde w_{\mathrm{cu}}(\xi),
&\xi\geq\xi_{\mathrm{ISCO}},
\end{cases}
\end{equation}
In the particular scenario with $\beta > 0$ and $\varepsilon_{\mathrm{max}} > w_{\mathrm{cu}}(\xi_\mathrm{tp})$, the upper bound is
\begin{equation}
\varepsilon_{\max}(\xi)=
\begin{cases}
w^{\mathrm{(upper)}}_{\mathrm{cu}}(\xi),
&\xi < \xi_1(\varepsilon_{\mathrm{max}}), \\
\varepsilon_{\max},
& \xi \geq \xi_1(\varepsilon_{\max}),
\end{cases}
\end{equation}
and simply $\varepsilon_{\max}(\xi) = \varepsilon_{\max}$, otherwise. These expressions are understood to apply only where the corresponding branches exist and where $\varepsilon_{\min}(\xi)\leq\varepsilon_{\max}$. In particular, the radial support of the disk depends on the chosen energy cutoff. Its inner boundary is determined by $w_{\mathrm{cu}}(\xi_{\mathrm{in}})=\varepsilon_{\max}$, or by $\xi_{\mathrm{in}} = \xi_{\mathrm{tp}}$ when $\beta > 0$ and $\varepsilon_{\mathrm{max}} > w_{\mathrm{cu}}(\xi_\mathrm{tp})$. For $\beta>0$, let $\varepsilon_*=\varepsilon_{\mathrm{sc}}(\tilde\xi)$. If $\varepsilon_{\mathrm{ISCO}}<\varepsilon_\mathrm{max}<\varepsilon_\ast$, the disk terminates before reaching $\widetilde\xi$, and its outer boundary is determined by $\tilde w_{\mathrm{cu}}(\xi_{\mathrm{out}})=\varepsilon_\mathrm{max}$. If $\varepsilon_\ast \leq \varepsilon_{\max} < 1$, the outer boundary is instead given by $\xi_{\mathrm{out}}=\xi_{\mathrm{sc}}(\varepsilon_\mathrm{max})$, as defined in Eq.\ \eqref{wsc}. For $\varepsilon_\mathrm{max} \geq 1$, the radial support is unbounded. For $\beta<0$, the outer boundary is $\xi_\mathrm{o}$, defined above.

\end{document}